\documentclass[11pt]{article}

\usepackage{graphicx}
\usepackage{subcaption}
\usepackage[export]{adjustbox}
\usepackage{amsmath}
\usepackage{amsfonts}
\usepackage{mathdots}
\usepackage{amssymb}
\usepackage{physics}
\usepackage{microtype}
\usepackage{bbold}
\usepackage{dsfont}
\usepackage{mathtools}
\usepackage{tikz}
\usepackage{simpler-wick}
\usepackage{comment}
\usepackage{cancel}
\usepackage{cite} 
\usepackage{enumitem}
\usepackage{xcolor}
\usepackage{mdframed}
\usepackage{xpatch}
\usepackage{etoolbox}
\usepackage[normalem]{ulem}
\usepackage[colorlinks=true, linkcolor=blue, filecolor=blue, urlcolor=black, citecolor=blue]{hyperref}

\newcommand{\knl}[3]{\mathbb{#1}_{\{#2\};\{#3\}}}

\allowdisplaybreaks
\numberwithin{equation}{section}

\newcommand{\mean}[1]{\overline{#1}}
\newcommand{\dash}{\hspace{0.2mm}'}

\newcommand\mtiny[1]{\mbox{\tiny\ensuremath{#1}}}
\newcommand\msmall[1]{\mbox{\small\ensuremath{#1}}}

\newcommand{\fker}[6]{
\mathbb{F}_{#1 #2}{\mtiny{\arraycolsep=0.2\arraycolsep\ensuremath{\begin{bmatrix}#4 & #3 \\ #5 & #6\end{bmatrix}}}}
}
\newcommand{\fkerbig}[6]{
\mathbb{F}_{#1 #2}{\msmall{\arraycolsep=0.2\arraycolsep\ensuremath{\begin{bmatrix}#4 & #3 \\ #5 & #6\end{bmatrix}}}}
}
\newcommand{\sker}[3]{
\mathbb{S}_{#1 #2}[#3]
}
\newcommand{\clo}{\mathcal{O}}

\newcommand{\bbf}{\mathbb{F}}
\newcommand{\bbs}{\mathbb{S}}

\newcommand{\e}{\textnormal{e}}

\newcommand{\bbi}{\mathbb{1}}

\newcommand{\bfP}{\mathbf{P}}
\newcommand{\bfp}{\mathbf{P}}
\newcommand{\MCG}{\mathrm{MCG}}
\definecolor{ForestGreen}{RGB}{34,139,34}

\renewcommand{\order}[1]{O\left(#1\right)}

\newcommand{\mlab}[1]{\;\big\lvert_{\text{#1}}}

\begin{document}
\thispagestyle{empty}

\phantom{*}

\vskip30mm

\begin{center}

{\huge Multiboundary wormholes and OPE statistics}

\vskip10mm

\normalsize
Jan de Boer, Diego Liska, Boris Post
\vskip 0.75cm
\small{\textit{Institute for Theoretical Physics,
University of Amsterdam,}}
\vskip -.25cm
\small{\textit{ PO Box 94485, 1090 GL Amsterdam, The Netherlands}}
\end{center}

\vspace{15mm}

\begin{abstract}
\normalsize
\noindent We derive higher moments in the statistical distribution of OPE coefficients in holographic 2D CFTs, and show that such moments correspond to multiboundary Euclidean wormholes in pure 3D gravity. The $n$th cyclic non-Gaussian contraction of heavy-heavy-light OPE coefficients follows from crossing symmetry of the thermal $n$-point function. We derive universal expressions for the cubic and quartic moments and demonstrate that their scaling with the microcanonical entropy agrees with a generalization of the Eigenstate Thermalization Hypothesis. Motivated by this result, we conjecture that the full statistical ensemble of OPE data is fixed by three premises: typicality, crossing symmetry and modular invariance. Together, these properties give predictions for non-factorizing observables, such as the generalized spectral form factor. Using the Virasoro TQFT, we match these connected averages to new on-shell wormhole topologies with multiple boundary components. Lastly, we study and clarify examples where the statistics of heavy operators are not universal and depend on the light operator spectrum. We give a gravitational interpretation to these corrections in terms of Wilson loops winding around non-trivial cycles in the bulk.

\end{abstract}

\newpage

\phantom{a}
\vspace{-4em}
{
\hypersetup{linkcolor=black}
\tableofcontents
}

\newpage

\section{Introduction}

Recent developments in low-dimensional models of gravity have established a connection between gravity and a `coarse-grained' version of the holographic correspondence.
This connection is especially clear in two dimensions, where quantum gravity is equivalent to random matrix theory (RMT) \cite{Saad:2019lba}. Insights into the statistical description of gravity largely arise from the study of Euclidean geometries with disconnected boundaries that are connected through the bulk. These configurations are 
known as Euclidean wormholes. 
Euclidean wormholes provide coarse-grained non-perturbative information about the black hole microstates of the dual system. However, they also lead to the so-called `factorization puzzle' \cite{Maldacena:2004rf}, as they hinder the expected factorization of multi-boundary quantities. 
One resolution of the factorization puzzle is to interpret semiclassical gravity as capturing only a statistical description of the true microscopic observables. Several attempts to construct such  descriptions include averaging over Hamiltonians
\cite{Stanford:2019vob,
Turiaci:2023jfa,
Witten:2020wvy,Altland:2020ccq,Altland:2022xqx,
Collier:2023cyw}, 
operators
\cite{Altland:2021rqn,
Belin:2020hea,
Chandra:2022bqq,
Collier:2023fwi,Collier:2024mgv,
Sasieta:2022ksu,
Balasubramanian:2022lnw2}, 
states
\cite{Freivogel:2021ivu,
Chandra:2022fwi,Chandra:2023rhx,
Chandra:2023dgq}, 
or any combination thereof 
\cite{Belin:2023efa,
Jafferis:2022wez,
Jafferis:2022uhu,deBoer:2023vsm}. 

The success of averaging and coarse-graining in quantum gravity stems from the fact that black holes exhibit a strong form of chaos, most clearly quantified by thermal out-of-time-order correlation functions \cite{Maldacena:2015waa}. Quantum chaotic systems exhibit a large degree of universality because they can be modeled by random matrices in sufficiently narrow energy bands. 
RMT allows one to make important statements about the spectrum and observables of 
these systems. It explains the observed eigenvalue repulsion, it underpins the assumption that eigenvectors are approximately random unit vectors, a concept known as \emph{typicality}, and it proposes that the matrix elements of certain operators can be treated as 
pseudorandom variables.  This last statement ultimately leads to the Eigenstate Thermalization Hypothesis (ETH), which is our current framework for understanding the emergence of thermal equilibrium in isolated quantum systems \cite{PhysRevE.50.888,PhysRevA.43.2046,Rigol:2007juv}. 

In this paper, we will connect the ETH directly to a bulk gravitational theory, using AdS/CFT. Specifically, we study Einstein gravity in three dimensions, coupled to massive point particles, which thanks to its exact solvability \cite{WITTEN198846} provides an excellent playground to test the ETH in the dual CFT$_{2}$. In its most familiar form, the ETH is an ansatz for the matrix elements of an operator $\clo$ in the energy basis of the Hamiltonian. It states that if  $\clo$ is a simple operator, then \vspace{2mm}
\begin{equation}\label{eq:intro_ETH}
    \bra{E_i}\clo\ket{E_j} =  g_1(\bar E) \delta_{ij} + g_2(\bar E,\omega)^{1/2} R_{ij},\vspace{2mm}
\end{equation}
where $g_1(\bar E)$ and $g_2(\bar E, \omega)$ are smooth functions of the mean energy $\bar E = (E_i+E_j)/2$  and energy difference $\omega=E_i-E_j$. The ansatz moreover asserts the exponential scaling of the function $g_2(\bar E,0) \sim \e^{-S(\bar E)}$, where $S(E)= \log \rho_0(E)$ is the coarse-grained microcanonical entropy. The $R_{ij}$ are pseudorandom variables whose variance has been normalized to one. 

Ordinarily, the statistics of $R_{ij}$ are taken to be Gaussian. Nonetheless, it is well-known that this can only be an approximation, as non-Gaussian contractions are needed to correctly reproduce the late-time behaviour of out-of-time-ordered correlators \cite{Foini:2018sdb,Murthy_2019,Jackson_2015,Mertens:2017mtv}. Using an argument based on typicality, it was shown in \cite{Foini:2018sdb} that the higher statistical moments of the matrix elements of $\clo$, for a cyclic pattern of index contractions, should scale as
\begin{equation}\label{eq:cyclicscaling}
    \mean{\clo_{12}\clo_{23}\dots\clo_{n1}} \coloneqq g_n(\bar E,\omega_1,\dots,\omega_{n-1})\sim \e^{-(n-1)S(\bar E)}.
\end{equation}
Here and in the rest of the paper, we use the simplified labels $1,2, \dots, n$ to denote the indices $i_1,i_2\dots i_n$ when writing matrix elements $\clo_{12} = \langle E_1 |\clo|E_2\rangle$. The indices in \eqref{eq:cyclicscaling} are all taken to be different, and $\bar E$ corresponds to the mean of the energies $E_1,\dots,E_n$. The version of the ETH that includes these connected cyclic moments is known as the \emph{generalized ETH}. 

The goal of this paper is to show that the generalized ETH, with $n\geq 2$, universally holds true for irrational 2D CFTs with a sparse light spectrum, a large gap above the vacuum, and only Virasoro symmetry, as long as the energies $E_i$ are sufficiently large. The effective statistical description of such theories is provided by 3D semiclassical gravity, which can be solved exactly, topology-by-topology, if we restrict to on-shell solutions of Einstein's equations \cite{Collier:2023fwi}.
Moreover, we aim to derive explicit expressions for the smooth functions $g_n$ in \eqref{eq:cyclicscaling}, using the conformal bootstrap.

To achieve this goal, we use the state-operator correspondence to reformulate the generalized ETH for CFT$_2$'s as a statement about the statistical distribution of OPE coefficients,\vspace{1mm}
\begin{equation}
\label{eq:def1}
   C_{1\clo 2} =  \bra{E_1}\clo(1)\ket{E_2} = \bra{0} \clo_1(\infty)\clo(1) \clo_2(0) \ket{0}.\vspace{1mm}
\end{equation}
In \cite{Cardy:2017qhl}, a universal asymptotic formula was derived for the product of two OPE coefficients, by generalizing Cardy's argument for the asymptotic density of states \cite{Cardy:1986ie}. Subsequently, in \cite{Collier:2019weq}, an improved formula for the OPE variance was derived from Virasoro crossing kernels. Taking this variance as an input, and assuming Gaussian statistics, \cite{Belin:2020hea} and \cite{Chandra:2022bqq} were able to match the predictions from the OPE ensemble to various Euclidean wormholes in 3D gravity. The analysis of non-Gaussian statistics was done in \cite{Belin:2021ryy,Anous:2021caj}, for which a partial interpretation in gravity was given recently in \cite{Collier:2024mgv}. Earlier studies on thermalization in CFT$_2$ include \cite{Lashkari:2016vgj,Turiaci_2016,Hikida_2018,Basu_2017,Kraus_2017}. 

In order to connect the statistical ensemble of OPE coefficients to the generalized ETH, we will study a class of higher statistical moments with the following cyclic index structure\vspace{2mm}
\begin{equation}\label{eq:Intro_Cyclic}
    \mean{C_{1\clo 2}C_{2\clo 3}\cdots C_{n\clo 1} 
    }\vspace{2mm},
\end{equation}
for a fixed external operator $\clo$ and heavy operators $\clo_1,\dots,\clo_n$. These moments are accessible from $n$-point correlation functions on the torus. Owing to 
the presence of Virasoro symmetry, 2D CFTs offer a rare opportunity to derive analytic formula's for these moments in an appropriate high-energy regime. The main tool to do so is the infinite set of consistency conditions imposed on any CFT$_2$, known as \emph{crossing symmetry}, as well as the associated crossing kernels. 

Diagrammatically, the sequence of crossing moves that relates the cyclic OPE contraction \eqref{eq:Intro_Cyclic} to a channel in which the identity module dominates at high energies, is given by
\begin{equation}\label{eq:intro_crossingmoves}
\vcenter{
\hbox{
\begin{tikzpicture}
\node at (0,0) {\includegraphics{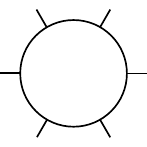}};
\node at (0,-1.1) {$\cdots$};
\end{tikzpicture}
}
}
\stackrel{\bbf} {\longrightarrow}
\vcenter{\hbox{
\begin{tikzpicture}
\node at (0,0) {\includegraphics{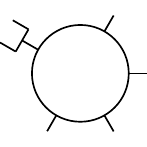}};
\node at (0.15,-1.1) {$\cdots$};
\end{tikzpicture}
}}
\stackrel{(\bbf)^{n-2}} {\relbar\!\!\relbar\!\!\longrightarrow}
\vcenter{\hbox{
\begin{tikzpicture}
\node at (0,0) {\includegraphics{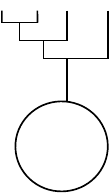}};
\node at (-0.8,1) {$\ddots$};
\end{tikzpicture}
}}
\stackrel{\bbs} {\longrightarrow}
\vcenter{\hbox{
\begin{tikzpicture}
\node at (0,0) {\includegraphics{img/OPE2.pdf}};
\node at (-0.8,1) {$\ddots$};
\end{tikzpicture}
}}\,.
\end{equation}
Here $\mathbb{F}$ and $\mathbb{S}$ denote the \emph{fusion kernel} and \emph{modular S-transform}, respectively. We will now describe the result of this procedure, and its implications for 3D gravity.

\subsection{Summary of the results}

In Section \ref{sec:typicality}, we explain how the assumption of typicality, when applied to CFT$_2$, fixes the allowed index contractions for any statistical model of OPE coefficients. In particular, we take into account the symmetries imposed by index permutations and reality conditions of OPE coefficients. Besides the index contractions, we also want to determine analytic results for the smooth functions $g_n$ in \eqref{eq:cyclicscaling}. To do so, we assume that the OPE ensemble is approximately crossing symmetric, which automatically forces the statistics to be non-Gaussian.\footnote{A recent investigation of approximate CFTs can be found in \cite{Belin:2023efa}. In the discussion section, we comment on the matrix-tensor model proposed there in more detail.} 

From the crossing equations induced by the crossing moves \eqref{eq:intro_crossingmoves}, we derive new universal asymptotic formulas for a simple class of cyclic index contractions, \eqref{eq:Intro_Cyclic}. In particular, we derive the cubic and quartic moments
\begin{align}
\label{eq:introthree}
  \mean{C_{1\clo 2}C_{2\clo 3} C_{3\clo 1}
    }\mlab{cyclic} &= C_{\clo\clo\clo} \left|
    C_0(P_1,P_3,P_{\clo})
    \frac{
    \fker{P_{\clo}}{P_2}{P_{\clo}}{P_1}{P_3}{P_{\clo}}
    }{\rho_0(P_2)}
    \right|^2 + \order{\e^{-4\pi (\alpha_\chi P + \bar \alpha_\chi \bar P)}}\\[1em]
\label{eq:introfour}
    \mean{C_{1\clo 2}C_{2\clo 3} C_{3\clo 4}C_{4\clo 1}
    }\mlab{cyclic} &=\! (-1)^{\sum_{i=1}^4J_i} \!\left|
    C_0(P_4,P_1,P_{\clo})
    C_0(P_4,P_3,P_{\clo})
    \frac{\fker{P_4}{P_2}{P_{\clo}}{P_3}{P_{\clo}}{P_1}}
    {\rho_0(P_2)}
    \right|^2 \!+\! \cdots
\end{align}
where the corrections to the second formula \eqref{eq:introfour} are the same order as the corrections to the first, \eqref{eq:introthree}. In other words, the corrections are exponentially suppressed at high energies by a parameter $\alpha_\chi$ that depends on the lightest primary field above the vacuum. These equations hold in the heavy limit $P\rightarrow\infty$, where the energy differences $P_i-P_j$ and the dimension of the external operator $P_\clo$ is kept fixed.

These moments are functions of the conformal weights $h_{1,2,3,4}$ and their antiholomorphic counterparts, parametrized in terms of the `Liouville momentum' $P^2 =h - \frac{c-1}{24}$. The notation $|\bullet|^2$ means a product of holomorphic and antiholomorphic contributions. The functions appearing in these formulas are all related to the two basic crossing kernels $\mathbb{F}$ and $\mathbb{S}$, which are the same for any irrational CFT with only Virasoro symmetry, and which are known in closed form \cite{Ponsot:1999uf,Ponsot:2000mt,Eberhardt:2023mrq}. The only non-universal datum is the OPE coefficient $C_{\clo\clo\clo}$, but this should be viewed as part of the input  together with $h_\clo$.\footnote{In particular, $C_{\clo\clo\clo}$ could be zero. A more general formula proportional to $C_{\clo_1\clo_2\clo_3}$ is given in Eq.\eqref{eq:cyclic33}.}

Importantly, having derived explicit expressions for $g_3$ and $g_4$ allows us to check that the entropic suppression indeed matches with the generalized ETH, \eqref{eq:cyclicscaling}, for $n=3,4$. The microcanonical entropy depends on the mean energy (conformal dimension) via the usual Cardy formula $S(\Delta) \approx \sqrt{\frac{c}{3}\Delta}$. We end Section \ref{sec:typicality} with a general argument showing that the large-$\Delta$ scaling of the $n$th cyclic contraction has the expected entropic suppression for any $n\geq 2$,\footnote{We exclude $n=1$, since for CFT$_2$ the average of a single OPE coefficient is already exponentially small \cite{Kraus_2017,Collier:2019weq}.}\vspace{2mm} 
\begin{equation}
    \mean{C_{1\clo 2}C_{2\clo 3}\cdots C_{n\clo 1}
    }\mlab{cyclic} \sim \e^{-(n-1)S(\Delta)}.
\end{equation}

The cyclic OPE contractions predict connected contributions to the average over a product of CFT observables. As an example, we study the product of one-point functions on the torus. In Section \ref{sec:gravity} we show that these connected averages can be matched holographically to Euclidean wormholes with multiple asymptotic boundaries. We construct the new wormhole topologies with three and four boundary components and compute their exact gravitational partition functions. Using the recently developed Virasoro TQFT \cite{Collier:2023fwi}, we find perfect agreement between the bulk and boundary computations. For example,
\begin{align}
C_{\clo\clo\clo}\,\overline{\langle\clo\rangle_{\tau_1,\bar\tau_1}\langle\clo\rangle_{\tau_2,\bar\tau_2}\langle\clo\rangle_{\tau_3,\bar\tau_3}} \,= \,Z_{\mathrm{grav}}\!\!\left(
\!\!\!
\vcenter{\hbox{\begin{tikzpicture}
\node at (0,0) {\includegraphics[width=2.5cm]{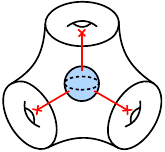}};
\end{tikzpicture}}}
\!\!\!\right),
\end{align}
where the left-hand side is a product of torus one-point functions,\footnote{The factor $C_{\clo\clo\clo}$ is an overall normalization, and in particular is not being averaged over.} with modular parameters $\tau_{1,2,3}$, averaged using the cubic moment \eqref{eq:introthree}. 
The right-hand side represents a wormhole with a boundary consisting of three punctured tori, and one bulk `vertex' or junction where three Wilson lines meet. The explicit bulk topology is constructed as a so-called \emph{Heegaard splitting} in Section \ref{sec:onepointWH}.

An analogous computation using the quartic moment \eqref{eq:introfour} gives a connected average of the product of four torus one-point functions, which matches to a four-boundary wormhole with a pair of bulk Wilson lines,
\begin{equation}
  \overline{\langle\clo\rangle_{\tau_1,\bar\tau_1}\langle\clo\rangle_{\tau_2,\bar\tau_2}\langle\clo\rangle_{\tau_3,\bar\tau_3}\langle\clo\rangle_{\tau_4,\bar\tau_4}} \,= \, Z_{\mathrm{grav}}\!\!\left(
\!\!\!
\vcenter{\hbox{\begin{tikzpicture}
\node at (0,0) {\includegraphics[width=2.5cm]{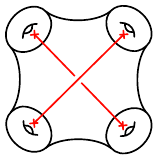}};
\end{tikzpicture}}}
\!\!\!\right).
\end{equation}
This bulk topology is constructed in Section \ref{sec:4bdyWH}. We would like to stress that the gravitational partition function on the right includes \emph{all perturbative orders} in $1/G_N$. The Virasoro TQFT automatically includes these loop corrections, and gives a versatile way to do the bulk computations without having to construct an explicit metric. Further subleading bulk saddles only give non-perturbative corrections in $G_N$. The left-hand side is computed using the quartic moment \eqref{eq:introfour}. In a CFT with a gap of $\order{c}$, the corrections to this result are also non-perturbative.

So far, we have discussed the generalized ETH, which is a statement about `heavy-light-heavy' OPE coefficients $C_{1\clo 2}$. The \emph{OPE randomness hypothesis} \cite{Belin:2020hea} is a generalization of the statistical description, that also includes `heavy-heavy-heavy' OPE coefficients $C_{123}$, where all conformal weights are above the black hole threshold. Their distribution is necessarily non-Gaussian, to ensure crossing symmetry of higher genus partition functions. We focus on the known non-Gaussianity, derived from a genus-three crossing equation, which is related to the Virasoro $6j$ symbol via
\begin{align}\label{eq:moments}
\overline{C_{123}C_{156}C_{264}C_{345}}  \,\big\vert_{6j} &=   \sqrt{\mathsf{C}_{123}\mathsf{C}_{156}\mathsf{C}_{264}\mathsf{C}_{345}} \,\begin{Bmatrix}
        1 & 2 & 3 \\ 4 & 5 & 6
    \end{Bmatrix}_{6j}\,,
\end{align}
where $\mathsf{C}_{123}$ is a shorthand notation for $\big|C_0(P_1,P_2,P_3)\big|^2$, and the function in brackets corresponds to the Virasoro $6j$ symbol. The novel contribution of this paper is to identify and compute the new Euclidean wormhole in 3D gravity that arises from the above contraction. Specifically, we compute the average of the product of two genus-two partition functions, using the contraction \eqref{eq:moments}, finding 
\begin{equation}
\label{eq:gentwointro}
\begin{split}
     \overline{Z_{g=2}(\Omega_1,\bar\Omega_1)Z_{g=2}(\Omega_2,\bar\Omega_2)} \,\big \vert_{6j} \,= \, 
Z_{\mathrm{grav}}\!\!\left(
\!\!\!
\vcenter{\hbox{\begin{tikzpicture}
\node at (0,0) {\includegraphics{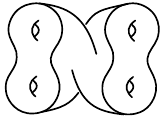}};
\node at (0,0) {$\gamma$};
\end{tikzpicture}}}
\!\!\!\right).
\end{split}
\end{equation}
On the right-hand side we drew a cartoon of the bulk wormhole geometry that contributes to this average. Its full topology is described in Section \ref{sec:genus2}. This manifold has two asymptotic genus-two boundaries, and the bulk consists of a pair of genus-two handlebodies glued together by an element $\gamma$ from the Moore-Seiberg group. Up to a phase, $\gamma$ is simply the fusion move $\mathbb{F}$, so we can think of the above wormhole as interpolating between the `dumbbell' and the `sunset' conformal block. Interestingly, we find that this non-Gaussian wormhole is of the same order as the Maldacena-Maoz wormhole \cite{Chandra:2022bqq} that describes the Gaussian OPE contraction.

Finally, in Section \ref{sec:lightmatter}, we comment on the effects of light matter on the ensemble of OPE coefficients. In the crossing equations of Section \ref{sec:typicality}, we only keep the contribution from $h=\bbi$ in suitable OPE channels. However, there are crossing equations in which the contribution of the identity module vanishes, and so the leading answer depends on the coupling of the lightest scalar. This phenomenon was observed in various places before \cite{Kraus_2017,Belin:2021ryy}. Our new insight is that the bulk interpretation of this phenomenon is given by Euclidean wormholes containing bulk Wilson loops that do not reach the asymptotic boundary. 
As an example, we study a genus-two crossing equation contributing to the average $\mean{C_{112}C_{233}}$ through a light state $\chi$, which is captured by the following two-boundary wormhole:
\begin{equation}
\mean{C_{112}C_{233}} \supset \left|\frac{
\sker{P_\chi}{P_3}{P_2}
\sker{P_\chi}{P_1}{P_2}
}{\rho_0(P_1)\rho_0(P_3)}C_0(P_2,P_\chi,P_\chi)\right|^2 = Z_{\mathrm{grav}}\!\left(\!
\vcenter{\hbox{\begin{tikzpicture}
\node at (0,0) {\includegraphics[width=4.2cm]{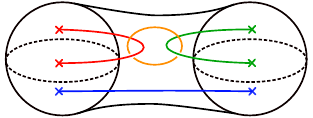}};
\end{tikzpicture}}}\! \right).
\end{equation}
The right-hand side shows a wormhole with a pair of three-punctured sphere boundaries. The Wilson loop corresponding to $\chi$ is drawn in orange. If we thicken the red, green and blue Wilson lines to solid cylinders, we can equivalently think of the complement of this geometry in $S^3$ as a genus-two handlebody with a $\chi$-loop wrapping a non-contractible cycle. A clarifying illustration of this geometrical set-up is shown in Figure \ref{fig:gravInt}. 

One may wonder if these light-matter corrections are ever relevant. It turns out that there are kinematic regimes where the contribution from the light fields cannot be ignored. To make this more precise, we end Section \ref{sec:lightmatter} with an explicit geometric analysis of the moduli space of genus-two surfaces, in a corner where an embedded cylinder becomes very long. In this `pinching limit', we show that the modular S-kernel no longer provides an exponential hierarchy (in $\e^S$) between the vacuum and the lightest non-identity state. 

We finish the paper with some conclusions and open questions in Section \ref{sec:discu}. Additionally, we present most of the technical details about the fusion and modular kernels in Appendix \ref{app:kernels}.

\newpage

\section{Typicality and the bootstrap}\label{sec:typicality} 
\vspace{-5pt}
Let us first describe the concept of typicality in generic many-body quantum chaotic systems, before applying it specifically to holographic CFT$_2$'s. For a review of this topic, see \cite{Jafferis:2022uhu} and \cite{Foini:2018sdb}. In essence,
typicality is an assumption that constrains the allowed non-Gaussian extensions of the ETH. The central idea behind typicality is that  simple operators should not be able to distinguish between nearby energy microstates. This idea can be expressed as the statement that the moments of the random variables $\clo_{12}\coloneqq \langle E_1 | \clo |E_2\rangle$ should be approximately invariant under unitary transformations that act within small microcanonical windows,\footnote{Note that this is only an approximation, which is used to extract the index structure of the ensemble. Imposing exact unitary invariance would render the functions $g_{\sigma}(E_1,\dots, E_n)$ exactly constant, see e.g. \cite{Wang:2023qon}.} 
\vspace{-5pt}
\begin{equation}
\label{eq:uni-trans}
    \mean{\clo_{12}\clo_{34}\cdots \clo_{(n-1)\,n}}  \stackrel{!}{=} \sum_{1',\dots, n'} U^{\dagger}_{11'}U_{22'}\cdots U^{\dagger}_{n-1\,(n-1)'}U_{n\,n'}
   \;
   \mean{\clo_{1'2'}\clo_{3'4'}\cdots \clo_{(n-1)'n'}}.
\end{equation}
Here $U$ is a block-diagonal unitary transformation that only mixes levels within windows of size $\e^{S(E)}$. The size of each block-matrix is small enough to keep intact the energy band structure of the system, but it contains exponentially many energy levels. The typicality assumption \eqref{eq:uni-trans} is a powerful assumption that fixes the index structure of the moments of the matrix elements:
\begin{equation}
\label{eq:genETH}
     \mean{\clo_{12}\clo_{34}\cdots \clo_{n-1\, n}} = \sum_{\sigma\in S_n}
     g_{\sigma}(E_1,\dots, E_n)\; \delta_{1\sigma(2)}
     \delta_{3 \sigma(4)}
     \cdots\delta_{(n-1) \sigma(n)}.
\end{equation}
The functions $g_{\sigma}(E_1,\dots E_n)$ are smooth functions of the energies $E_{i}$, and the sum is over all the permutations $\sigma$ of the even indices. One can check that this ansatz is invariant under \eqref{eq:uni-trans} as long as the functions $g_{\sigma}(E_1,\dots E_n)$ are approximately constant within the microcanonical window mixing the energy eigenstates. These functions are theory dependent and can be extracted from single-trace observables such as thermal correlation functions of the type
\begin{equation}
    \mean{\Tr (\clo \,\e^{-\beta_1 H}\cdots \clo\, \e^{-\beta_n H})} = \sum_{1,\dots, n} \e^{-\sum_i\beta_{i} E_{i}}\; \mean{ \clo_{12}\clo_{23}\cdots\clo_{n1}}
\end{equation}
for a collection of (possibly complex) parameters $\beta_1,\dots,\beta_n$. From the bulk point of view, such single-trace observables correspond to single-boundary gravitational amplitudes. Notice that the above thermal $n$-point function leads to a \emph{cyclic} contraction of the indices of $\clo$.  

The statement of the \emph{generalized} ETH can now be phrased in terms of such cyclic contractions: it is the hypothesis that in chaotic quantum systems, the $n^{\mathrm{th}}$ cyclic (connected) moment scales with the microcanonical entropy as \cite{Foini:2018sdb}
\begin{equation}
    \mean{\clo_{12}\clo_{23}\dots \clo_{n1}}\mlab{cyclic} \eqqcolon g_n(E_1,\dots, E_n) \sim  \e^{-(n-1)S(\bar E)}.
\end{equation}
In this expression, the indices $1,\dots n$ are all taken to be different; we have also introduced the notation $g_n(E_1,\dots, E_n)$ to describe these moments. Note that for $n=2$, we recover the standard ETH ansatz \eqref{eq:intro_ETH}. If one further assumes that products of thermal correlation functions factorize up to some small corrections,\enlargethispage{\baselineskip}
\begin{multline}
\label{eq:expfact}
\mean{\Tr (\clo \e^{-\beta_1 H}\cdots \clo \e^{-\beta_n H})
\Tr (\clo \e^{- \beta_{n+1} H}\cdots \clo \e^{- \beta_{n+m} H})
} \\[1em]
\approx 
\mean{\Tr (\clo \e^{-\beta_1 H}\cdots \clo \e^{-\beta_n H})}\;\;\mean{
\Tr (\clo \e^{-\beta_{n+1} H}\cdots \clo \e^{-\beta_{n+m} H})
},
\end{multline}
then one can show that the 
cyclic moments $g_n(E_1,\dots,E_n)$ are enough to describe, with exponential accuracy, all other contractions $g_\sigma(E_1,\dots,E_n)$ that define the ensemble of matrix elements $\clo_{12}$. The crucial insight of this result is that the functions $g_\sigma$ can be expressed as products of the functions $g_n$. Put differently, \eqref{eq:expfact} implies that the moments of matrix elements approximately factorize, e.g.
\begin{equation}
    \mean{\clo_{12}\clo_{21}\clo_{34}\clo_{45}\clo_{53}} \;\approx\;
\mean{\clo_{12}\clo_{21}}\;\;\mean{\clo_{34}\clo_{45}\clo_{53}}.
\end{equation}
As explained in \cite{Pappalardi:2022aaz}, this is intimately related to the theory of free probability.

\subsection{Typicality in conformal field theories}

We will now extend the above discussion to two-dimensional conformal field theories with only Virasoro symmetry. However, before addressing the technical aspects of typicality in CFTs, we have to establish some notation. We label primary states by the left- and right-moving conformal weights $h,\bar h$, which determine the scaling dimension $\Delta$ and spin $J$ of the corresponding operator as
\begin{equation}
    \Delta = h + \bar h, \quad J = h-\bar h.
\end{equation}
It will be useful to introduce the so-called Liouville parametrization 
\begin{equation}
\label{eq:Liouville_par}
     h = \frac{c-1}{24}+P^2, \quad \alpha = \frac{Q}{2} + i P
\end{equation}
where we refer to $P$ as the Liouville momentum, and $Q$ is related to the central charge $c$ as
\begin{equation}
\label{eq:Liouville_par2}
    c = 1+6 Q^2 = 1 + 6\left(b+b^{-1}\right)^2.
\end{equation}
For the application to holography, we will be interested in the regime  where $c\geq 25$, and work with the convention that $b < 1$ and $\Re(P)\geq 0$. The central charge is related to the 3D gravitational coupling by the Brown-Henneaux relation $c = \tfrac{3}{2G_{\mathrm{N}}}$ \cite{Brown:1986nw}.

We will distinguish between two different 
regimes for the conformal weights and Liouville momenta, which we refer to as the `discrete' part of the spectrum and the `continuum':
\begin{equation}
\begin{array}{rccc}
   \text{Discrete:}&
   \quad 0 \leq h < \tfrac{c-1}{24},&
   \quad 0\leq \alpha \leq Q/2,&
   \quad  P \in i (0,Q/2],\\[1em]
   \text{Continuum:} &
   \quad  h \geq \tfrac{c-1}{24}, &
   \quad \alpha = Q/2 + i \mathbb{R}_{\geq0},&
   \quad  P \in \mathbb{R}_{\geq0}~.
\end{array}
\end{equation} 
This terminology is inspired by analogy to the continuum of BTZ black hole solutions in 3D gravity, which are labelled by their mass $M\geq M_0$ and angular momentum. In the CFT, the black hole threshold $M_0$ corresponds  to the scaling dimension $\Delta=\frac{c-1}{12}$, or equivalently $P,\bar P=0$. 

Evidently, a single instance of the dual  microscopic unitary, compact CFT has a fully discrete spectrum, even above the threshold. However, for holographic CFTs with a weakly coupled gravitational dual, the spectrum should be dense enough above threshold so that, to a first approximation, we can smear the energy levels over small microcanonical windows. This leads to an effective continuum of states, with a leading spectral density given by the (extended) Cardy formula\vspace{1mm}
\begin{equation}\label{eq:rho0function}
    \rho_0(P,\bar P) = \rho_0(P)\rho_0(\bar P), \quad \mathrm{where}\quad \rho_0(P) = 4\sqrt{2} \sinh(2\pi b P)\sinh(2\pi b^{-1}P).\vspace{1mm}
\end{equation}
This density is determined by imposing modular invariance of the torus partition function \cite{Cardy:1986ie,Collier:2019weq}.\footnote{Note that this formula has been derived by only requiring $S$-invariance of the partition function. Imposing the full $PSL(2,\mathbb{Z})$-invariance should give corrections to the spectral density $\rho_0(P,\bar P)$. An attempt at a fully $PSL(2,\mathbb{Z})$ invariant density, inspired by gravity, is the Maloney-Witten-Keller density $\rho_{\mathrm{MWK}}(P,\bar P)$ \cite{Maloney_2010,Keller:2014xba}. Famously, this leads to negativities in the spectrum, which may be treated in various ways \cite{Benjamin:2020mfz,Maxfield:2020ale,DiUbaldo:2023hkc}.} For large enough $c$, it is expected to hold all the way down to $P=0$. For large momenta, the spectral density \eqref{eq:rho0function} exhibits the Cardy growth $\rho_0(P)\sim \e^{2\pi Q P}$.

We call operators with conformal weights in the discrete range `light' (L), and above the black hole threshold `heavy' (H). Note that in this terminology, even sub-threshold operators with conformal dimension scaling with $c$ are called `light', even though they lead to backreaction on the geometry. We make the heavy/light distinction in our notation for the OPE coefficients, by using numeral indices for heavy operators and the label $\clo_i$ for light operators, e.g.
\begin{equation}
   C_{123} \quad ( \mathrm{HHH}),\quad   C_{12\clo_3} \quad (\mathrm{HHL}),\quad  C_{1\clo_2 \clo_3} \quad (\mathrm{HLL}).
\end{equation}
Moreover, the vacuum state $(h=\bar h=0)$ will be denoted by $\bbi$. It is generally believed that any holographic CFT$_2$ should have a \emph{sparse} spectrum of light states, with a large gap between the vacuum and the lightest non-vacuum primary \cite{Hartman:2014oaa}. We will always assume this to be the case in what follows.

As explained in the introduction, the operator-state correspondence allows one to reformulate the eigenstate thermalization hypothesis in CFT$_2$ as a statement about the statistical distribution of heavy-heavy-light OPE coefficients $C_{1\clo_2 3}$. The proposal of \cite{Belin:2020hea} is to extend this statistical description of the OPE data also to the HLL and HHH regimes. Our goal is to determine the precise statistics of the ensemble of OPE data compatible with a holographic description in terms of pure 3D gravity. In other words, we want to obtain all the moments
\begin{equation}
    \mean{C_{123}C_{456}\cdots C_{(n-2)\,(n-1)\,n}}\,,
\end{equation}
and similar expressions for when a subset of the indices is light. In what follows, the overline notation will refer to an average over the heavy indices only.

To do so, we need to determine both the \emph{index structure} of the possible contraction patterns and the \emph{smooth functions} $g_\mu$ that contain the dependence of the moments on the band structure of the Hamiltonian $L_0 + \bar L_0$. We will argue that both properties are fixed by imposing:  
\begin{enumerate}
    \item Typicality,
    \item Crossing symmetry and modular invariance.  
\end{enumerate}

Let us first explain the typicality argument. 
We will restrict ourselves to CFTs whose operators have integer spins $J  \in \mathbb{Z}$. These CFTs have correlation functions that are invariant under the permutations of the fields. It can then be shown from the consistency of the three-point function that the structure constants must satisfy 
\begin{equation}
\label{eq:condition1}
    C_{\sigma(1)\sigma(2)\sigma(3)} = \text{sgn}(\sigma)^{J_1+J_2+J_3} C_{123}.
\end{equation}
Moreover, the additional Hermiticity property $C_{123} = C_{321}^*$ implies that OPE coefficients are real numbers if the total spin is even, and pure imaginary if odd. To respect \eqref{eq:condition1}, the most natural approach is to consider unitary transformations $T(\Delta,J)$ that only mix states with the same spin $J$ around a small microcanonical window surrounding $\Delta$.\footnote{See e.g. \cite{Haehl:2023tkr} for further arguments why random matrix universality is expected to hold within each spin sector of a CFT$_2$, at least in the near-extremal limit.} The reality conditions imply that $T(\Delta,J)$ must be a real orthogonal matrix.\footnote{
In \cite{Yan:2023rjh}, it was argued that the relevant ensemble for capturing the spectral statistics in parity-invariant CFTs is the GOE. This is consistent with the symmetries that we observe in OPE coefficients. 
} Because of the state-operator correspondence, these transformations should act on both states and OPE coefficients in an obvious way:
\vspace{1mm}
\begin{equation}
    \ket{\Delta_1,J} = \sum_{2} T_{12}(\Delta_1,J) \ket{\Delta_2,J}, \quad C_{123}=\sum_{1'2'3'}T_{11'}T_{22'}T_{33'} C_{1'2'3'}.
\end{equation}
The statement of typicality for CFTs is then that the averages of arbitrary products of OPE coefficients are invariant under these orthogonal transformations
\begin{equation}
\label{eq:typasum}
\mean{C_{123}C_{456}\cdots C_{(n-2)\,(n-1)\,n}} \stackrel{!}{=} \sum_{1',\dots, n'}  T_{11'}(\Delta_1,J)\cdots T_{nn'}(\Delta_n,J) \;\mean{C_{1'2'3'}\cdots C_{(n-2)'\,(n-1)'\,n'}}.
\end{equation}
This assumption puts a strong constraint on the allowed index contractions. In particular, it fixes the indices to be pairwise contracted. That is, if we have $m$ OPE coefficients, with a total of $n=3m$ indices, then there are smooth functions $g_\mu$ such that\vspace{2mm}
\begin{equation}\label{eq:pairwise}
    \mean{C_{123}C_{456}\dots C_{(n-2)\,(n-1)\,n}} = \sum_{\mu\in \mathcal{P}} 
    g_{\mu}(P_1,\bar P_1,\dots, P_{n},\bar P_{n}) \,
    \delta_{\mu_1}\dots \delta_{\mu_{n/2}},
\end{equation}
where the sum runs over the set $\mathcal{P}$ of all pairings {\small $\mu = (\mu_1,\dots,\mu_{n/2}), \mu_i = (a,b)$}\,,  of the set of $n$ elements. The functions {\small $g_\mu$} depend on the mean scaling dimension {\small $\Delta_{1\cdots n} = \tfrac{1}{n}(\Delta_1+\dots+\Delta_n)$}, their differences {\small $\omega_{ij}=\Delta_i-\Delta_j$}, and the spins {\small $J_1,\dots,J_n$}. They are assumed to be `smooth enough', by which we mean that they are approximately constant in sufficiently narrow microcanonical windows. 
The analysis when some of the indices are light (and not averaged over) follows the same steps. The conclusion is that the index structure contracts heavy indices in pairs. If the number of heavy indices is odd the average is zero.

As an example, consider $n=6$. The sum on the right-hand side of \eqref{eq:pairwise} yields the following set of contributions to the second moment of OPE coefficients:\vspace{2mm}
\begin{equation}\label{eq:OPEvariance}
\mean{C_{123}C_{456}} = \delta_{14}\delta_{25}\delta_{36} \, g_{\mu_1}(P_1,\bar P_1,P_2,\bar P_2,P_3,\bar P_3) + \delta_{12}\delta_{34}\delta_{56} \, g_{\mu_2}(P_1,\bar P_1,P_3,\bar P_3,P_6,\bar P_6) + \text{perm}.\vspace{2mm}
\end{equation}
where the permutations come with signs according to \eqref{eq:condition1}. The second contraction, associated to the averaged value $\mean{C_{113}C_{366}}$, is an example of a non-diagonal contraction that is allowed by typicality. Indeed, we will show in Section \ref{sec:gravity} that a non-zero function exists for precisely this non-diagonal contraction.

Having fixed the index structure, we turn our attention to the smooth functions $g_\mu$. Remarkably, they are \emph{fully determined by modular invariance and crossing symmetry} for any holographic CFT$_2$. Specifically, we impose the (non-rational extension of) the Moore-Seiberg consistency conditions \cite{Moore:1988qv}. This statement is so remarkable because for most chaotic systems obeying ETH the $g_\mu$ are not known analytically -- although numerical results can be obtained, e.g. for the SYK model \cite{Sonner:2017hxc}.

Taking as an example again the variance \eqref{eq:OPEvariance}, the function $g_{\mu_1}$ can be determined by imposing the invariance of the genus-two partition function under the following change of OPE channel:\footnote{The diagrammatic notation will be explained in Section \ref{sec:3pointfunction}.}
\begin{equation}
\label{eq:croseqgen2}
    Z_{g=2}\left(\begin{tikzpicture}[x=0.75pt,y=0.75pt,yscale=0.8,xscale=0.8,baseline={([yshift=-.5ex]current bounding box.center)}]
\node at (0,0) [midway]{\includegraphics{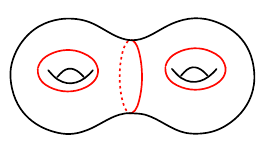}};
\end{tikzpicture}\right) = \,Z_{g=2}\left(\begin{tikzpicture}[x=0.75pt,y=0.75pt,yscale=0.8,xscale=0.8,baseline={([yshift=-.5ex]current bounding box.center)}]
\node at (0,0) [midway]{\includegraphics{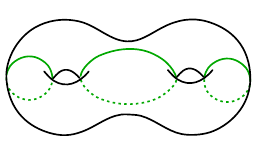}};
\end{tikzpicture}\right).
\end{equation}
One can think of this requirement as the analogue of  modular invariance of the torus partition function, generalized to higher genus \cite{Cardy:2017qhl}. As shown in \cite{Collier:2019weq}, the above genus-two crossing equation leads to a universal formula for the variance of OPE coefficients:\vspace{1.5mm}
\begin{equation}
\label{eq:Gauss}
  \mean{C_{123}C_{456}} =
  \sum_{\sigma \in S^3}\delta_{1\sigma(4)}\delta_{2\sigma(5)}\delta_{3\sigma(6)}\, \text{sgn}(\sigma)^{J_1+J_2+J_3} \,C_0(P_1,P_2,P_3)C_0(\bar P_1,\bar P_2,\bar P_3) +\cdots .\vspace{1mm}
\end{equation}
The totally symmetric function $C_0(P_1,P_2,P_3)$ is equivalent to the DOZZ formula for the three-point function in Liouville theory. It has an explicit analytic expression as a product of Barnes double gamma functions\footnote{For a definition and discussion of the properties of $\Gamma_b(x)$ we recommend references \cite{Collier:2018exn,Eberhardt:2023mrq}.}:
\begin{equation}\label{eq:C0function}
    C_0(P_1,P_2,P_3)= \frac{\Gamma_b(2Q)\Gamma_b(\frac{Q}{2}\pm i P_1 \pm i P_2 \pm i P_3)}{\sqrt{2}\Gamma_b(Q)^3\prod_{k=1}^3\Gamma_b(Q\pm 2i P_k)}.\vspace{2mm}
\end{equation}
The notation $\pm$ denotes a product over all possible sign choices in the formula. The terms $+\cdots$ contain the non-diagonal contraction $g_{\mu_2}$ in \eqref{eq:OPEvariance}. In principle, these other contractions can be fixed by imposing more and more crossing equations. For example, the leading contribution to the function $g_{\mu_2}$ can be derived from a genus-two crossing equation similar to \eqref{eq:croseqgen2}, but now expanded in a dumbbell channel  where the two `bells' are linked.\footnote{We thank Scott Collier for explaining this to us.}

Moreover, one can consider crossing equations for higher genus surfaces, and include punctures for operator insertions. This has produced several interesting formulas for the non-Gaussian statistics of OPE coefficients, see e.g. \cite{Collier:2019weq, Belin:2021ryy, Anous:2021caj,Kraus_2017}. For example, the genus-three crossing equation
\begin{equation}
    Z_{g=3}\left(\begin{tikzpicture}[x=0.75pt,y=0.75pt,yscale=0.75,xscale=0.75,baseline={([yshift=-.5ex]current bounding box.center)}]
\node at (0,0) [midway]{\includegraphics{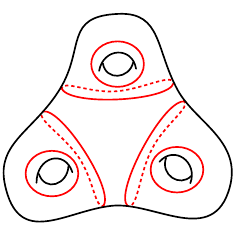}};
\end{tikzpicture}\right) = \,Z_{g=3}\left(\begin{tikzpicture}[x=0.75pt,y=0.75pt,yscale=0.8,xscale=0.8,baseline={([yshift=-.5ex]current bounding box.center)}]
\node at (0,0) [midway]{\includegraphics{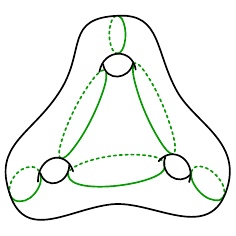}};
\end{tikzpicture}\right),
\end{equation}
leads to a non-Gaussianity that is again universal in the high-energy sector of the CFT. This non-Gaussianity can be fully expressed in terms of Virasoro crossing kernels:\enlargethispage{0.5\baselineskip}
\begin{align}
\label{eq:non-Gauss}
   \!\!\mean{C_{123}C_{156}C_{264}C_{345}} \mlab{6\it j} &=\! \left|C_0(P_1,P_5,P_6)C_0(P_3,P_4,P_5)\frac{
    \fker{P_5}{P_2}{P_6}{P_4}{P_1}{P_3}
    }{\rho_0(P_2)}\right|^2
\end{align}
where by $|\bullet|^2$ we mean the product of the holomorphic and antiholomorphic counterpart of this \pagebreak
expression. We will use the notation $\lvert_{6j}$ when referring to this particular contraction, because it is proportional to the Virasoro $6j$ symbol (recall equation \eqref{eq:moments} in the introduction). This symbol enjoys tetrahedral symmetry and will play an important role later in Section \ref{sec:genOPE}.\footnote{The $6j$ contraction also appears in the tensor-matrix model described in \cite{Belin:2023efa} as the quartic interaction for $C_{ijk}$.} 

Again, the right-hand side of the quartic moment \eqref{eq:non-Gauss} is a completely known meromorphic function of its complex variables $P_1,\dots,P_6$. We have already introduced the functions $\rho_0$ and $C_0$, so the only new ingredient is the so-called Virasoro fusion kernel $\bbf_{PP'}$, whose definition can be found in Appendix \ref{app:kernels}. The main point for now is that $\rho_0$, $C_0$ and $\mathbb{F}$ are all examples of so-called crossing kernels. These objects naturally appear in the representation theory of the Virasoro algebra. Roughly speaking, they are the \emph{change-of-basis transformations} between the different possible channel decompositions of a given Virasoro conformal block. A thorough overview of crossing kernels can be found in \cite{Eberhardt:2023mrq}.

In order to make the comparison to the ETH in generic many-body quantum chaotic systems, we would like to extract the scaling of the smooth functions {\small $g_\mu$} with the microcanonical entropy {\small $S_0(P,\bar P) = \log \rho_0(P,\bar P)$}. We will show, for cyclic contractions of heavy-heavy-light OPE coefficients, that the entropy scaling precisely matches the generalized ETH prediction for $n\geq 2$:\vspace{2mm}
\begin{equation}\label{eq:entropyscaling}
    \mean{C_{1\clo_1 2}C_{2\clo_2 3}\cdots C_{n\clo_n 1}
    }\mlab{cyclic} \coloneqq g_n(\mathbf{P},\bar{\mathbf{P}};\mathbf{P}_\clo,\bar{\mathbf{P}}_\clo) \sim \e^{-(n-1)S_0(P,\bar P)}.\vspace{2mm}
\end{equation}
The only exception is for $n=1$, as it was noted already in \cite{Kraus_2017,Collier:2019weq} that the average of a single OPE coefficient is exponentially small. This is not so relevant for our discussion, since we can always subtract the one-point function from $C_{123}$ and work with variables with zero mean.

In the rest of this section, we will bootstrap explicit formulas for the smooth functions $g_n$ for $n=3,4$, and show that the exponential hierarchy \eqref{eq:entropyscaling} holds true. We will then describe a recursive procedure for general $n$, from which we conclude that \eqref{eq:entropyscaling} is satisfied for all $n$. We want to emphasize that this scaling is true when the dimension of $\clo$ is below the black hole threshold. In general, the entropy scaling will be different when all indices are heavy, due to exponential enhancements from the crossing kernels when all internal momenta are taken to be large. 

\subsection{Thermal three-point function}\label{sec:3pointfunction}
In this section, we will harness the constraints imposed by crossing symmetry and modular invariance on the torus three-point function. We aim to extract a formula for the averaged cyclic contraction 
\begin{equation}
\label{eq:cyclic3}
   \mean{ C_{1\clo_12}C_{2\clo_23}C_{3\clo_31}},
\end{equation}
where $\clo_1, \clo_2$ and $\clo_3$ are three fixed primary operators, and $h_1,h_2,h_3$ are heavy. The definition of the average $\mean{\;\bullet\;}$ will be made more precise towards the end of this section. 

 To study this contraction, consider the unnormalized torus three-point function:\vspace{2mm}
\begin{equation}
\label{eq:tor3pt}
    \big\langle \clo_1(1,1)\clo_2(z_2,\bar z_2)\clo_3(z_3,\bar z_3) \big\rangle_{\tau,\bar\tau} =  \Tr \left(q^{L_0-\frac{c}{24}}\bar q^{\bar L_0-\frac{c}{24}}\clo_1(1,1)\clo_2(z_2,\bar z_2)\clo_3(z_3,\bar z_3)\right), \vspace{2mm}
\end{equation}
where $q = \exp(2\pi i\tau)$. For simplicity we have set $z_1,\bar z_1 = 1$. Inserting a resolution of the \pagebreak identity in between the operators $\clo_1$ and $\clo_2$ and the operators $\clo_2$ and $\clo_3$ leads to the following expression for this three-point function
\begin{equation}
\label{eq:decneck}
 \sum_{1,2,3} C_{1\clo_12}C_{2\clo_23}C_{3\clo_31} \;\mathcal{F}^{\mathcal{N}}(\mathbf{P};\mathbf{P}_{\clo};\Omega)
 \bar{\mathcal{F}}^{\mathcal{N}}(\bar{ \mathbf{P}};\bar{\mathbf{P}}_{\clo};\bar \Omega).
\end{equation}
The sum is only over primary operators, as we have resummed the contribution of descendant states into the conformal blocks $\mathcal{F}^{\mathcal{N}}(\mathbf{P};\mathbf{P}_{\clo};\Omega)$.
In our notation for the conformal blocks, $\mathbf{P}=(P_1,P_2,P_3)$ labels the internal momenta and $\mathbf{P}_{\clo} = (P_{\clo_1},P_{\clo_2},P_{\clo_3})$ labels the momenta of the external operators; the moduli $(z_{2}, z_3,\tau)$ are collectively denoted by $\Omega$. As before, we used the notation $\sum_{1} \equiv \sum_{h_1,\bar h_1}$ for the sum over all the primary states in the spectrum. 

It is often useful to represent conformal blocks pictorially, for instance
\begin{equation}
\label{eq:neclace}
  \mathcal{F}^{\mathcal{N}}(\bfP;\bfP_\clo;\Omega) =   
   \vcenter{
   \hbox{\begin{tikzpicture}
   \node at (0,0) {\includegraphics{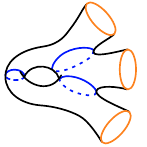}};
   \node at (1,1){$\clo_1$};
   \node at (1.4,0) {$\clo_2$};
   \node at (1.1,-1.1){$\clo_3$};
   \node at (-1.4,0){\scriptsize{$P_1$}};
   \node at (0.10,0.59){\scriptsize{$P_2$}};
    \node at (0.10,-0.51){\scriptsize{$P_3$}};
   \end{tikzpicture}}
   }.
\end{equation}
The blue lines in the diagram specify the pair-of-pants decomposition of the Riemann surface and label the internal momenta. A Riemann surface can be decomposed into pairs of pants by choosing a basis of $3g-3+n$ non-intersecting cycles on the surface. Different pants decompositions correspond to the different channels into which one can write the correlation function \eqref{eq:tor3pt}. The decomposition shown here corresponds to the so-called \emph{necklace channel} $\mathcal{N}$, which arises when a resolution of the identity is inserted between adjacent external operators.

The second channel we will be considering is the one where we first perform the OPE of $\clo_1$ and $\clo_2$, and then the OPE between the resulting operator and $\clo_3$. In this channel, the torus three-point function takes the form 
\begin{equation}
\label{eq:decope}
    \sum_{1',2',3'} C_{2'\clo_1\clo_2}C_{3'2'\clo_3}C_{1'3'1'} \mathcal{F}^{\,\text{OPE}}(\bfP\dash;\bfP_{\clo};\Omega)
    \bar{\mathcal{F}}^{\,\text{OPE}}(\bar \bfP\dash;\bar \bfP_{\clo};\bar \Omega).
\end{equation}
The pair-of-pants decomposition corresponding to this channel is given by 
\begin{equation}
  \mathcal{F}^{\text{OPE}}(\bfP\dash;\bfP_{\clo};\Omega) =   
   \vcenter{
   \hbox{\begin{tikzpicture}
   \node at (0,0) {\includegraphics{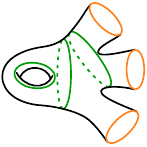}};
   \node at (1,1){$\clo_1$};
   \node at (1.5,0) {$\clo_2$};
   \node at (1.2,-1.2){$\clo_3$};
   \node at (-1.5,0){\scriptsize{$P_1'$}};
   \node at (0.26,0.23){\scriptsize{$P_2'$}};
    \node at (0.10,-0.51){\scriptsize{$P_3'$}};
   \end{tikzpicture}}
   }.
\end{equation}
We refer to this channel as the \emph{OPE channel} since it was obtained by applying the OPE consecutively to the operators inside the correlation function.

It turns out that to extract an asymptotic formula for the cyclic contraction \eqref{eq:cyclic3}, we only need to know how the two bases of conformal blocks in the different channels are related to each other. This relationship is expressed as the following integral transform
\begin{equation}\label{eq:relker}
    \mathcal{F}^{\,\text{OPE}}(\bfP';\bfP_{\clo};\Omega) = \int_{0}^{\infty} \dd^3 P \; \knl{K}{\bfP'}{\bfP}\;
    \mathcal{F}^{\,\mathcal{N}}(\bfP;\bfP_{\clo};\Omega),
\end{equation}
where $\dd^3 P = \dd P_1\dd P_2\dd P_3$.
The function $\knl{K}{\bfP'}{\bfP}$ is known as a crossing kernel. Note that the range of the integral is over real momenta $P \in \mathbb{R}_{\geq 0}$, corresponding to $h \geq \frac{c-1}{24}$. This range of momenta gives a complete basis of the Hilbert space of conformal blocks in a given channel. The crossing kernels have been constructed and are known in closed form \cite{Ponsot:1999uf}; we will briefly explain their construction in the second part of this section. 

To derive the asymptotic formula, the first step is to write the expressions \eqref{eq:decneck} and \eqref{eq:decope} as integrals over a distribution:
\begin{align}
 \big\langle \clo_1(1,1)&\clo_2(z_2,\bar z_2)\clo_3(z_3,\bar z_3) \big\rangle_{\tau,\bar\tau} \nonumber\\[0.8em] \label{eq:t3p1} 
 &= \int \dd^3 P \dd^3 \bar P \,\rho_0(\bfP,\bar \bfP)
    \;\mathcal{Q}^{\mathcal{N}}(\bfP,\bar\bfP)\;
    \mathcal{F}^{\mathcal{N}}(\bfP;\bfP_{\clo};\Omega)
    \bar{\mathcal{F}}^{\mathcal{N}}(\bar \bfP;\bar \bfP_{\clo};\Omega)\\[0.8em] \label{eq:t3p2} 
 &= \int \dd^3 P' \dd^3 \bar P'\rho_0(\bfP',\bar \bfP')
    \mathcal{Q}^{\text{OPE}}(\bfP',\bar \bfP')\;
    \mathcal{F}^{\text{OPE}}(\bfP';\bfP_{\clo};\Omega)
    \bar{\mathcal{F}}^{\text{OPE}}(\bar \bfP';\bar \bfP_{\clo};\Omega),
\end{align}
where $\rho_0(\bfP,\bar \bfP) = \prod_{i=1}^3 \rho_0(P_i,\bar P_i)$. These equations are meant to be exact, meaning that the distributions $Q^{\mathcal{N}}$ and $Q^{\text{OPE}}$ are a sum of delta functions weighted by OPE coefficients:
\begin{align}
Q^{\mathcal{N}}(\bfp,\bar \bfp) &= \frac{1}{\rho_0(\bfp,\bar \bfp)}
\sum_{1',2',3'}
C_{1'\clo_12'}
C_{2'\clo_23'}
C_{3'\clo_31'}
\prod_{i=1}^3
\delta(P_i - P'_i)\delta(\bar P_i - \bar P'_i)\\
Q^{\text{OPE}}(\bfp,\bar \bfp) &= \frac{1}{\rho_0(\bfp,\bar \bfp)}
\sum_{1',2',3'}
C_{2'\clo_1\clo_2}
C_{3'2'\clo_3}
C_{1'3'1'}
\prod_{i=1}^3
\delta(P_i - P'_i)\delta(\bar P_i - \bar P'_i).
\end{align}
Recall that $P_i$ is simply a parametrization of the conformal weight $h_i$, which runs over the full unitary CFT spectrum. The factor of $\rho_0^{-1}$ is a convenient normalization.

The next step is to insert the integral expression \eqref{eq:relker} into \eqref{eq:t3p2}. Changing the order of the integrals and comparing the result with \eqref{eq:t3p1} yields the relation:\footnote{
In \eqref{eq:t3p2}, the $P$ integral is along a contour $\mathcal{C}$ that includes the light operator dimensions, while to compare the conformal blocks, one needs to integrate only over the real line $P\in \mathbb{R}$. To arrive at \eqref{eq:crosseq3}, we used that the contour $\mathcal{C}$ can be smoothly deformed to $\mathbb{R}$ as long as we avoid the poles of $\mathbb{K}$. See \cite{Collier:2018exn} for a discussion of the analytic structure of the crossing kernels.}
\begin{align}
\label{eq:crosseq3}
    \mathcal{Q}^{\mathcal{N}}(\bfp,\bar \bfp) &= \frac{1}{\rho_0(\bfp,\bar \bfp)}
    \int \dd^3 P'\dd^3 \bar P' \rho_0(\bfP',\bar \bfP')\;
   \knl{K}{\bfP'}{\bfP}
   \knl{K}{\bar \bfP'}{\bar \bfP}
   \mathcal{Q}^{\text{OPE}}(\bfp',\bar \bfp')\\[1em]
   &= \frac{1}{\rho_0(\bfp,\bar \bfp)}\sum_{1'2'3'} C_{2'\clo_1\clo_2}C_{3'2'\clo_3}
   C_{1'3'1'} 
   \;\knl{K}{\bfp'}{\bfp}
   \knl{K}{\bar \bfp'}{\bar \bfp}.\label{eq:crosseq3-1}
\end{align}
The step from the first to the second line is simply the definition of the distribution $Q^{\text{OPE}}(\bfp',\bar \bfp')$.
Equation \eqref{eq:crosseq3} is known as a crossing equation. Importantly, this equation only involves the crossing kernel $\mathbb{K}$, and there are no conformal blocks. 

As we will show shortly, in the heavy limit defined as 
\begin{equation}
\label{eq:heavy1}
    P_{1,2,3} = P + \delta_{1,2,3}, \quad \text{and} \quad P \,\gg\, c,\, \delta_i,\, P_{\clo_i}\vspace{3mm}
\end{equation}
the dominant contribution to the right-hand side of \eqref{eq:crosseq3-1} comes from the terms in the sum with $1',3' = \bbi$ and $2' = \clo_3$. This leads to the following asymptotic formula for $\mathcal{Q}^{\mathcal{N}}(\bfp,\bar\bfp)$ as $P$ approaches infinity:\vspace{2mm}
\begin{equation}
\label{eq:result1}
 \mathcal{Q}^{\mathcal{N}}(\bfp,\bar \bfp) =  C_{\clo_1\clo_2\clo_3}
   \frac{\knl{K}{\bbi,P_{\clo_3},\bbi}{P_1,P_2,P_3}}{\rho_0(P_1)\rho_0(P_2)\rho_0(P_3)}
   \frac{\knl{K}{\bbi,\bar P_{\clo_3},\bbi}{\bar P_1,\bar P_2,\bar P_3}}{\rho_0(\bar P_1)\rho_0(\bar P_2)\rho_0(\bar P_3)} + \cdots\vspace{2mm}
\end{equation}
The terms in $+\dots$ correspond to non-perturbative corrections, which will be quantified in the subsequent paragraphs. An important fact about the Virasoro crossing kernels $\mathbb{K}$ is that they are meromorphic functions of the momenta $P,\bar P$. Therefore, we can perform a small microcanonical smearing in the arguments of the distribution $\mathcal{Q}^{\mathcal{N}}(\mathbf{P},\bar{\mathbf{P}})$. For a sufficiently dense set of OPE coefficients of the underlying chaotic CFT, this smearing turns $\mathcal{Q}^{\mathcal{N}}$ into a smooth function, which we take to be the definition of the average of OPE coefficients:
\begin{equation}
\label{eq:CFT3pointresult}
    \mean{C_{1\clo_12}C_{2\clo_23}C_{3\clo_31}} \coloneqq \frac{1}{N_{\delta,\bar\delta}(P,\bar P)}
    \int_{P-\delta}^{P+ \delta}\dd P'\int_{\bar P-\bar \delta}^{\bar P+\bar \delta}  \dd\bar P'
    \;\rho_0(\bfp',\bar \bfp') 
    \mathcal{Q}^{\mathcal{N}}(\bfp',\bar \bfP'),
\end{equation}
where $N_{\delta,\bar\delta}(P,\bar P)$ is the number of operators within the microcanonical window defined by $P\pm \delta$ and $\bar P\pm \bar \delta$. The energy window  $\delta,\bar \delta$ should be such that $N_{\delta,\bar\delta}(P,\bar P) \gg 1$.
The size of $\delta,\bar \delta$ is theory dependent, in particular, for a generic, chaotic theory the expectation is that we only need to average over a small energy window containing $\e^{S(P,\bar P)}$ states. Tauberian theorems have often been used to investigate the robustness of formulas like \eqref{eq:CFT3pointresult}, and to understand the smearing and the size of the microcanonical window \cite{Mukhametzhanov:2019pzy,Das:2017vej,Pal:2019zzr,Das:2020uax,Pal:2019yhz}. 
These theorems strictly speaking only apply when the quantity that is averaged over is positive definite \cite{Qiao:2017xif}. However, one can quantify deviations from the average by considering the squared of these quantities, which can be accessed through higher moments.\footnote{We thank Sridip Pal for discussions on this topic.} We leave the precise details of this aspect to future study.

\begin{figure}
\centering
\begin{tikzpicture}
\node at (0,0) [midway] {\includegraphics{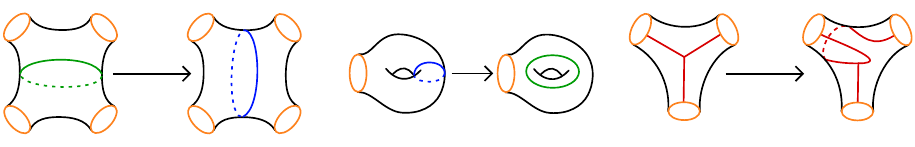}};
\node at (-7.9,1.3) [above] {a)}; 
\node at (-2,1.3) [above] {b)}; 
\node at (2.8,1.3) [above] {c)}; 
\node at (-5.15,0) [above] {$\mathbb{F}$};   
\node at (0.2,0) [above] {$\mathbb{S}$}; 
\node at (5.2,0) [above] {$\mathbb{B}$}; 
\node at (-5.75,0.8) [above] {$1$};
\node at (-7.7,0.8) [above] {$2$};
\node at (-5.8,-0.9) [below] {$4$};
\node at (-7.7,-0.9) [below] {$3$};
\node at (-6.7,-0.25) [above] {$P$};
\node at (-3.6,-0.25) [above] {$P'$};
\node at (-2.65,0.8) [above] {$1$};
\node at (-4.6,0.8) [above] {$2$};
\node at (-2.7,-0.9) [below] {$4$};
\node at (-4.6,-0.9) [below] {$3$};
\node at (-1.7,-0.4) [below] {$1$};
\node at (0.8,-0.4) [below] {$1$};
\node at (-0.7,0.1) [above] {$P$};
\node at (1.3,0.1) [above] {$P'$};
\node at (2.9,0.8) [above] {$1$};
\node at (4.75,0.8) [above] {$2$};
\node at (3.85,-0.8) [below] {$3$};
\node at (5.8,0.8) [above] {$2$};
\node at (7.75,0.8) [above] {$1$};
\node at (6.8,-0.8) [below] {$3$};
\end{tikzpicture}
\caption{\small The three basic crossing moves that relate conformal block decompositions in different channels. From left to right, the moves are called fusion, modular S-transform and braiding.}
\label{fig:moves}
\end{figure}
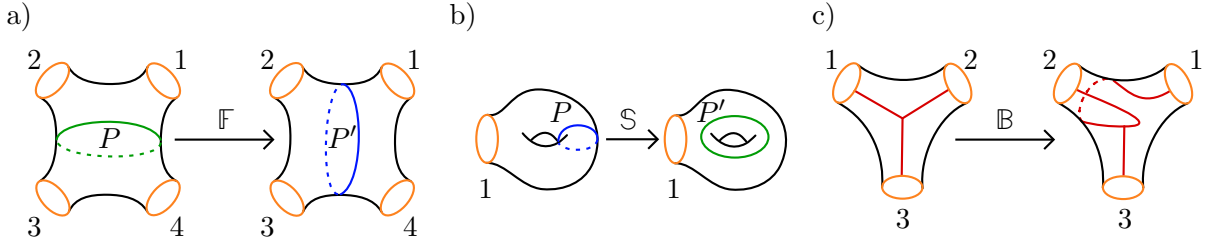

To complete the calculation, we need to build the crossing kernel $\mathbb{K}$, which connects the two conformal block decompositions. A generic crossing kernel can be built from three elementary crossing moves that go under the name of fusion $\mathbb{F}$, modular S-transform $\mathbb{S}$, and braiding $\mathbb{B}$ (see Figure \ref{fig:moves}). These kernels are known analytically and their properties and relations are summarized in Appendix \ref{app:kernels}. The process of building generic crossing kernels from these basic moves is known as the Moore-Seiberg construction \cite{Moore:1988qv}. \pagebreak We can use these moves to relate the conformal blocks in the necklace and OPE channel as follows:\vspace{3mm}
\begin{align}
\vcenter{\hbox{
\begin{tikzpicture}
\node at (0,0) {\includegraphics{img/BlockT3PT2.pdf}};
\node at (1,1){$\clo_1$};
\node at (1.5,0) {$\clo_2$};
\node at (1.2,-1.2){$\clo_3$};
\node at (-1.5,0){\scriptsize{$P_1'$}};
\node at (0.26,0.23){\scriptsize{$P_2'$}};
\node at (0.10,-0.51){\scriptsize{$P_3'$}};
\end{tikzpicture}
}} \!\! &=
\int_0^\infty\dd P_1\;
\sker{P_1'}{P_1}{P_3'}
\vcenter{\hbox{
\begin{tikzpicture}
\node at (0,0) {\includegraphics{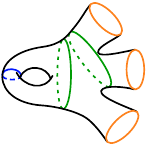}};
\node at (1,1){$\clo_1$};
\node at (1.5,0) {$\clo_2$};
\node at (1.2,-1.2){$\clo_3$};
\node at (0.26,0.23){\scriptsize{$P_2'$}};
\node at (0.10,-0.51){\scriptsize{$P_3'$}};
\node at (-1.4,0){\scriptsize{$P_1$}};
\end{tikzpicture}
}} \\ \notag
&=
\int_0^\infty\dd P_1\dd P_3\;
\sker{P_1'}{P_1}{P_3'}
\fker{P_3'}{P_3}{P_2'}{P_1}{P_1}{P_{\clo_3}}
\vcenter{\hbox{
\begin{tikzpicture}
\node at (0,0) {\includegraphics{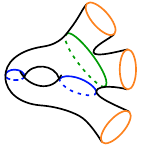}};
\node at (1,1){$\clo_1$};
\node at (1.5,0) {$\clo_2$};
\node at (1.2,-1.2){$\clo_3$};
\node at (0.15,0.23){\scriptsize{$P_2'$}};
\node at (-1.4,0){\scriptsize{$P_1$}};
\node at (0.10,-0.51){\scriptsize{$P_3$}};
\end{tikzpicture}
}} \\ \notag
&= 
\int_0^\infty\dd P_1\dd P_3\dd P_2
\;\sker{P_1'}{P_1}{P_3'}
\fker{P_3'}{P_3}{P_2'}{P_1}{P_1}{P_{\clo_3}}
\fker{P_2'}{P_2}{P_{\clo_1}}{P_1}{P_3}{P_{\clo_2}}
\vcenter{\hbox{
\begin{tikzpicture}
\node at (0,0) {\includegraphics{img/BlockT3PT1.pdf}};
\node at (1,1){$\clo_1$};
\node at (1.5,0) {$\clo_2$};
\node at (1.2,-1.2){$\clo_3$};
\node at (-1.4,0){\scriptsize{$P_1$}};
\node at (0.10,0.59){\scriptsize{$P_2$}};
\node at (0.10,-0.51){\scriptsize{$P_3$}};
\end{tikzpicture}
}}\!\!.
\end{align}
From this expression, we read off the crossing kernel $\mathbb{K}$ to be\vspace{2mm}
\begin{equation}
\knl{K}{P_1',P_2',P_3'}{P_1,P_2,P_3}
=\sker{P_1'}{P_1}{P_3'}\,
\fker{P_3'}{P_3}{P_2'}{P_1}{P_1}{P_{\clo_3}}
\fker{P_2'}{P_2}{P_{\clo_1}}{P_1}{P_3}{P_{\clo_2}}.\vspace{2mm}
\end{equation}
In the heavy limit described by \eqref{eq:heavy1}, the modular  S-kernel is such that asymptotically as $P_1\rightarrow\infty$,
\begin{equation}
\label{eq:supression1}
    \frac{\sker{P_1'}{P_1}{P_3'}}{\sker{\bbi}{P_1}{\bbi}}
     \sim \e^{-4\pi \alpha_1' P_1}, 
\end{equation}
See Appendix \ref{app:kernels} for a derivation of this statement. The analogous relation also holds for the antiholomorphic counterpart of \eqref{eq:supression1}. This result implies that the leading contribution to the sum in \eqref{eq:crosseq3} is given by the vacuum $1'=\bbi$, which corresponds to $\alpha_1' = 0$. The fusion kernels do not lead to an exponential suppression for $\alpha_2'$ and $\alpha_3'$ analogous to \eqref{eq:supression1}, however, by virtue of the OPE coefficients $C_{\bbi 3' \bbi} = \delta_{3'\bbi}$ and  $C_{\bbi2'\clo_3} = \delta_{2'\clo_3}$ we must also set $3'$ to the identity and $2'$ to $\clo_3$. Our  analysis shows that the corrections to \eqref{eq:result1} are exponentially suppressed by a factor of $\exp[-4\pi \left( \alpha_\chi P + \bar \alpha_{\chi} \bar P\right)]$, where $\chi$ is the lightest operator in the theory that is not the identity. 

Plugging our result for the crossing kernel into \eqref{eq:result1}, we have derived the following expression for the cyclic contraction with three external operators
\begin{equation}
\label{eq:cyclic33}
\overline{C_{1\clo_12}C_{2\clo_23}C_{3\clo_31}} = C_{\clo_1\clo_2\clo_3}\left| C_0(P_1,P_3,P_{\clo_3})
\frac{\fker{P_{\clo_3}}{P_2}{P_{\clo_1}}{P_1}{P_3}{P_{\clo_2}}}{\rho_0(P_2)}\right|^2 +\order{\e^{-4\pi (\alpha_\chi P + \bar \alpha_\chi \bar P)}}\,.
\end{equation}
Interestingly, the result depends explicitly on the OPE coefficient $C_{\clo_1\clo_2\clo_3}$. In Section \ref{sec:gravity}, we will interpret this dependence from a gravitational point of view. The left-hand side of this equation is clearly invariant under the simultaneous cyclic permutation $(1\,2\,3)$ and $(\clo_1\,\clo_2\,\clo_3)$. Although not immediately obvious, this permutation symmetry is present on the right-hand side as well, where it follows from the tetrahedral symmetry of the $\bbf$ kernel \eqref{eq:tetrahedral_sym}.

An important check of our result is the scaling of the leading term with the Cardy entropy. Using the asymptotic expansions in Appendix \ref{app:kernels}, we find that in the large-$P$ limit, this contraction scales as 
\begin{equation}
\overline{C_{1\clo_12}C_{2\clo_23}C_{3\clo_31}} \sim \e^{-2 S_0(P,\bar P)}, \quad \text{with}\quad S_0(P,\bar P) \coloneqq\log \rho_0(P,\bar P).  
\end{equation}
This is precisely the scaling expected from the framework of generalized ETH. Remarkably, we were able to find this answer using only the relation between Virasoro conformal blocks in \eqref{eq:relker}. This should be seen as evidence that chaotic CFTs are consistent with the generalized version of the ETH ansatz. 

\subsection{Thermal four-point function}
\label{sec:4-pt}

We can derive higher-point correlation functions in the ensemble of OPE coefficients by examining thermal $n$-point functions. 
The strategy is essentially the same as for the three-point function: first, write the thermal correlator in the necklace and OPE channels, then derive a crossing equation using the kernel that relates the two different bases of Virasoro conformal blocks, and finally, consider a suitable large-$P$ limit where an identity exchange dominates in the dual channel. 

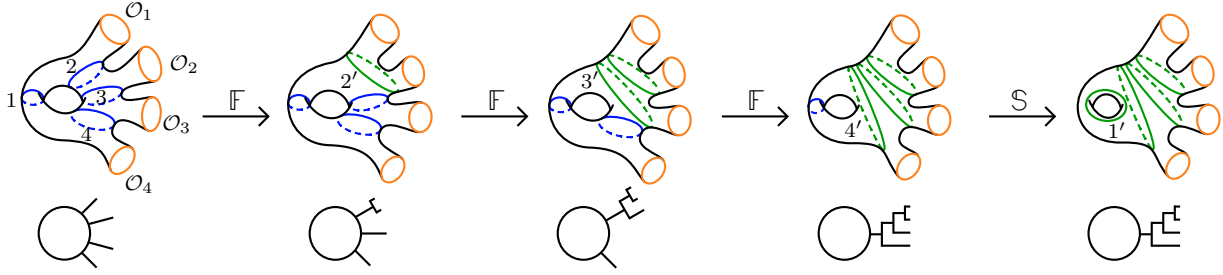
\begin{figure}
\centering
\begin{tikzpicture}
\node at (0,0) [midway]{\includegraphics{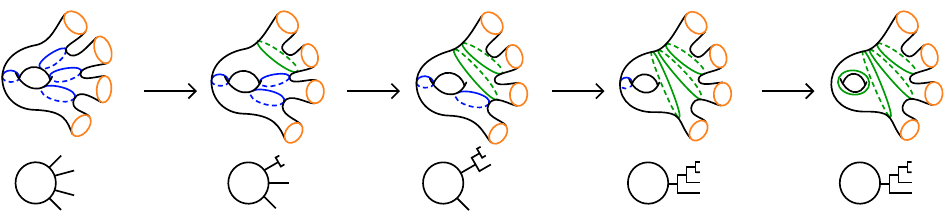}};
\node at (-5.1,0.2) [above] {$\mathbb{F}$};  
\node at (-1.65,0.2) [above] {$\mathbb{F}$}; 
\node at (1.8,0.2) [above] {$\mathbb{F}$}; 
\node at (5.3,0.2) [above] {$\mathbb{S}$}; 
\node at (-6.4,1.4) [above] {\scriptsize{$\clo_1$}};
\node at (-5.8,0.7) [above] {\scriptsize{$\clo_2$}};
\node at (-5.9,0.4) [below] {\scriptsize{$\clo_3$}};
\node at (-6.4,-0.4) [below]{\scriptsize{$\clo_4$}};
\node at (-8.1,0.25) [above] {\scriptsize{$1$}};
\node at (-7.3,0.7) [above] {\scriptsize{$2$}};
\node at (-6.9,0.28) [above] {\scriptsize{$3$}};
\node at (-7.1,-0.2) [above] {\scriptsize{$4$}};
\node at (-3.6,0.5) [above] {\scriptsize{$2'$}};
\node at (-0.4,0.5) [above] {\scriptsize{$3'$}};
\node at (3.1,-0.15) [above] {\scriptsize{$4'$}};
\node at (6.6,-0.17) [above] {\scriptsize{$1'$}};
\end{tikzpicture}
\caption{\small The sequence of moves relating the necklace and OPE channel for the torus four-point function. Depicted below each surface is the corresponding trivalent diagram representing the pair-of-pants decomposition.}
\label{fig:moves4pt}
\end{figure}

There is a crucial difference, however, between $n = 3$ and $n > 3$. When $n > 3$, the resulting asymptotic formula is not dominated by a single term but consists of a sum over the spectrum of the CFT$_2$. The purpose of this section is to understand this sum when $n = 4$.
The sequence of crossing moves we consider is shown in Figure \ref{fig:moves4pt}. 
Following the same steps as in the previous section, the crossing equation is given by
\begin{gather}
\label{eq:crosseq4pt}
\mathcal{Q}_4^{\mathcal{N}}(\bfP,\bar \bfp) = 
\frac{1}{\rho_0(\bfp,\bar \bfp)}
\sum_{1'\dots 4'} 
C_{2'\clo_1\clo_2}
C_{3'2'\clo_3}
C_{4'3'\clo_4}
C_{1'4'1'} 
\times
\knl{K}{\bfp'}{\bfp}\;
\knl{K}{\bar \bfp'}{\bar \bfp},
\end{gather}
where now $\mathbf{P}=(P_1,P_2,P_3,P_4)$. From the sequence of crossing transformations shown in Figure \ref{fig:moves4pt}, we can read off the crossing kernel between the necklace and OPE channel blocks to be  
\begin{equation}\label{eq:4pointcrossingkernel}
\knl{K}{P_1',\dots,P_4'}{P_1,\dots,P_4} 
=
\fker{P_2'}{P_2}{P_3}{P_{\clo_2}}{P_{\clo_1}}{P_1}     
\fker{P_3'}{P_3}{P_4}{P_{\clo_3}}{P_2'}{P_1}
\fker{P_4'}{P_4}{P_1}{P_{\clo_4}}{P_3'}{P_1}
\sker{P_1'}{P_1}{P_4'}.
\end{equation}

We now want to simplify the right-hand side of the crossing equation \eqref{eq:crosseq4pt} by using the large-$P$ asymptotics of the crossing kernel. Fortunately, we can use the same property \eqref{eq:supression1} of the
modular kernel $\mathbb{S}$ that we used before. The appearance of $\mathbb{S}$ in the crossing kernel \eqref{eq:4pointcrossingkernel} organizes the sum over the spectrum into an exponential hierarchy. So the dominant contribution to the crossing equation in \eqref{eq:crosseq4pt} is given by the identity exchange of $1'$ with $\alpha_1' = 0$.  Furthermore, the selection rules given by $C_{\bbi 4'\bbi}$ and $C_{\bbi 3' \clo_4}$ set $4'=\bbi$ and $3'=\clo_4$. 

Hence the resulting expression for the  necklace OPE density is given by
\begin{multline}
\label{eq:cyclic4pt}
\mathcal{Q}_4^{\mathcal{N}}(\bfP,\bar \bfP) = \frac{1}{\prod_{j=2}^{4}\rho_0(P_j,\bar P_j)}
\sum_{2'} 
C_{2'\clo_1\clo_2}
C_{\clo_4\,2'\,\clo_3}\left|\fker{P_2'}{P_2}{P_3}{P_{\clo_2}}{P_{\clo_1}}{P_1}     
\fker{P_{\clo_4}}{P_3}{P_4}{P_{\clo_3}}{P_2'}{P_1}
\fker{\bbi}{P_4}{P_1}{P_{\clo_4}}{P_{\clo_4}}{P_1}\right|^2 \\
\;
+
  \order{\e^{ -4\pi\alpha_{\chi}P-4\pi\bar{\alpha}_{\chi}\bar{P}}}.
\end{multline}
As before, we define the average in terms of the OPE density $\mathcal{Q}^{\mathcal{N}}$, after a suitable smearing in the conformal weights: \vspace{2mm}
\begin{equation}
\label{eq:cyclic4}
Q_{4}^{\mathcal{N}}(\bfP,\bar \bfP)
\;\; \xrightarrow{\;\;\text{smearing}\;\;} \;\;
\mean{
C_{1\clo_12}
C_{2\clo_23}
C_{3\clo_34}
C_{4\clo_41}
}.\vspace{2mm}
\end{equation}
The result \eqref{eq:cyclic4pt} should be seen as an asymptotic formula for the averaged value of four OPE coefficients with a cyclic index structure. 

Unlike the analog expression for three external operators, this time \eqref{eq:cyclic4pt} contains an explicit sum over the spectrum of the CFT$_2$ weighted by a product of fusion kernels. We will now show that this sum leads to the following three important contractions:
\begin{equation}
\label{contractions}
\begin{split}
\mean{C_{1\clo_12}C_{2\clo_23}C_{3\clo_34}C_{4\clo_41}} 
\,\,=\!\!
&\;\,\phantom{+}
\delta_{\clo_1\clo_2}
\delta_{\clo_3\clo_4}
\left|\frac{\delta(P_1-P_3)}{\rho_0(P_1)} C_0(P_{\clo_1},P_1,P_2)C_0(P_{\clo_3},P_3,P_4)\right|^2
\\[10pt] &+
\delta_{\clo_1\clo_4} 
\delta_{\clo_2\clo_3}
\left|\frac{\delta(P_2-P_4)}{\rho_0(P_2)}C_0(P_{\clo_1},P_1,P_4)C_0(P_{\clo_2},P_2,P_3)\right|^2
\\[10pt] &
+ \Big|g_4(P_1,P_2,P_3,P_4)\Big|^2\;
+ \cdots
\end{split}
\end{equation}
The first two terms correspond to the Gaussian part of the statistics. The last contraction, $g_4(P_1,\dots, P_4)$, is the cyclic non-Gaussianity that appears in generalized ETH. The corrections $+\cdots$ correspond to exponentially suppressed exchanges with $1'\neq \bbi$.

The first Gaussian contraction is reproduced by the crossing equation via the exchange of the identity operator in the sum of \eqref{eq:cyclic4pt}, $2'=\bbi$. This identity exchange leads to the expression\vspace{1mm}
\begin{equation}
\label{eq:gaus1}
\mean{
C_{1\clo_12}
C_{2\clo_23}
C_{3\clo_34}
C_{4\clo_41}
}\supset
\delta_{\clo_1\clo_2}\delta_{\clo_3\clo_4}\left|
\frac{\delta(P_1-P_3)}{\rho_0(P_1)} C_0(P_{\clo_3},P_3,P_4)C_0(P_{\clo_1},P_1,P_2)\right|^2.\vspace{1mm}
\end{equation}
The Kronecker delta functions are due to the OPE coefficients $C_{2'\clo_1\clo_2}$ and $C_{2'\clo_3\clo_4}$, and the continuous delta function arises from the fusion kernel
\begin{equation}
    \fker{P_{\clo_3}}{P_3}{P_{\clo_3}}{P_4}{\bbi}{P_1} = \delta(P_1-P_3).
\end{equation}

The second contraction in \eqref{contractions} is not related to the exchange of a single state, but it is given by the sum over states in the heavy part of the spectrum, $P_2' \geq 0$. For these heavy exchanges, we can approximate the product of OPE coefficients by their corresponding averaged value. Again, this approximation is justified as long as the heavy spectrum is sufficiently dense. There are two contractions to consider, which differ by a phase factor:\vspace{1mm}
\begin{equation}
\label{eq:gausCont}
\mean{C_{2'\clo_1\clo_2}C_{\clo_42'\clo_3}}  = 
\Big(
\delta_{\clo_1\clo_4}\delta_{\clo_2\clo_3} + (-1)^{J_2'+J_{\clo_2}+J_{\clo_1}} \delta_{\clo_1\clo_3}\delta_{\clo_2\clo_4}\Big)
\big|C_0(P_2',P_{\clo_1},P_{\clo_2})\big|^2.\vspace{1mm}
\end{equation}
Let us consider the first contraction in \eqref{eq:gausCont}. If we substitute it into \eqref{eq:cyclic4pt}, and take the continuum approximation 
$
    \sum_{2'\neq \bbi}\rightarrow\int \dd P_2'\dd \bar P_2'\,\rho_0(P_2,\bar P_2),
$
then we find an integral expression for equation \eqref{eq:cyclic4pt}. 
The $P_2',\bar P_2'$ integrals can be performed exactly, using an integral identity known as the \emph{pentagon identity}; see equation \eqref{eq:penta} in Appendix \ref{app:identities_WH}. The result is the second Gaussian contraction:\vspace{1mm}
\begin{equation}
\label{eq:gauss2}
\mean{
C_{1\clo_12}
C_{2\clo_23}
C_{3\clo_34}
C_{4\clo_41}
}  \supset 
\delta_{\clo_1\clo_4}\delta_{\clo_2\clo_3}
\left|\frac{\delta(P_2-P_4)}
{\rho_0(P_2)}
C_0(P_4,P_{\clo_4},P_1)C_0(P_2,P_{\clo_2},P_3)
\right|^2. \vspace{1mm}
\end{equation}
Note that this time the Wick contraction gives a different pairing of the external operators.

Next, consider the second contraction in \eqref{eq:gausCont}. As we will see, it leads to a contribution to the non-Gaussianity $g_4(P_1,\dots,\bar P_4)$. The integral expression that results from this contraction has additional phases because of the permutation of the indices,\vspace{2mm}
\begin{equation}
    \delta_{\clo_1\clo_3}
    \delta_{\clo_2\clo_4}
    \left|\frac{
    C_0(P_4,P_{\clo_2},P_1)
    }
    {\rho_0(P_2)\rho_0(P_3)} 
    \int \dd P_2'\;
    \e^{i\pi\left(h_2' + h_{\clo_2}+h_{\clo_1}\right)}
    \fker{\bbi}{P_2'}{P_{\clo_2}}{P_{\clo_1}}{P_{\clo_1}}{P_{\clo_2}}
    \fker{P_{\clo_2}}{P_3}{P_4}{P_{\clo_1}}{P_2'}{P_1}
    \fker{P_2'}{P_2}{P_3}{P_{\clo_2}}{P_{\clo_1}}{P_1}\right|^2.\vspace{2mm}
\end{equation}
Here we chose the branch $(-1)^{J_2'}=e^{i\pi(h_2'-\bar h_2')}$, but the integral does not depend on this choice. The integral over $P_2'$ can be solved using the integral identities satisfied by $\mathbb{B}$ and $\mathbb{F}$ arising from the Moore-Seiberg consistency conditions. In Appendix \ref{app:identities_WH}, we show that the pentagon identity can be combined with the so-called \emph{hexagon identity} to give a simple expression for the $P_2'$ integral; see \eqref{app:hexapenta}. The result is
\begin{multline}
\label{eq:nongauss}
\mean{
C_{1\clo_12}
C_{2\clo_23}
C_{3\clo_34}
C_{4\clo_41}
}  
\supset 
\delta_{\clo_1\clo_3}
\delta_{\clo_2\clo_4}
(-1)^{J_1+J_2+J_3+J_4}\\ \times\left|
    C_0(P_4,P_1,P_{\clo_2})
    C_0(P_4,P_3,P_{\clo_1})
    \frac{\fker{P_4}{P_2}{P_{\clo_2}}{P_3}{P_{\clo_1}}{P_1}}
    {\rho_0(P_2)}
    \right|^2.
\end{multline}
 This is a universal contribution to the function $g_4(P_1,\dots,P_4)$ that defines the cyclic non-Gaussianity in \eqref{eq:cyclic4pt}. Interestingly, comparing this result to the $6j$ contraction in \eqref{eq:non-Gauss}, we see that the two expressions are structurally very similar. The phase factors in \eqref{eq:nongauss} can be absorbed by a rearrangement of the indices of the OPE coefficients, using \eqref{eq:condition1}. In doing so, the $6j$ contraction can be analytically continued to the above formula \eqref{eq:nongauss} by continuing two of its heavy momenta to imaginary values $P\to P_\clo \in i\mathbb{R}$. 
 
Lastly, let us derive the entropy suppression of the quartic cyclic contraction, and check whether it agrees with the prediction of the generalized ETH. Using the asymptotic formulas for the crossing kernels given in Appendix \ref{app:asymptotic}, it is easy to see from \eqref{eq:nongauss} that 
\begin{equation}
|g_4(P_1,\dots, P_4)|^2 =  \overline{C_{1\clo_12}C_{2\clo_23}C_{3\clo_34}C_{4\clo_41}}\mlab{connected} \sim \e^{-3 S_0(P,\bar P)}
\end{equation}
thanks to the factor of $1/\rho_0$ in each $C_0$ function (see equation \eqref{eq:FC_0}). This scaling with the Cardy entropy is precisely
the answer predicted by the generalized ETH.

\subsection{Generalization to higher moments}

The analysis for higher moments is conceptually straightforward, as it follows the same steps as in the previous sections. To derive an expression for the $n$th cyclic contraction, we write down a crossing equation for the $n$-punctured torus. Fundamentally, the crossing equation describes a relationship between two OPE densities: the necklace density $\mathcal{Q}^\mathcal{N}_n(\bfp,\bar \bfp)$, and a second simpler density $\mathcal{P}_{n-3}(\bfp,\bar \bfp)$. 
In this section, we clarify this relationship and explain how it determines the cyclic moments of the ensemble. We start with their definitions:
\begin{align}
\mathcal{Q}^\mathcal{N}_n(\bfp_n,\bar \bfp_n) &\coloneqq \frac{1}{\rho_0(\bfP_n,\bar \bfp_n)}\sum_{1',\dots, n'} C_{1'\clo_12'}C_{2'\clo_23'}\cdots C_{n'\clo_n 1'}\; 
\prod_{i=1}^{n}
\delta(P_i - P'_i)\delta(\bar P_i - \bar P'_i)
, \\[1em]
    \mathcal{P}_{n-3}(\bfp_{n-3},\bar \bfp_{n-3}) &\coloneqq 
    \frac{1}{ \rho_0(\bfP_{n-3},\bar \bfp_{n-3})}
    \sum_{1',\dots, (n-3)'} 
    C_{\clo_1\clo_2\,1'}C_{1'\,\clo_3\,2'}\cdots C_{(n-3)'\,\clo_{n-1}\clo_n}
    \\[-9pt] \notag & \phantom{.}\hspace{6.3cm} \times
    \prod_{i=1}^{n-3}
\delta(P_i - P'_i)\delta(\bar P_i - \bar P'_i), 
\end{align}
where we have use the notation $\bfp_k = (P_1,\dots, P_k)$. It is often more useful to represent these densities using trivalent diagrams. To take into account the orientation of the OPE coefficients, we read each vertex in a clockwise orientation. Diagrammatically, we have 
\begin{align}
\label{eq:qdiagram}
\mathcal{Q}^\mathcal{N}_n(\bfp_n,\bar \bfp_n) &= 
\vcenter{\hbox{
\begin{tikzpicture}
\node at (0,0) {\includegraphics{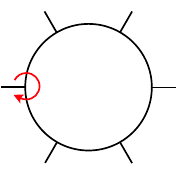}};
\node at (-1.8,0){$\clo_1$};
\node at (-0.9,1.62){$\clo_2$};
\node at (0.9,1.62){$\clo_3$};
\node at (1.8,0){$\clo_4$};
\node at (0.9,-1.62){$\clo_5$};
\node at (-0.9,-1.62){$\clo_n$};
\node at (-1.26,0.7){\scriptsize{$P_2$}};
\node at (0,1.4){\scriptsize{$P_3$}};
\node at (1.26,0.7){\scriptsize{$P_4$}};
\node at (1.26,-0.7){\scriptsize{$P_5$}};
\node at (-0,-1.4){$\dots$};
\node at (-1.26,-0.7){\scriptsize{$P_1$}};
\end{tikzpicture}
}} \\ 
\mathcal{P}_{n-3}(\bfp_{n-3},\bar \bfp_{n-3}) &=
\vcenter{\hbox{
\begin{tikzpicture}
\node at (0,0) {\includegraphics{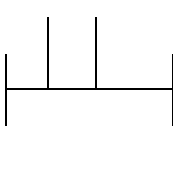}};
\node at (-0.8,2){$\phantom{.}$};
\node at (-1.7,-1.){$\clo_1$};
\node at (-1.7,1.){$\clo_2$};
\node at (-0.75,1.5){$\clo_3$};
\node at (0.15,1.5){$\clo_4$};
\node at (0.8,0.3){$\dots$};
\node at (1.7,1){$\clo_{n-1}$};
\node at (1.7,-1){$\clo_{n}$};
\node at (-0.97,0.4){\scriptsize{$P_1$}};
\node at (-0.3,0.4){\scriptsize{$P_2$}};
\end{tikzpicture}
}},
\end{align}
where the red arrow indicates the orientation of the diagram and we have omitted the antiholomorphic labels for clarity. The connection between these two diagrams is given by a sequence of $\mathbb{F}$ transforms followed by a single $\mathbb{S}$ move. The moves can be depicted as follows
\begin{equation}
\label{eq:movves}
\vcenter{
\hbox{
\begin{tikzpicture}
\node at (0,0) {\includegraphics{img/QN2.pdf}};
\node at (0,-1.1) {$\cdots$};
\node at (-1,-0.65) {\scriptsize{$P_1$}};
\node at (-1,0.65) {\scriptsize{$P_2$}};
\node at (-0.8,-1.5) {$\clo_n$};
\node at (-1.65,0) {$\clo_1$};
\end{tikzpicture}
}
}
\stackrel{\bbf} {\longrightarrow}
\vcenter{\hbox{
\begin{tikzpicture}
\node at (0,0) {\includegraphics{img/QN3.pdf}};
\node at (-0.5,0.9) {\scriptsize{$P_2'$}};
\node at (0.15,-1.1) {$\cdots$};
\node at (-0.8,-0.5) {\scriptsize{$P_1$}};
\end{tikzpicture}
}}
\stackrel{(\bbf)^{n-2}} {\relbar\!\!\relbar\!\!\longrightarrow}
\vcenter{\hbox{
\begin{tikzpicture}
\node at (0,0) {\includegraphics{img/OPE2.pdf}};
\node at (-0.1,0.2) {\scriptsize{$P_n'$}};
\node at (-0.8,1) {$\ddots$};
\node at (-1,-0.75) {\scriptsize{$P_1$}};
\end{tikzpicture}
}}
\stackrel{\bbs} {\longrightarrow}
\vcenter{\hbox{
\begin{tikzpicture}
\node at (0,0) {\includegraphics{img/OPE2.pdf}};
\node at (-0.8,1) {$\ddots$};
\node at (-1,-0.75) {\scriptsize{$P_1'$}};
\end{tikzpicture}
}}\,.
\end{equation}
The sequence begins with an $\mathbb{F}$ transform acting on the internal momentum labeled by $P_2,\bar P_2$, followed by  $P_3,\bar P_3$ and so on, until we reach $P_n,\bar P_n$. At this stage we can apply a modular $\mathbb{S}$ transform to $P_1,\bar P_1$. This completes the sequence of moves and yields a channel that contains the density $\mathcal{P}_{n-3}(\bfp,\bar \bfp)$ and an additional loop. The crossing equation that follows from \eqref{eq:movves} is
\begin{multline}
   \rho_0(\bfp_n,\bar \bfp_n)\; \mathcal{Q}^{\mathcal{N}}_n(\bfP_n,\bar \bfP_n) \\[-4pt] =\!\! \int\!\! \dd^n P'\dd^n \bar P' \;\rho_0(\bfp'_n,\bar \bfp'_n) \left(\!\!\!\!
\vcenter{\hbox{
\begin{tikzpicture}
\node at (0,0) {\includegraphics{img/OPE2.pdf}};
\node at (-1,1.7) {$\clo_1$};
\node at (-0.3,1.7) {$\clo_2$};
\node at (0.3,1.7) {$\dots$};
\node at (0.95,1.7) {$\clo_n$};
\node at (-0.9,0.9) {\scriptsize{$P_2'$}};
\node at (-0.55,0.65) {\scriptsize{$\ddots$}};
\node at (-0.15,0.25) {\scriptsize{$P_n'$}};
\node at (0.55,-0.85) {\scriptsize{$P_1'$}};
\end{tikzpicture}
}}
\!\!\!\!\!\right)
\left|\fker{P_2'}{P_2}{P_3}{P_{\clo_2}}{P_{\clo_1}}{P_1}
\!\dots
\fker{P_n'}{P_n}{P_1}{P_{\clo_n}}{P_{(n-1)'}}{P_1}
\sker{P_1'}{P_1}{P_n'}
\right|^2\,.
\end{multline}
Thanks to the $\mathbb{S}$ transform in the final step of the sequence, 
the above integral is dominated by the exchange $1' = \bbi$ in the large-$P$ limit where $P_i = P+\delta_i$. The hierarchy is again given by the property \eqref{eq:supression1} of the modular $\mathbb{S}$ transform. When $1'=\bbi$, the selection rules given by the OPE coefficients $C_{1'n'1'}$ and $C_{(n-1)'\clo_n n'}$ inside the dual density set the exchanges $n' = \bbi$ and $(n-1)' = \clo_n$. 

Consequently, the formula when $1' = \bbi$ is given by
\begin{multline}
\label{eq:corss}
\frac{\rho_0(\bfP_n,\bar \bfP_n)}{\rho_0(P_1,\bar P_1)}\mathcal{Q}^{\mathcal{N}}_n(\bfP_n,\bar \bfP_n)  = \int \dd^{n-3} P'\dd^{n-3} \bar P' \;
\rho_0(\bfp_{n-3}',\bar \bfp_{n-3}') \\ \times
\mathcal{P}_{n-3}(\bfP_{n-3}',\bar \bfP_{n-3}')
\left|
\fker{P_2'}{P_2}{P_3}{P_{\clo_2}}{P_{\clo_1}}{P_1}
\!\cdots
\fker{\bbi}{P_n}{P_1}{P_{\clo_n}}{P_{\clo_n}}{P_1}\right|^2\, + \order{\e^{-4\pi \alpha_\chi P-4\pi \bar\alpha_\chi \bar P}}.\vspace{2mm}
\end{multline} 
On the right-hand side of this equation, the indices run from $2'$ to $(n-2)'$, i.e. $\bfP_{n-3}' = (P_2',\dots, P_{n-2}')$. Here we have also evaluated the modular kernel $\sker{P_1'}{P_1}{P_n'}$ at $P_1'=P_n'=\bbi$ and divided the left- and right-hand side by a factor of $\rho_0(P_1,\bar P_1)$. 

Identifying the density $\mathcal{P}_{n-3}(\bfp'_{n-3},\bar \bfp'_{n-3})$ with an average of OPE coefficients, this crossing equation takes into account all the possible contractions of \vspace{2mm}
\begin{equation}
\label{eq:density6p}
  \mathcal{P}_{n-3}(\bfP_{n-3}',\bar \bfp_{n-3}') \;\;\xrightarrow{\;\text{smearing}\;}\;\; \mean{
  C_{{\clo_1\clo_2\,2'}}
  C_{2'\clo_3\,3'}
  C_{3'\clo_4\,4'}
  \dots
  C_{(n-2)'\,\clo_{n-1}\clo_{n}}
  }.\vspace{2mm}
\end{equation} 
Furthermore, aside from the Gaussian contractions, we also expect non-Gaussianities inside this distribution. This gives a recursive procedure for obtaining the $n$th moment in terms of lower moments. 

To completely solve for the moments of \eqref{eq:density6p} is outside the scope of this paper, however, it is possible to find such moments using crossing kernels. Similar contractions have been studied in the literature, see e.g. \cite{Anous:2021caj}, and the idea is again to relate a given density to a simpler one involving a smaller number of OPE coefficients. For the purposes of this paper, it suffices to estimate the order of magnitude for the $n$th cyclic contraction, which we argue to be\vspace{1mm}
\begin{equation}\label{eq:nthcyclic}
|g_n(P_1,\dots P_n)|^2 = \mean{C_{1\clo_12}
C_{2\clo_23}\cdots
C_{n \clo_n1}}\mlab{cyclic} \sim \e^{-(n-1)S_0(P,\bar P)}.\vspace{1mm}
\end{equation}
Namely, in equation \eqref{eq:corss}, the left-hand side is explicitly multiplied by a factor of 
\begin{equation}
    \e^{(n-1)S_0(P,\bar P)} \approx \e^{2\pi (n-1)(P+ \bar P) Q}.
\end{equation}
Meanwhile, the right-hand side grows at most like a power of $P, \bar P$. This is because, in this limit, the fusion kernels are not exponentially dependent on $P,\bar P$ but grow at most like a power of $P,\bar P$. See also Appendix \ref{app:asymptotic}.
The saddle-point of this integral, if it exists, is determined by the $P,\bar P$-independent component 
\begin{equation}
    \rho_0(\bfp_{n-3}',\bar \bfp_{n-3}') \mathcal{P}_{n-3}(\bfP_{n-3}',\bar \bfP_{n-3}').
\end{equation}
together with the $P,\bar P$-independent part of the crossing kernels. Evaluating the integrand on the saddle $P'_\star,\bar P'_\star$, we get at most a power-law scaling in $P,\bar P$ from the crossing kernels. The result is that, in order for the right-hand side of \eqref{eq:corss} to be of the same order as the left-hand side, the cyclic contraction must be of the order of $\e^{-(n-1)S_0(P,\bar P)}$, as claimed.

\section{Match to gravity}
\label{sec:gravity}

In the previous section, we have derived expressions for heavy-heavy-light statistical moments by imposing crossing symmetry of the boundary CFT$_2$ on the torus with multiple punctures. In this section, we will match these connected contributions to new multi-boundary Euclidean wormholes in the dual description. To do so, we use the description of AdS$_3$ quantum gravity with massive point particles in terms of Virasoro TQFT as explained and developed in the recent works \cite{Collier:2023fwi,Collier:2024mgv}. To be self-contained, we briefly review the main elements of this construction. 

\subsection{Bulk Hilbert space and Virasoro TQFT}
To determine the bulk Hilbert space of pure 3D gravity with negative cosmological constant on a 3-manifold $M$, consider a spatial slice $\Sigma$, which we take to be of genus $g$ and having $n$ punctures. Then the gravitational phase space consists of two copies of Teichm\"uller space $\mathcal{T}\times\overline{\mathcal{T}}$ of  $\Sigma$ \cite{universalphasespace,Krasnov:2005dm}.\footnote{One way to see this is to rewrite the Einstein-Hilbert action in the first order formalism as two copies of $PSL(2,\mathbb{R})$ Chern-Simons theory \cite{WITTEN198846}. The space of classical solutions is then a subspace of the space of flat connections $A,\bar A$. This subspace is precisely the Teichm\"uller component $\mathcal{T}\times \overline{\mathcal{T}}$, consisting of the flat connections that give rise to a non-degenerate metric $g$.}
Each copy of Teichm\"uller space can be quantized using geometric quantization, by introducing the line bundle of Virasoro conformal blocks on $\mathcal{T}$ \cite{Witten:1988hf}. The quantum wavefunctions are holomorphic sections of this bundle \cite{Verlinde:1989ua}. 
Since the classical phase space of 3D gravity is two copies of Teichm\"uller space, the quantum Hilbert space  consists of two copies of the space of Virasoro conformal blocks
\begin{equation}\label{eq:Hilbert_space}
    \mathcal{H}(\Sigma) = \mathcal{H}_{\mathrm{Vir}}(\Sigma)\otimes \overline{\mathcal{H}_{\mathrm{Vir}}}(\Sigma)\,.
\end{equation}
We denote states in $\mathcal{H}_\Sigma$ by ket vectors, labelled by a choice of pair-of-pants decomposition $\mathcal{C}$ of the surface $\Sigma$. A continuous basis of states is spanned by
\begin{equation}\label{eq:TQFTbasis}
    \ket{\mathcal{F}^{\,\mathcal{C}}_{g,n}(\mathbf{P};\mathbf{P}_{\clo})}\otimes \ket{\bar{\mathcal{F}}^{\,\mathcal{C}}_{g,n}(\bar{\mathbf{P}};\bar{\mathbf{P}}_{\clo})} \,.
\end{equation}
The `internal momenta' $\mathbf{P} = (P_1,\dots,P_{3g-3+n})$ are real non-negative parameters associated to the set of non-intersecting cycles in the pair-of-pants decomposition $\mathcal{C}$, while the `external momenta' $\bfP_\clo = (P_{\clo_1},\dots,P_{\clo_n})$ are associated to the punctures of $\Sigma$. The left- and right-moving conformal blocks of the previous section are recovered from these ket vectors via the wavefunction basis
\begin{equation}
   \mathcal{F}^{\,\mathcal{C}}_{g,n}(\mathbf{P};\mathbf{P}_\clo;\Omega) = \bra{\Omega}\ket{\mathcal{F}^{\,\mathcal{C}}_{g,n}(\mathbf{P};\mathbf{P}_\clo)},
\end{equation}
where $\Omega$ collectively denote the moduli of $\Sigma$. 

To make $\mathcal{H}_\Sigma$ into a Hilbert space, we need to provide an inner product. The inner product on the space of conformal blocks $\mathcal{H}_{\mathrm{Vir}}$ was determined in \cite{Collier:2023fwi} for $2g-2+n >0$\footnote{The Virasoro TQFT inner product on non-hyperbolic surfaces is ill-defined, which means that the sphere and the torus without punctures can not be used as an intermediate slice $\Sigma$.}, with the simple form
\begin{equation}\label{eq:innerproduct}
    \braket{\mathcal{F}^{\,\mathcal{C}}_{g,n}(\mathbf{P};\mathbf{P}_{\clo})}{\mathcal{F}^{\,\mathcal{C}}_{g,n}(\mathbf{P}\dash;\mathbf{P}_{\clo})} = \frac{\delta^{3g-3+n}(\mathbf{P}-\mathbf{P}\dash)}{\rho_{g,n}(\mathbf{P})}\,,\vspace{2mm}
\end{equation}
and a similar expression for the right-moving sector with all $\bfP$ replaced by $\bar{\bfP}$. The normalization factor is given by
\begin{equation}\label{eq:normalization}
    \rho_{g,n}(\mathbf{P}) = \prod_{\substack{\mathrm{cuffs}\\ a}}\rho_0(P_a) \prod_{\substack{\mathrm{pairs\,of\,pants} \\ (i,j,k)}}C_0(P_i,P_j,P_k),
\end{equation}
where the functions $\rho_0(P)$ and $C_0(P_i,P_j,P_k)$ have been defined in \eqref{eq:rho0function} and \eqref{eq:C0function}, respectively. One can think of this factor as the Feynman amplitude associated to the dual graph of $\mathcal{C}$, where each pair-of-pants corresponds to a trivalent vertex, and each cuff to a propagator. 

The Hilbert space of Virasoro conformal blocks carries a projective unitary representation of the Moore-Seiberg groupoid, which is generated by the elementary crossing moves depicted in Figure \ref{fig:moves}, together with all the Dehn twists around the cycles in $\mathcal{C}$ \cite{Moore:1988qv}. This means that an element $\gamma_{12}$ that maps the cycles in $\mathcal{C}_1$ to those in $\mathcal{C}_2$, can be represented by a unitary operator\footnote{A generating set for the unitary operators $\mathbb{U}(\gamma)$ can be found in Appendix \ref{app:kernels}. The generators are the $\mathbb{S}$-transform of the once-punctured torus \eqref{eq:app_mod}, the $\mathbb{F}$-transform of the four-punctured sphere \eqref{eq:app_fusion}, and the braiding move $\mathbb{B}$ \eqref{eq:app_braid}, which were constructed explicitly in \cite{Ponsot:1999uf,Ponsot:2000mt}.}
\begin{equation}
  \ket{\mathcal{F}^{\,\mathcal{C}_2}_{g,n}(\mathbf{P};\mathbf{P}_{\clo})} = \mathbb{U}(\gamma_{12})\cdot \ket{\mathcal{F}^{\,\mathcal{C}_1}_{g,n}(\mathbf{P};\mathbf{P}_{\clo})} .
\end{equation}
The fact that $\mathbb{U}^\dagger \mathbb{U}$ acts as the identity (up to a phase) then shows that the normalization factor $\rho_{g,n}(\bfP)$ in \eqref{eq:normalization}
is independent of the choice of pair-of-pants decomposition.

Having described the space of states, one can now use the fact that pure 3D gravity is a topological theory to compute amplitudes for any 3-manifold $M$. Namely, one postulates the usual TQFT axioms for cutting and glueing along intermediate slices $\Sigma_i$, associating a Hilbert space $\mathcal{H}_{\Sigma_i}$ defined by \eqref{eq:Hilbert_space} to each slice. The partition function $| Z_{\mathrm{Vir}}(M)|^2$ on a 3-manifold with $n$ boundaries $\partial M = \bigcup_{k=1}^n\Sigma_k$ is then a state in the product Hilbert space 
\begin{equation}
\mathcal{H}\Big(
{\textstyle\bigcup_{k=1}^n\Sigma_k}
\Big) =  \bigotimes_{k=1}^n\,\mathcal{H}_{\mathrm{Vir}}(\Sigma_k) \!\otimes \overline{\mathcal{H}_{\mathrm{Vir}}}(\Sigma_k)
\end{equation}
whose expansion coefficients in the basis \eqref{eq:TQFTbasis} are determined by the glueing prescription of the bulk 3-manifold. This construction was named the Virasoro TQFT \cite{Collier:2023fwi}.\footnote{The Virasoro TQFT shares some similarities to the state-integral models based on quantum Teichm\"uller theory \cite{EllegaardAndersen:2011vps,andersen2013new, andersen2018teichmuller,Hikami2001,Hikami_2007}. However, the latter requires a triangulation of the hyperbolic 3-manifold into ideal tetrahedra, which is not needed to define the Virasoro TQFT.} 

To obtain the full 3D gravity partition function from the Virasoro TQFT on a manifold $M$ with boundary $\partial M$, one still has to sum over the modular images of the boundary mapping class group, quotiented by the bulk mapping class group of $M$, 
\begin{equation}\label{eq:fullgravityPI}
    Z_{\mathrm{grav}}(M) = \sum_{\gamma \in \frac{\mathrm{MCG}(\partial M)}{\mathrm{MCG}(M)}} |Z_{\mathrm{Vir}}(M^\gamma)|^2.
\end{equation}
The quotient is well defined because for hyperbolic 3-manifolds the bulk mapping class group maps injectively into the boundary mapping class group. Again, we refer to \cite{Collier:2023fwi} for the detailed prescription and subtleties. Instead, we now work out a simple example of the above formalism which will serve as the blueprint of the more general wormhole computations in Sections \ref{sec:multibdy} and \ref{sec:4bdyWH}.

\subsection{Example: torus one-point wormhole}\label{sec:onepointWH}
\begin{figure}
    \centering
    \includegraphics[width=0.95\textwidth]{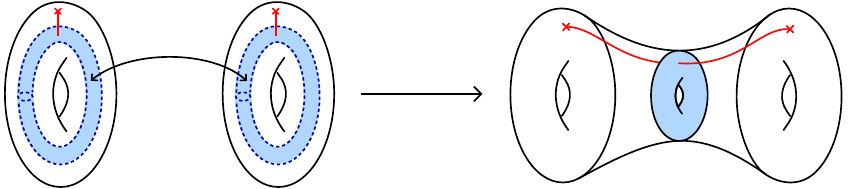}
    \caption{{\small Glueing along the inner torus boundaries in blue produces the punctured-torus wormhole.}}
    \label{fig:MMwormhole}
\end{figure}

Consider a 3-manifold with two asymptotic boundaries that are each a once-punctured torus. The simplest topology connecting the two boundaries is the Maldacena-Maoz wormhole \cite{Maldacena:2004rf} (generalized to punctured surfaces in \cite{Krasnov:2005dm}). It is topologically $\Sigma_{1,1}\times I$, and can be constructed by taking two solid tori, carving out a solid torus from the interior of each, and glueing the two manifolds along their inner torus boundaries. We add  boundary punctures, which extend into the bulk as trajectories of conical defects. The procedure is illustrated in Figure \ref{fig:MMwormhole}. 

In the Virasoro TQFT, this produces a state 
\begin{equation}
    \int_0^\infty \dd P \dd P\dash \braket{\mathcal{F}^{\,\mathrm{inner}}_{1,1}(P;P_\clo)}{\mathcal{F}^{\,\mathrm{inner}}_{1,1}(P\dash;P_\clo)} \,
    \ket{
    \widehat{\mathcal{F}}^{\,\mathrm{outer}}_{1,1}(P;P_\clo)
    }
    \otimes
    \ket{\widehat{\mathcal{F}}^{\,\mathrm{outer}}_{1,1}(P\dash;P_\clo)},
\end{equation}
where the hatted kets are normalized by the factor in \eqref{eq:normalization},
\begin{equation}\label{eq:normnotation}
    \ket{
   \widehat{\mathcal{F}}_{g,n}^{\,\mathcal{C}}(\bfp;\bfp_{\clo})}
   \coloneqq
   \rho_{g,n}(\bfp)
   \ket{
   \mathcal{F}_{g,n}^{\,\mathcal{C}}(\bfp;\bfp_{\clo})}.
\end{equation}
In the case of the one-holed torus, we have one internal cycle $P$, one vertex, and one external momentum $P_\clo$ (recall Figure \ref{fig:moves}b), so the normalization factor evaluates to
\begin{equation}\label{eq:torusnorm}
    \rho_{1,1}(P) =\rho_0(P)C_0(P,P,P_\clo).
\end{equation}

Using the inner product \eqref{eq:innerproduct} to evaluate the overlap between the states on the inner boundary and going to the wavefunction basis $\langle \tau_1|\otimes \langle \tau_2 |$ gives the following amplitude for this particular wormhole topology:
\begin{equation}\label{eq:torusonepointwh}
    Z_{\mathrm{Vir}}(\tau_1,\tau_2) =\int_0^\infty \dd P \,\rho_0(P)C_0(P,P,P_\clo) \,\mathcal{F}_{1,1}(P;P_\clo;\tau_1)\mathcal{F}_{1,1}(P;P_\clo;\tau_2).
\end{equation}
We have dropped the superscript `outer', as the inner boundaries have now been glued together to form the bottleneck of the wormhole. The torus conformal blocks are given by a trace over the descendants of a primary state with holomorphic conformal weight $h=\frac{c-1}{24}+P^2$,
\begin{equation}
    \mathcal{F}_{1,1}(P;P_\clo;\tau) = \Tr_h(q^{L_0-\frac{c}{24}}\clo).
\end{equation}

The full gravity partition function is then given by the product of the left- and right-moving sector, summed over boundary mapping class group transformations as in \eqref{eq:fullgravityPI}. In the example of the Maldacena-Maoz wormhole that we are considering, this modular sum is only over relative modular transformations of one of the boundary tori:
\begin{equation}\label{eq:torusonepointgrav}
    Z_{\mathrm{grav}}(\tau_1,\bar\tau_1;\tau_2,\bar\tau_2) = \sum_{\gamma \in PSL(2,\mathbb{Z})} \left | \int_0^\infty \dd P \,\rho_0(P)C_0(P,P,P_\clo) \,\mathcal{F}_{1,1}(P;P_\clo;\tau_1)\mathcal{F}_{1,1}(P;P_\clo;\gamma\cdot \tau_2) \right|^2 
\end{equation}
The reason to include a Poincar\'e sum over only the mapping class group of \emph{one} of the torus boundaries, is that the TQFT amplitude \eqref{eq:torusonepointwh} is already invariant under simultaneous boundary modular transformations 
\begin{equation}
    \tau_1\to\gamma\cdot \tau_1,\quad \tau_2\to \gamma^{-1}\cdot \tau_2,
\end{equation}
similarly to what has been found for the torus wormhole in \cite{Cotler:2020ugk}. This is should be seen as part of the bulk mapping class group, which is gauged. To see the invariance explicitly, recall that the $T$ transformation acts by multiplication by a phase, while $T^{-1}$ acts with the complex conjugate. To prove invariance under the $S$ transform, we need to use the following identity \cite{Collier:2019weq}:
\begin{equation}
    \rho_0(P)C_0(P,P,P_\clo)\mathbb{S}_{P'P}[P_\clo] =  \rho_0(P')C_0(P',P',P_\clo)\mathbb{S}_{PP'}[P_\clo],
\end{equation}
together with the idempotency relation satisfied by the $S$-kernel\footnote{Recall that $[S] = [S^{-1}]$ as conjugacy classes in $PSL(2,\mathbb{Z})$, while $\mathbb{S}^{-1} = e^{-i\pi \Delta_\clo}\mathbb{S}$ as operators in the TQFT.},
\begin{equation}
    \int_0^\infty \dd P\,\mathbb{S}_{P_1P}[P_\clo]\mathbb{S}_{PP_2}[P_\clo] = e^{i\pi \Delta_{\clo}}\delta(P_1-P_2).
\end{equation}
It is then an easy exercise to show that indeed 
\begin{equation}
\begin{split}
   & Z_{\mathrm{Vir}}(S\cdot \tau_1,S^{-1}\cdot \tau_2)=   Z_{\mathrm{Vir}}( \tau_1, \tau_2).
    \end{split}
\end{equation}

Let us remark that the sum over relative $PSL(2,\mathbb{Z})$ images in \eqref{eq:torusonepointgrav} can be understood as a sum over bulk topologies. Namely, we could have glued the inner boundaries in Figure \ref{fig:MMwormhole} with a relative mapping class group transformation $\gamma$ of the splitting surface $\Sigma_{1,1}$. Since the Virasoro TQFT carries a projective representation of the mapping class group, the TQFT partition function on this new wormhole topology is found to be
\vspace{2mm}
\begin{equation}\label{eq:PSLwormhole}
    \int_0^\infty \dd P \dd P\dash \bra{\mathcal{F}^{\,\mathrm{inner}}_{1,1}(P;P_\clo)}\mathbb{U}(\gamma)\ket{\mathcal{F}^{\,\mathrm{inner}}_{1,1}(P\dash;P_\clo)} \,\ket{\widehat{\mathcal{F}}^{\,\mathrm{outer}}_{1,1}(P;P_\clo)} \otimes \ket{\widehat{\mathcal{F}}^{\,\mathrm{outer}}_{1,1}(P\dash;P_\clo)}.
\end{equation}

A general element $\gamma\in PSL(2,\mathbb{Z})$ can be decomposed as some word in the generators $S$ and $T$, which in turn are represented by the $\mathbb{S}$-kernel and $\mathbb{T}$ acting on the blocks. Evaluating the integrals in the same way as before, we see that each term in the modular sum \eqref{eq:torusonepointgrav} arises from a wormhole of this type. This defines a $PSL(2,\mathbb{Z})$ family of Maldacena-Maoz wormholes, related to the `diagonal' wormhole $(\gamma = \mathbb{1})$ by Dehn surgery.\footnote{In Section \ref{sec:genus2}, we will see an example of a similar construction where higher genus wormholes of the form \eqref{eq:PSLwormhole} are not part of the boundary modular sum. The reason that it does hold for the once-punctured torus wormhole is that the mapping class group of the torus coincides with its set of Moore-Seiberg transformations. For more general higher genus surfaces this is no longer true.} The simplest non-trivial example, $\gamma = S$, corresponds to a wormhole which is topologically the punctured Hopf link, described in more detail in \cite{Collier:2024mgv}.

Finally, we note that the gravitational computation \eqref{eq:torusonepointgrav} can be matched to the ETH prediction in the boundary CFT$_2$, which is given by the connected average of a  product of thermal one-point functions. Suppose we use the Gaussian ansatz\vspace{1mm}
\begin{equation}\label{eq:Gaussian_ansatz_1point}
    \overline{C_{1\clo 1}C_{2\clo 2}} \,\supset\, \delta_{12} \,C_0(P_1,P_1,P_\clo)C_0(\bar P_1,\bar P_1,\bar P_\clo)\vspace{1mm}
\end{equation}
for the universal OPE two-point density. This is the leading contraction in the OPE ensemble described in Section \ref{sec:typicality}. Then we obtain for the average product of torus one-point functions\vspace{2mm}
\begin{align}\label{eq:twopointCFT}
    \overline{\langle \clo \rangle_{\tau_1}\langle \clo \rangle_{\tau_2}} &= \sum_{1,2}\overline{C_{1\clo 1}C_{2\clo 2}}\,\mathcal{F}_{1,1}(P_1;P_\clo;\tau_1)\bar{\mathcal{F}}_{1,1}(\bar P_1;\bar P_\clo;\bar\tau_1)\,\mathcal{F}_{1,1}(P_2;P_\clo;\tau_2)\bar{\mathcal{F}}_{1,1}(\bar P_2;\bar P_\clo;\bar\tau_2) \nonumber \\ 
    &\approx \left |\int_0^\infty \dd P \rho_0(P) C_0(P,P,P_\clo) \mathcal{F}_{1,1}(P;P_\clo;\tau_1)\mathcal{F}_{1,1}(P;P_\clo;\tau_2) \right|^2 
\end{align}
where in the second line we made the continuum approximation 
$\sum_{h,\bar h} \approx \int \dd P\dd \bar P \rho_0(P)\rho_0(\bar P)$ for the heavy states. We also used that $C_{\mathbb{1}\mathcal{O}\mathbb{1}}=0$ and assumed that the light spectrum was sparse enough to neglect its contribution to the sum over states. Note that the integral is strictly above the black hole threshold $h= \frac{c-1}{24}+P^2$, $P\geq 0$, consistent with \cite{Schlenker_2022}. In a previous work \cite{deBoer:2023vsm}, we showed that the saddle point of the integral is also above the black hole threshold when the conformal dimension of $\clo$ is in the conical defect range. 

We see that the above Gaussian approximation matches to the diagonal wormhole ($\gamma =1$) in the modular sum \eqref{eq:torusonepointgrav}. In order to see the modular images, which we argued to come from wormholes with non-trivial bulk topology, we have to add corrections to the Gaussian ansatz \eqref{eq:Gaussian_ansatz_1point}. For example, in order to reproduce the contribution coming from twisting the torus wormhole with a relative $S$-transform (i.e. $\gamma = S$), we need to add the off-diagonal contribution
\begin{equation}
\label{eq:non-diagonal}
    \overline{C_{1\clo 1}C_{2\clo 2}} \,\supset \, \left|\frac{C_0(P_1,P_1,P_\clo)\mathbb{S}_{P_1P_{2}}[P_\clo]}{\rho_0(P_{2})}\right|^2, \qquad (P_1\neq P_2) 
\end{equation}
Indeed, one easily checks that this reproduces \eqref{eq:PSLwormhole} with $\mathbb{U}(\gamma) = \mathbb{S}$. This same non-Gaussianity was recently argued for in \cite{Collier:2024mgv} by studying a three-punctured sphere wormhole with a tangle of conical defect lines. It can also be derived from a genus-two crossing equation in a modified dumbbell channel, where the two dumbbells are linked. Note that this non-Gaussianity is exponentially suppressed compared to the Gaussian contraction \eqref{eq:Gaussian_ansatz_1point}, due to the factor $\rho_0(P)^{-1}$, in accordance with the ETH expectation.

\subsection{Three-boundary wormhole}\label{sec:multibdy}

In order to extend the above discussion to the case of generalized ETH, 
we now set out to construct new wormhole topologies with multiple asymptotic boundaries. In this section, we assume each boundary component is a once-punctured torus, with an external operator $\clo$ inserted at the puncture. The boundary operator insertions correspond to massive heavy particles propagating in the bulk; these insertions make the geometry on-shell, as opposed to the case of multiple-boundary torus wormholes without punctures, which are always off-shell geometries \cite{Maloney_2010}.

We have seen in the previous section that higher, non-Gaussian, statistical moments may depend on non-universal data, like the structure constant {\small $C_{\clo\clo\clo}$} appearing in \eqref{eq:result1}. We will regard these `light data' as input in the gravitational computation. The bulk Virasoro TQFT will then only make use of the heavy intermediate states. The first non-trivial example will be the three-boundary amplitude obtained in \eqref{eq:result1}. 

\begin{figure}
    \centering
    \includegraphics[width=0.95\textwidth]{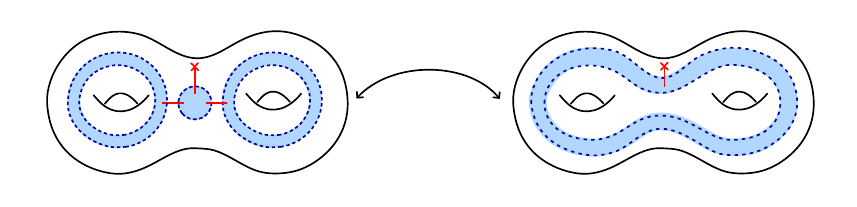}
    \caption{{\small Heegaard splitting of the three-boundary wormhole.}}
    \label{fig:genus2identify}
\end{figure}
The three-boundary wormhole topology {$M_3$} can be described  by the following Heegaard splitting,
illustrated in Figure \ref{fig:genus2identify}. Namely, take a pair of genus-two handlebodies {\small $M_{2,1}^{\mathsf{L}}$} and {\small $M_{2,1}^{\mathsf{R}}$}, with punctured boundary surfaces {\small $\Sigma_{2,1}^{\mathsf{L}}$} and {\small $\Sigma_{2,1}^{\mathsf{R}}$}. Carve out a pair of solid tori from {\small $M_{2,1}^{\mathsf{L}}$}, each passing through a different handle of {\small $M_{2,1}^{\mathsf{L}}$}. Carve out a single solid torus from {\small $M_{2,1}^{\mathsf{R}}$}, going through both handles of {\small $M_{2,1}^{\mathsf{R}}$}. This produces two compression bodies {\small $X^{\mathsf{L,R}}$} with boundary
 \begin{equation}
   \partial X^{\mathsf{L}} = \Sigma_{2,1}^{\mathsf{L}} \cup \Sigma^{\mathsf{L}_1}_{1,1} \cup \Sigma^{\mathsf{L}_2}_{1,1},\quad \partial X^{\mathsf{R}} = \Sigma_{2,1}^{\mathsf{R}} \cup \Sigma_{1,1}^{\mathsf{R}}.
 \end{equation}
 
Now, connect the puncture of {\small $\Sigma_{2,1}^{\mathsf{R}}$} with a bulk Wilson line to the puncture of the boundary torus {\small $\Sigma_{1,1}^{\mathsf{R}}$}, and connect the puncture of {\small $\Sigma_{2,1}^{\mathsf{L}}$} to the punctures of {\small $\Sigma^{\mathsf{L}_1}_{1,1}$} and {\small $\Sigma^{\mathsf{L}_2}_{1,1}$} via the three-point vertex of Wilson lines. We can deal with three-point vertices in the Virasoro TQFT by blowing up the vertex to a three-punctured sphere, with the correct normalization \cite{Collier:2023fwi}
\begin{equation}\label{eq:3pointnorm}
\vcenter{
\hbox{
\begin{tikzpicture}
\node at (0,0) {\includegraphics{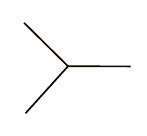}};
\node at (1.35,0) {$P_{\clo_1}$};
\node at (-1.2,1) {$P_{\clo_2}$};
\node at (-1.1,-0.9) {$P_{\clo_3}$};
\end{tikzpicture}
}
}
\quad \longrightarrow \quad
\frac{1}{C_0(P_{\clo_1},P_{\clo_2},P_{\clo_3})}
\vcenter{
\hbox{\begin{tikzpicture}
\node at (0,0) {\includegraphics{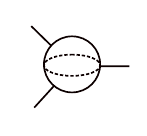}};
\node at (1.4,0) {$P_{\clo_1}$};
\node at (-1,1.) {$P_{\clo_2}$};
\node at (-0.7,-1) {$P_{\clo_3}$};
\end{tikzpicture}
}
}. 
\end{equation}
The normalization is chosen such that the three-punctured sphere wormhole evaluates to the $C_0(P_{\clo_1},P_{\clo_2},P_{\clo_3})$ formula. 
See also \cite{Chang:2016ftb,Berenstein_2023} for the semiclassical computation of a three-point vertex of conical defect lines in AdS$_3$. For simplicity, we take all external operators equal, $P_{\clo_i} = P_{\clo}$. 

As a next step, we glue the genus-two boundaries to each other. In principle, the identification can be made with any element $\gamma$ of the mapping class group of the genus-two splitting surface. For now, we take the identity element $\gamma = 1$, commenting on the more general case later. The glueing is represented in the Virasoro TQFT as an overlap between punctured genus-two conformal blocks, with the following assignment of internal momenta:
\begin{equation}\label{eq:genus2overlap}
\braket{\mathcal{F}^{\,\mathcal{C}}_{2,1}}{\mathcal{F}^{\,\mathcal{C}\dash}_{2,1}
} = 
\Bigg \langle
\vcenter{
\hbox{
\begin{tikzpicture}
\node at (0,0.4) {\includegraphics{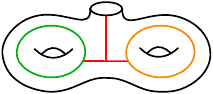}
    };
\node at (2,0) {$P_2$};
\node at (-2,0) {$P_1$};
\node at (0,1.4) {$P_\clo$};
\end{tikzpicture}}}
\Bigg\vert
\vcenter{
\hbox{\begin{tikzpicture}
\node at (0,0.4) {\includegraphics{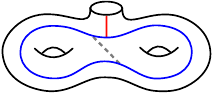}};
\node at (2,0) {$P_3$};
\node at (0,1.4) {$P_\clo$};
\node at (-0.25,0.35) {$\mathbb{1}$};
    \end{tikzpicture}}
}
\Bigg\rangle\,.
\end{equation}

\begin{figure}
    \centering
    \includegraphics[width=4.5cm]{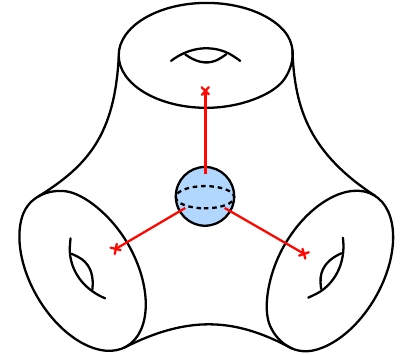}\vspace{0.2cm}
    \caption{{\small The Heegaard splitting described in this section leads to a three-boundary torus wormhole, whose conical defect lines meet in a bulk three-point vertex. The bulk has non-trivial topology (not shown in this figure), arising from the glueing of genus-two handlebodies.}}
    \label{fig:3boundaryWH}
\end{figure}

The resulting 3-manifold $M_3$ after glueing has three once-punctured torus boundaries, and one bulk three-point vertex (which, as explained above, can be seen as a three-punctured sphere boundary with a different normalization).\footnote{We thank Lorenz Eberhardt for insightful discussions on this construction.} This is illustrated in Figure \ref{fig:3boundaryWH}. 
Hence the Virasoro TQFT partition function is a state in $\mathcal{H}_{0,3} \otimes \mathcal{H}_{1,1}^{\otimes 3}$, whose expansion coefficients are determined by the overlap of conformal blocks defined in equation \eqref{eq:genus2overlap},
\begin{equation}
\ket{Z_{\mathrm{Vir}}(M_3)\vphantom{F^2}} =
     \int_0^\infty \dd P_1\dd P_2\dd P_3 \,\braket{\mathcal{F}^{\,\mathcal{C}}_{2,1}}{\mathcal{F}^{\,\mathcal{C}\dash}_{2,1}}\,  {\ket{\widehat{\mathcal{F}}_{0,3}(P_\clo,P_\clo,P_\clo)}} \bigotimes_{i=1}^3
     {\ket{\widehat{\mathcal{F}}_{1,1}(P_i;P_\clo)}}.
\end{equation}
The notation for the normalization factors of the hatted states is the same as in \eqref{eq:normnotation}. The overlap can be computed explicitly by applying crossing moves on the conformal blocks that bring the bra and ket to the same channel,
\tikzset{every picture/.style={line width=0.75pt}}
\begin{equation}
    \begin{tikzpicture}[x=0.75pt,y=0.75pt,yscale=-1,xscale=1,baseline={([yshift=-.5ex]current bounding box.center)}]
\draw   (250,210) .. controls (250,198.95) and (258.95,190) .. (270,190) .. controls (281.05,190) and (290,198.95) .. (290,210) .. controls (290,221.05) and (281.05,230) .. (270,230) .. controls (258.95,230) and (250,221.05) .. (250,210) -- cycle ;
\draw    (290,210) -- (330,210) ;
\draw   (330,210) .. controls (330,198.95) and (338.95,190) .. (350,190) .. controls (361.05,190) and (370,198.95) .. (370,210) .. controls (370,221.05) and (361.05,230) .. (350,230) .. controls (338.95,230) and (330,221.05) .. (330,210) -- cycle ;
\draw    (350,170) -- (350,190) ;
\draw [shift={(350,170)}, rotate = 90] [color={rgb, 255:red, 0; green, 0; blue, 0 }  ][fill={rgb, 255:red, 0; green, 0; blue, 0 }  ][line width=0.75]      (0, 0) circle [x radius= 1.34, y radius= 1.34]   ;
\draw   (68,210) .. controls (68,198.95) and (90.39,190) .. (118,190) .. controls (145.61,190) and (168,198.95) .. (168,210) .. controls (168,221.05) and (145.61,230) .. (118,230) .. controls (90.39,230) and (68,221.05) .. (68,210) -- cycle ;
\draw  [dash pattern={on 1.5pt off 1.5pt on 1.5pt off 1.5pt}]  (101.33,191.17) -- (134.3,228.83) ;
\draw    (118,170) -- (118,190) ;
\draw [shift={(118,170)}, rotate = 90] [color={rgb, 255:red, 0; green, 0; blue, 0 }  ][fill={rgb, 255:red, 0; green, 0; blue, 0 }  ][line width=0.75]      (0, 0) circle [x radius= 1.34, y radius= 1.34]   ;
\draw    (180,210) -- (217,210) ;
\draw [shift={(220,210)}, rotate = 180] [fill={rgb, 255:red, 0; green, 0; blue, 0 }  ][line width=0.08]  [draw opacity=0] (6.25,-3) -- (0,0) -- (6.25,3) -- cycle    ;
\draw   (470,210) .. controls (470,198.95) and (478.95,190) .. (490,190) .. controls (501.05,190) and (510,198.95) .. (510,210) .. controls (510,221.05) and (501.05,230) .. (490,230) .. controls (478.95,230) and (470,221.05) .. (470,210) -- cycle ;
\draw    (510,210) -- (550,210) ;
\draw   (550,210) .. controls (550,198.95) and (558.95,190) .. (570,190) .. controls (581.05,190) and (590,198.95) .. (590,210) .. controls (590,221.05) and (581.05,230) .. (570,230) .. controls (558.95,230) and (550,221.05) .. (550,210) -- cycle ;
\draw    (530,180) -- (530,210) ;
\draw [shift={(530,180)}, rotate = 90] [color={rgb, 255:red, 0; green, 0; blue, 0 }  ][fill={rgb, 255:red, 0; green, 0; blue, 0 }  ][line width=0.75]      (0, 0) circle [x radius= 1.34, y radius= 1.34]   ;
\draw    (400,210) -- (437,210) ;
\draw [shift={(440,210)}, rotate = 180] [fill={rgb, 255:red, 0; green, 0; blue, 0 }  ][line width=0.08]  [draw opacity=0] (6.25,-3) -- (0,0) -- (6.25,3) -- cycle    ;
\draw (333,166.4) node [anchor=north west][inner sep=0.75pt]  [font=\footnotesize]  {$\clo$};
\draw (307,196.4) node [anchor=north west][inner sep=0.75pt]  [font=\footnotesize]  {$P$};
\draw (233,202.4) node [anchor=north west][inner sep=0.75pt]  [font=\footnotesize]  {$P_{3}$};
\draw (373,202.4) node [anchor=north west][inner sep=0.75pt]  [font=\footnotesize]  {$P_{3}$};
\draw (101,166.4) node [anchor=north west][inner sep=0.75pt]  [font=\footnotesize]  {$\clo$};
\draw (105,206.4) node [anchor=north west][inner sep=0.75pt]  [font=\footnotesize]  {$\bbi$};
\draw (51,202.4) node [anchor=north west][inner sep=0.75pt]  [font=\footnotesize]  {$P_{3}$};
\draw (184,190.4) node [anchor=north west][inner sep=0.75pt]  [font=\small]  {$\mathbb{F}_{\bbi P}$};
\draw (513,172.4) node [anchor=north west][inner sep=0.75pt]  [font=\footnotesize]  {$\clo$};
\draw (514.2,196.4) node [anchor=north west][inner sep=0.75pt]  [font=\footnotesize]  {$P$};
\draw (453,202.4) node [anchor=north west][inner sep=0.75pt]  [font=\footnotesize]  {$P_{3}$};
\draw (593,202.4) node [anchor=north west][inner sep=0.75pt]  [font=\footnotesize]  {$P_{3}$};
\draw (400,190.4) node [anchor=north west][inner sep=0.75pt]  [font=\small]  {$\mathbb{F}_{P_3P'}$};
\draw (533.8,195.4) node [anchor=north west][inner sep=0.75pt]  [font=\footnotesize]  {$P'$};
\end{tikzpicture}\vspace{0.5cm}
\end{equation}
We applied the fusion move on two internal edges to transform the sunset channel into the dumbbell channel with the correct position of the external momentum. Next, we can use the inner product  on the space of conformal blocks given by \eqref{eq:innerproduct} to compute the overlap
\begin{equation}
    \begin{split}
    \braket{\mathcal{F}^{\,\mathcal{C}}_{2,1}}{\mathcal{F}^{\,\mathcal{C}\dash}_{2,1}} &= \int_0^\infty \dd P\dd P' \,\fkerbig{\mathbb{1}}{P}{P_3}{P_3}{P_3}{P_3}\fkerbig{P_3}{P'}{P_\clo}{P}{P_3}{P_3}\braket{
\vcenter{
\hbox{
\begin{tikzpicture}
\node at (0,0) {\includegraphics{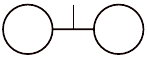}};
\node at (0.77,0) {\scriptsize{$P_2$}};
\node at (-0.75,0) {\scriptsize{$P_1$}};
\node at (-0.21,-0.3) {\scriptsize{$P_\clo$}};
\node at (0.25,-0.3) {\scriptsize{$P_\clo$}};
\node at (0,0.62) {\scriptsize{$P_\clo$}};
\end{tikzpicture}}}}{
\vcenter{
\hbox{\begin{tikzpicture}
\node at (0,0) {\includegraphics{img/genus2diagram2m.pdf}};
\node at (-0.75,0) {\scriptsize{$P_3$}};
\node at (0.77,0) {\scriptsize{$P_3$}};
\node at (0,0.62) {\scriptsize{$P_\clo$}};
\node at (-0.2,-0.3) {\scriptsize{$P$}};
\node at (0.2,-0.3) {\scriptsize{$P'$}};
\end{tikzpicture}}}} \\[1.4em]
&=
\frac{\delta(P_1-P_3)\delta(P_2-P_3)}{\rho_0(P_3)^2\rho_0(P_\clo)^2C_0(P_3,P_3,P_\clo)^2C_0(P_\clo,P_\clo,P_\clo)} \, \fkerbig{\mathbb{1}}{P_\clo}{P_3}{P_3}{P_3}{P_3}\fkerbig{P_3}{P_\clo}{P_\clo}{P_\clo}{P_3}{P_3}.\\[1em]
\end{split}
\end{equation}
Plugging this into our formula for $Z_{\mathrm{Vir}}(M_3)$, we obtain a single integral over a variable $P_3$, which we relabel as $P$. If we also plug in the normalization factors of the torus one-point block \eqref{eq:torusnorm} and the three-point vertex \eqref{eq:3pointnorm}, we obtain after some simplifications:
\begin{multline}
    \ket{Z_{\mathrm{Vir}}(M_3)\vphantom{F^2}} = C_0(P_\clo,P_\clo,P_\clo)\int_0^\infty \dd P  \,C_0(P,P,P_\clo) \,\fker{P_\clo}{P}{P}{P_\clo}{P_\clo}{P} \\ \times  
    \big |\mathcal{F}_{0,3}(P_\clo,P_\clo,P_\clo)\big\rangle \otimes \big|\mathcal{F}_{1,1}(P;P_\clo)\big\rangle^{\otimes 3}.\vspace{1mm}
\end{multline}
In arriving at this answer, we used the relation \eqref{eq:FC_0} between the fusion kernel $\mathbb{F}_{\mathbb{1}P}$ and the three-point function $C_0$, as well as the identity satisfied by the fusion kernel under swapping the internal indices:
\begin{equation}
    \frac{\fker{P}{P_\clo}{P_\clo}{P_\clo}{P}{P}}{\rho_0(P_\clo)C_0(P,P,P_\clo)C_0(P_\clo,P_\clo,P_\clo)} =\frac{\fker{P_\clo}{P}{P}{P_\clo}{P_\clo}{P}}{\rho_0(P)C_0(P,P_\clo,P)C_0(P,P_\clo,P)}\,.
\end{equation}

Having found the Virasoro TQFT partition function, we can write down the gravity amplitude for the three-boundary wormhole constructed in this section. Namely, we multiply by the right-moving component and go to the wavefunction basis \enlargethispage{-1\baselineskip} \pagebreak for the dependence of the conformal blocks on the  moduli.  So, before summing over mapping class group images, the gravity partition function is:
\begin{multline}\label{eq:3bdywormhole}
Z_{\mathrm{grav}}\!\!\left(
\!\!\!
\vcenter{\hbox{\begin{tikzpicture}
\node at (0,0) {\includegraphics{img/3bdyWH2.pdf}};
\end{tikzpicture}}}
\!\!\!\right) 
\!\!=\! 
\left|C_0(P_\clo,P_\clo,P_\clo)  \!\int_0^\infty \!\!\dd P\, C_0(P,P,P_\clo) 
\fker{P_\clo}{P}{P}{P_\clo}{P_\clo}{P}\prod_{i=1}^3\mathcal{F}_{1,1}(P;P_\clo;\tau_i)\right|^2. 
\end{multline}
Here we used that the three-punctured sphere has no moduli and is normalized to $\braket{\Omega}{\Omega}=1$. The three-point function of the external operator has factored out, and the integral over the torus one-point blocks is weighted by two crossing kernels. 
To get the full gravity answer, we still have to sum the result \eqref{eq:3bdywormhole} over the modular images of each of the boundary tori, up to large bulk diffeomorphisms; but we will defer this point to the discussion section. 

\subsubsection*{CFT$_2$ prediction}
We now check that the three-boundary wormhole amplitude \eqref{eq:3bdywormhole} matches to the CFT$_2$ ensemble prediction. In Section \ref{sec:3pointfunction}, we found the formula for the cubic moment of heavy-heavy-light OPE coefficients, \eqref{eq:CFT3pointresult}. We can use this to compute the average product of thermal one-point functions in the generalized OPE ensemble,
\begin{align}\label{eq:CFT3pointprediction}
    \overline{\langle\clo\rangle_{\tau_1}\langle\clo\rangle_{\tau_2}\langle\clo\rangle_{\tau_3}} &= \sum_{h,\bar h} \overline{C_{h\clo h}C_{h\clo h}C_{h\clo h}} \prod_{i=1}^3\Big|\mathcal{F}_{1,1}(h;h_\clo;\tau_i)\Big|^2 \\[1em]
    &\approx C_{\clo\clo\clo}\left |\int_0^\infty \dd P \rho_0(P) \frac{C_0(P,P,P_\clo)\fker{P_\clo}{P}{P}{P_\clo}{P_\clo}{P}}{\rho_0(P)} \prod_{i=1}^3\mathcal{F}_{1,1}(P;P_\clo;\tau_i)\right |^2.
\end{align}
As before, we approximated the sum over states by keeping only the states above the $\frac{c-1}{24}$ threshold, for which we can replace the density of states by the universal smooth density  $\rho_0(P)$. We see that in order to match to the gravity answer, we have to multiply both sides of equation \eqref{eq:CFT3pointprediction} by $C_{\clo\clo\clo}$, and then identify 
\begin{equation}
    C_{\clo\clo\clo}^{\,2} = C_0(P_\clo,P_\clo,P_\clo) C_0(\bar P_\clo,\bar P_\clo,\bar P_\clo).\vspace{1mm}
\end{equation}
This is an analytic continuation of the universal result for the square of heavy OPE coefficients to the conical defect regime, where $h_\clo$ is smaller than $\frac{c-1}{24}$ but does scale with $c$. On the left-hand side of the equation, this means that the OPE average simply factorizes between the heavy and the light operators:\vspace{2mm}
\begin{equation}
\overline{C_{\clo\clo\clo}\langle\clo\rangle_{\tau_1}\langle\clo\rangle_{\tau_2}\langle\clo\rangle_{\tau_3}} = C_{\clo\clo\clo}\,\overline{\langle\clo\rangle_{\tau_1}\langle\clo\rangle_{\tau_2}\langle\clo\rangle_{\tau_3}}.\vspace{2mm}
\end{equation}
Again, this is expected based on the idea that the sub-threshold states should not be averaged over. The ensemble should only treat the heavy OPE indices statistically, while the external light operators are fixed to a non-random input value. 

An important check to make on the wormhole amplitude \eqref{eq:3bdywormhole} is that the integral has a saddle point  above the black hole threshold. The saddle-point equation can be accessed at high temperatures by using the large-$h$ asymptotics of the crossing kernels, derived in \cite{Collier:2018exn,Collier:2019weq}:
\begin{align}
\log C_0(P,P,P_\clo) & \sim - \sqrt{\frac{c}{6}h}+h_\clo\log h+O(1),\\
  \log \fkerbig{P_\clo}{P}{P}{P_\clo}{P_\clo}{P} &\sim \frac{1}{2} h_\clo \log h + O(1), \\
  \mathcal{F}_{1,1}(h;h_\clo;\tau) &\sim \e^{-\beta\left(h-\frac{c}{24}\right)}\left[\,\frac{\e^{-\frac{\beta}{24}}}{\eta(\frac{i\beta}{2\pi})} + \order{h^{-1}}\right],
\end{align}
which hold asymptotically as $h\to \infty$ with $h_\clo$ fixed. The saddle-point equation is therefore 
\begin{equation}
    \frac{d}{dh}\left[\tfrac{3}{2}h_\clo \log h -  \sqrt{\frac{c}{6}h} -3\beta h\right]=0.
\end{equation}
Here we took all $\tau_i = \frac{i\beta}{2\pi}$ equal to simplify the analysis. This is solved by
\begin{equation}
    h_* = \frac{c}{24}\left[\frac{h_\clo}{h_c}+ \frac{1-\sqrt{1+36\beta^2 (h_\clo/h_c)}}{18\beta^2}\right], \quad h_c \equiv \frac{\beta c}{12}.
\end{equation}
Expanding the square root to second order in $\beta^2$, and demanding that the saddle point is above the black hole threshold, one finds the constraint
\begin{equation}
    h_\clo > \frac{c}{36}.
\end{equation}
 This linear scaling with $c$ (but still below threshold) is expected for external operators $\clo$ that create a conical defect in the bulk \cite{Chandra:2022bqq}. In this regime, the three-boundary wormhole partition function has a saddle point above the black hole threshold. The saddle-point action $e^{-S_{\mathrm{on-shell}}}$ is positive and of the order $S(h^*)\sim c $, in line with \cite{Schlenker:2022dyo}.

The fact that the gravity partition function \eqref{eq:3bdywormhole} matches the CFT$_2$ ensemble answer using the cubic non-Gaussianity derived in Section \ref{sec:3pointfunction} is a non-trivial check that the gravitational theory correctly captures the statistics of the boundary OPE data. In the following subsection, we give another example of a wormhole that captures a non-Gaussian moment in the OPE ensemble.

\subsection{Four-boundary wormhole}\label{sec:4bdyWH}

As a next application, we construct the four-boundary wormhole $M_4$ that contributes to the connected OPE average of the product of four torus one-point functions \vspace{1mm}
\begin{equation} \overline{\langle\clo\rangle_{\tau_1}\langle\clo\rangle_{\tau_2}\langle\clo\rangle_{\tau_3}\langle\clo\rangle_{\tau_4}}.
\end{equation}

\begin{figure}
    \centering
\begin{tikzpicture}
    \node at (0,0) {\includegraphics[width=0.8\textwidth]{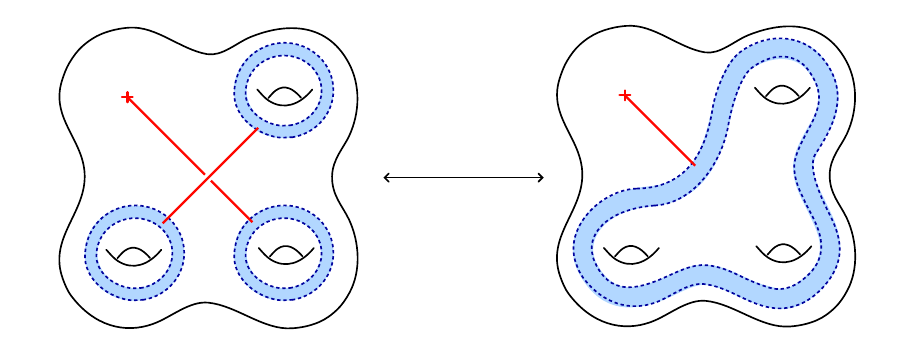}};
    \node at (0,0.5) {$\gamma$};
\end{tikzpicture}
    \caption{{\small Heegaard splitting of the four-boundary wormhole.}}
    \label{fig:genus3identify}
\end{figure}

The wormhole can be described topologically by the genus-three Heegaard splitting shown in Figure \ref{fig:genus3identify}. 
Namely, take a pair of once-punctured genus-three handlebodies {\small $M_{3,1}^{\mathsf{L}}$} and {\small $M_{3,1}^{\mathsf{R}}$}. Denote their boundary surfaces by {\small $\Sigma_{3,1}^{\mathsf{L}}$} and {\small $
\Sigma_{3,1}^{\mathsf{R}}$}, respectively. Drill out three solid tori from {\small $M_{3,1}^{\mathsf{L}}$}, each passing through a different handle. Also drill out a single solid torus from {\small $M_{3,1}^{\mathsf{R}}$}, going around all handles of {\small $M_{3,1}^{\mathsf{R}}$}. The result of this procedure is two compression bodies {\small $X^{\mathsf{L,R}}$}, whose boundary is
 \begin{equation}
   \partial X^{\mathsf{L}} = \Sigma_{3,1}^{\mathsf{L}} \cup \Sigma^{\mathsf{L}_1}_{1,1} \cup \Sigma^{\mathsf{L}_2}_{1,1}\cup \Sigma^{\mathsf{L}_3}_{1,1},\quad \partial X^{\mathsf{R}} = \Sigma_{3,1}^{\mathsf{R}} \cup \Sigma_{1,1}^{\mathsf{R}}.
 \end{equation}
 
Next, connect the punctures of {\small $\Sigma_{1,1}^{\mathsf{L_1}}$} and {\small $\Sigma_{1,1}^{\mathsf{L_2}}$} by a Wilson line through the bulk of the left handlebody. Also connect the punctures of {\small $\Sigma_{1,1}^{\mathsf{L_3}}$} and {\small $\Sigma_{3,1}^{\mathsf{L}}$}, as well as the punctures of {\small $\Sigma_{3,1}^{\mathsf{R}} $} and {\small $ \Sigma_{1,1}^{\mathsf{R}}$}, by bulk Wilson lines (drawn in red in Figure \ref{fig:genus3identify}). Finally, identify the genus-three boundaries with opposite orientation, twisted by a mapping class group element $\gamma \in \mathrm{MCG}(\Sigma_{3,1})$:
\begin{equation}
  \Sigma_{3,1}^{\mathsf{L}} \sim_\gamma  \overline{\Sigma}_{3,1}^{\mathsf{R}}.
\end{equation}
We will take $\gamma = 1$ for now, and comment on the more general case $\gamma \neq 1$ in the discussion.

To fully specify the handlebodies, we have to assign a set of cycles on the boundary that become contractible in the bulk. Since the Virasoro TQFT assigns a state to each boundary given by the corresponding Virasoro conformal block, specifying a set of (non-)contractible cycles is equivalent to specifying the channel decomposition $\mathcal{C}$ of the genus-three conformal block. The specific choice of $\mathcal{C}_{\mathsf{L}}$, $\mathcal{C}_{\mathsf{R}}$ that determines the Heegaard splitting in Figure \ref{fig:genus3identify} is given by the following overlap
\begin{equation}\label{eq:genus3overlap}
\braket{\mathcal{F}^{\,\mathcal{C}_{\mathsf{L}}}_{3,1}}{\mathcal{F}^{\,\mathcal{C}_{\mathsf{R}}}_{3,1}
} = 
    \Bigg \langle
    \vcenter{
    \hbox{\begin{tikzpicture}
    \node at (0,0) {\includegraphics{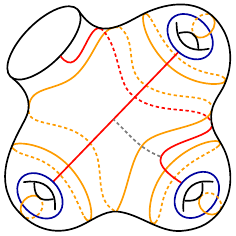}};
    \node at (0.4,-0.2) {$\bbi$};
    \node at (-0.5,-0.2) {$\clo$};
    \node at (-1.1,1.2) {$\clo$};
    \node at (1.75,1.2) {$1$};
    \node at (1.75,-0.85) {$2$};
    \node at (-1.7,-0.85) {$3$};
\end{tikzpicture}}}
\Bigg\vert \vcenter{\hbox{
\begin{tikzpicture}
\node at (0,0){\includegraphics{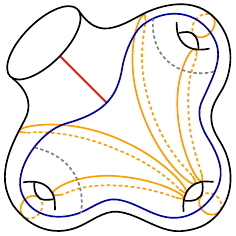}};
    \node at (-1.1,1.2) {$\clo$};
    \node at (-1.1,-0.85) {$\bbi$};
    \node at (0.7,0.85) {$\bbi$};
    \node at (-0.1,0.85) {$4$};
\end{tikzpicture}}
    }\Bigg \rangle.
\end{equation}
The dual cycles, which specify the pair-of-pants decomposition of the genus-three surface, are drawn in yellow. The identity exchanges are drawn as grey dotted lines. We have specified the Liouville momenta using the simplified notation $1\equiv P_1$ and similarly for $P_2,P_3,P_4$.  

The four-boundary wormhole partition function is then a state in the product Hilbert space $\mathcal{H}_{1,1}^{\otimes 4}$, with expansion coefficients determined by the above overlap:
\begin{equation}\label{eq:WHfourboundary}
     \ket{Z_{\mathrm{Vir}}(M_4)\vphantom{F^2}} = \int_0^\infty \dd^4 P \,\braket{\mathcal{F}^{\,\mathcal{C}_{\mathsf{L}}}_{3,1}}{\mathcal{F}^{\,\mathcal{C}_{\mathsf{R}}}_{3,1}
}
     \,\bigotimes_{i=1}^4{\ket{\widehat{\mathcal{F}}_{1,1}(P_i;P_\clo)}}.
\end{equation}
The normalization of the hatted states is given by \eqref{eq:torusnorm} as before. We will compute the overlap by applying a sequence of crossing moves to bring the bra and ket to the same channel, and then use the inner product \eqref{eq:innerproduct}. 

First, we apply a fusion move on the interior four-holed sphere of $\mathcal{C}_{\mathsf{L}}$ to get rid of the over-under crossing:
\tikzset{every picture/.style={line width=0.75pt}}
\begin{equation}
\begin{tikzpicture}[x=0.75pt,y=0.75pt,yscale=-0.5,xscale=0.5,baseline={([yshift=-.5ex]current bounding box.center)}]
\draw    (120,90) -- (167.17,136.83) ;
\draw    (220,90) -- (120,190) ;
\draw    (173.33,143.22) -- (220,190) ;
\draw (90,82.4) node [anchor=north west][inner sep=0.75pt]    {$\clo$};
\draw (90,182.4) node [anchor=north west][inner sep=0.75pt]    {$\clo$};
\draw (221,82.4) node [anchor=north west][inner sep=0.75pt]    {$\clo$};
\draw (221,182.4) node [anchor=north west][inner sep=0.75pt]    {$\clo$};
\end{tikzpicture} \quad=\quad 
\begin{tikzpicture}[x=0.75pt,y=0.75pt,yscale=-0.5,xscale=0.5,baseline={([yshift=-.5ex]current bounding box.center)}] 
\draw    (120,90) -- (167.17,136.83) ;
\draw    (220,90) -- (120,190) ;
\draw    (173.33,143.22) -- (220,190) ;
\draw  [dash pattern={on 1.5pt off 1.5pt on 1.5pt off 1.5pt}]  (190,120) -- (190,160) ;
\draw (90,82.4) node [anchor=north west][inner sep=0.75pt]    {$\clo$};
\draw (90,182.4) node [anchor=north west][inner sep=0.75pt]    {$\clo$};
\draw (221,82.4) node [anchor=north west][inner sep=0.75pt]    {$\clo$};
\draw (221,182.4) node [anchor=north west][inner sep=0.75pt]    {$\clo$};
\draw (197,132.4) node [anchor=north west][inner sep=0.75pt]    {$\bbi$};
\end{tikzpicture} \quad = \quad \int_0^\infty \dd P\,\fker{\bbi}{P}{P_\clo}{P_\clo}{P_\clo}{P_\clo}\mathbb{B}^P_{P_\clo P_\clo}\, \begin{tikzpicture}[x=0.75pt,y=0.75pt,yscale=-0.5,xscale=0.5,baseline={([yshift=-.5ex]current bounding box.center)}]
\draw    (160,160) -- (220,160) ;
\draw    (140,110) -- (160,160) ; 
\draw    (160,160) -- (140,210) ; 
\draw    (240,110) -- (220,160) ;
\draw    (220,160) -- (240,210) ;
\draw (110,102.4) node [anchor=north west][inner sep=0.75pt]    {$\clo$};
\draw (110,202.4) node [anchor=north west][inner sep=0.75pt]    {$\clo$};
\draw (241,102.4) node [anchor=north west][inner sep=0.75pt]    {$\clo$};
\draw (241,202.4) node [anchor=north west][inner sep=0.75pt]    {$\clo$};
\draw (180,130.4) node [anchor=north west][inner sep=0.75pt]    {$P$};
\end{tikzpicture}
\end{equation}
where the braiding phase is given by {\small $\mathbb{B}_{P_\clo P_\clo}^P = e^{-\pi i (h_P-2h_\clo)}$}. We then apply a fusion move on one of the legs of the four-holed sphere to bring the bra state into the following form:
\begin{align}
    \mathcal{F}^{\,\mathcal{C}_{\mathsf{L}}}_{3,1} &= \int_0^\infty \dd P\,\fker{\bbi}{P}{P_\clo}{P_\clo}{P_\clo}{P_\clo}\mathbb{B}^P_{P_\clo P_\clo}\, \begin{tikzpicture}[x=0.75pt,y=0.75pt,yscale=-0.7,xscale=0.7,baseline={([yshift=-.5ex]current bounding box.center)}] 
\draw    (160,160) -- (220,160) ;
\draw    (150,130) -- (160,160) ;
\draw [shift={(150,130)}, rotate = 71.57] [color={rgb, 255:red, 0; green, 0; blue, 0 }  ][fill={rgb, 255:red, 0; green, 0; blue, 0 }  ][line width=0.75]      (0, 0) circle [x radius= 1.34, y radius= 1.34]   ;
\draw    (160,160) -- (150,190) ;
\draw    (230,130) -- (220,160) ;
\draw    (220,160) -- (230,190) ; 
\draw   (221.2,116) .. controls (221.2,107.72) and (227.92,101) .. (236.2,101) .. controls (244.48,101) and (251.2,107.72) .. (251.2,116) .. controls (251.2,124.28) and (244.48,131) .. (236.2,131) .. controls (227.92,131) and (221.2,124.28) .. (221.2,116) -- cycle ;
\draw   (222,203.2) .. controls (222,194.92) and (228.72,188.2) .. (237,188.2) .. controls (245.28,188.2) and (252,194.92) .. (252,203.2) .. controls (252,211.48) and (245.28,218.2) .. (237,218.2) .. controls (228.72,218.2) and (222,211.48) .. (222,203.2) -- cycle ;
\draw   (129.13,203.36) .. controls (129.13,195.07) and (135.85,188.36) .. (144.13,188.36) .. controls (152.42,188.36) and (159.13,195.07) .. (159.13,203.36) .. controls (159.13,211.64) and (152.42,218.36) .. (144.13,218.36) .. controls (135.85,218.36) and (129.13,211.64) .. (129.13,203.36) -- cycle ;
\draw (130.09,122.69) node [anchor=north west][inner sep=0.75pt]  [font=\footnotesize]  {$\clo$};
\draw (136,162.4) node [anchor=north west][inner sep=0.75pt]  [font=\footnotesize]  {$\clo$};
\draw (226,142.4) node [anchor=north west][inner sep=0.75pt]  [font=\footnotesize]  {$\clo$};
\draw (226,160.4) node [anchor=north west][inner sep=0.75pt]  [font=\footnotesize]  {$\clo$};
\draw (183,141.4) node [anchor=north west][inner sep=0.75pt]  [font=\footnotesize]  {$P$};
\draw (253,109.07) node [anchor=north west][inner sep=0.75pt]  [font=\footnotesize]  {$P_{1}$};
\draw (254.11,197.07) node [anchor=north west][inner sep=0.75pt]  [font=\footnotesize]  {$P_{2}$};
\draw (160.11,197.29) node [anchor=north west][inner sep=0.75pt]  [font=\footnotesize]  {$P_{3}$};
\end{tikzpicture}
 \\
    &= \int_0^\infty \dd P\dd P'\,\fker{\bbi}{P}{P_\clo}{P_\clo}{P_\clo}{P_\clo}\mathbb{B}^P_{P_\clo P_\clo}\,\fker{P_\clo}{P'}{P_\clo}{P_3}{P_3}{P} \hspace{2mm} \begin{tikzpicture}[x=0.75pt,y=0.75pt,yscale=-0.7,xscale=0.7,baseline={([yshift=-.5ex]current bounding box.center)}] 
\draw    (210,180) -- (240,180) ;
\draw    (160,180) -- (180,180) ;
\draw [shift={(160,180)}, rotate = 0] [color={rgb, 255:red, 0; green, 0; blue, 0 }  ][fill={rgb, 255:red, 0; green, 0; blue, 0 }  ][line width=0.75]      (0, 0) circle [x radius= 1.34, y radius= 1.34]   ;
\draw    (250,150) -- (240,180) ; 
\draw    (240,180) -- (250,210) ;
\draw   (241.2,136) .. controls (241.2,127.72) and (247.92,121) .. (256.2,121) .. controls (264.48,121) and (271.2,127.72) .. (271.2,136) .. controls (271.2,144.28) and (264.48,151) .. (256.2,151) .. controls (247.92,151) and (241.2,144.28) .. (241.2,136) -- cycle ; 
\draw   (242,223.2) .. controls (242,214.92) and (248.72,208.2) .. (257,208.2) .. controls (265.28,208.2) and (272,214.92) .. (272,223.2) .. controls (272,231.48) and (265.28,238.2) .. (257,238.2) .. controls (248.72,238.2) and (242,231.48) .. (242,223.2) -- cycle ; 
\draw   (180,180) .. controls (180,171.72) and (186.72,165) .. (195,165) .. controls (203.28,165) and (210,171.72) .. (210,180) .. controls (210,188.28) and (203.28,195) .. (195,195) .. controls (186.72,195) and (180,188.28) .. (180,180) -- cycle ;
\draw (155.43,160.83) node [anchor=north west][inner sep=0.75pt]  [font=\footnotesize]  {$\clo$};
\draw (246,160.4) node [anchor=north west][inner sep=0.75pt]  [font=\footnotesize]  {$\clo$};
\draw (246,182.4) node [anchor=north west][inner sep=0.75pt]  [font=\footnotesize]  {$\clo$};
\draw (217,160.4) node [anchor=north west][inner sep=0.75pt]  [font=\footnotesize]  {$P$};
\draw (273,129.07) node [anchor=north west][inner sep=0.75pt]  [font=\footnotesize]  {$P_{1}$};
\draw (274.11,217.07) node [anchor=north west][inner sep=0.75pt]  [font=\footnotesize]  {$P_{2}$};
\draw (187.29,197.83) node [anchor=north west][inner sep=0.75pt]  [font=\footnotesize]  {$P_{3}$};
\draw (188.71,145.54) node [anchor=north west][inner sep=0.75pt]  [font=\footnotesize]  {$P'$};
\end{tikzpicture} .\label{eq:leftket}
\end{align}
Next, we turn to the ket state. We again perform a sequence of fusion moves to bring the conformal block into the same channel as \eqref{eq:leftket}. The result is
\begin{align}
    \mathcal{F}^{\,\mathcal{C}_{\mathsf{R}}}_{3,1}  = \,\begin{tikzpicture}[x=0.75pt,y=0.75pt,yscale=-0.7,xscale=0.7,baseline={([yshift=-2ex]current bounding box.center)}] 
\draw    (295,185) -- (295,205) ;
\draw [shift={(295,185)}, rotate = 90] [color={rgb, 255:red, 0; green, 0; blue, 0 }  ][fill={rgb, 255:red, 0; green, 0; blue, 0 }  ][line width=0.75]      (0, 0) circle [x radius= 1.34, y radius= 1.34]   ; 
\draw   (241.25,225) .. controls (241.25,213.95) and (265.31,205) .. (295,205) .. controls (324.69,205) and (348.75,213.95) .. (348.75,225) .. controls (348.75,236.05) and (324.69,245) .. (295,245) .. controls (265.31,245) and (241.25,236.05) .. (241.25,225) -- cycle ;
\draw  [dash pattern={on 1.5pt off 1.5pt on 1.5pt off 1.5pt}]  (270.1,207.5) -- (270.1,242.5) ;
\draw  [dash pattern={on 1.5pt off 1.5pt on 1.5pt off 1.5pt}]  (320,207.24) -- (320,242.65) ;
\draw (272.8,178) node [anchor=north west][inner sep=0.75pt]  [font=\footnotesize]  {$\clo$};
\draw (216.65,217.36) node [anchor=north west][inner sep=0.75pt]  [font=\footnotesize]  {$P_{4}$};
\draw (271,217.9) node [anchor=north west][inner sep=0.75pt]  [font=\footnotesize]  {$\bbi$};
\draw (321,217.9) node [anchor=north west][inner sep=0.75pt]  [font=\footnotesize]  {$\bbi$};
\end{tikzpicture}\, &= \int_0^\infty \dd s_1\dd s_2 \,\fker{\bbi}{s_1}{P_4}{P_4}{P_4}{P_4} \fker{\bbi}{s_2}{P_4}{P_4}{P_4}{P_4}  \,\,\begin{tikzpicture}[x=0.75pt,y=0.75pt,yscale=-0.7,xscale=0.7,baseline={([yshift=-.5ex]current bounding box.center)}]
\draw    (230,200) -- (260,200) ;
\draw    (275,165) -- (275,185) ;
\draw [shift={(275,165)}, rotate = 90] [color={rgb, 255:red, 0; green, 0; blue, 0 }  ][fill={rgb, 255:red, 0; green, 0; blue, 0 }  ][line width=0.75]      (0, 0) circle [x radius= 1.34, y radius= 1.34]   ; 
\draw    (320,200) -- (290,200) ;
\draw   (260,200) .. controls (260,191.72) and (266.72,185) .. (275,185) .. controls (283.28,185) and (290,191.72) .. (290,200) .. controls (290,208.28) and (283.28,215) .. (275,215) .. controls (266.72,215) and (260,208.28) .. (260,200) -- cycle ;
\draw   (320,200) .. controls (320,191.72) and (326.72,185) .. (335,185) .. controls (343.28,185) and (350,191.72) .. (350,200) .. controls (350,208.28) and (343.28,215) .. (335,215) .. controls (326.72,215) and (320,208.28) .. (320,200) -- cycle ;
\draw   (200,200) .. controls (200,191.72) and (206.72,185) .. (215,185) .. controls (223.28,185) and (230,191.72) .. (230,200) .. controls (230,208.28) and (223.28,215) .. (215,215) .. controls (206.72,215) and (200,208.28) .. (200,200) -- cycle ;
\draw (252.8,158) node [anchor=north west][inner sep=0.75pt]  [font=\footnotesize]  {$\clo$};
\draw (238,184.4) node [anchor=north west][inner sep=0.75pt]  [font=\footnotesize]  {$s_{1}$};
\draw (296,184.4) node [anchor=north west][inner sep=0.75pt]  [font=\footnotesize]  {$s_{2}$};
\draw (208.49,217.03) node [anchor=north west][inner sep=0.75pt]  [font=\footnotesize]  {$P_{4}$};
\draw (269.8,217.2) node [anchor=north west][inner sep=0.75pt]  [font=\footnotesize]  {$P_{4}$};
\draw (329.8,217.6) node [anchor=north west][inner sep=0.75pt]  [font=\footnotesize]  {$P_{4}$};
\end{tikzpicture} \\
    &= \int_0^\infty \dd s_1\dd s_2 \dd s_3 \,\fker{\bbi}{s_1}{P_4}{P_4}{P_4}{P_4} \fker{\bbi}{s_2}{P_4}{P_4}{P_4}{P_4} \fker{P_4}{s_3}{P_4}{P_4}{s_1}{s_2}\,\begin{tikzpicture}[x=0.75pt,y=0.75pt,yscale=-0.7,xscale=0.7,baseline={([yshift=-.5ex]current bounding box.center)}] 
\draw    (210,180) -- (240,180) ;
\draw    (160,180) -- (180,180) ;
\draw [shift={(160,180)}, rotate = 0] [color={rgb, 255:red, 0; green, 0; blue, 0 }  ][fill={rgb, 255:red, 0; green, 0; blue, 0 }  ][line width=0.75]      (0, 0) circle [x radius= 1.34, y radius= 1.34]   ;
\draw    (250,150) -- (240,180) ; 
\draw    (240,180) -- (250,210) ;
\draw   (241.2,136) .. controls (241.2,127.72) and (247.92,121) .. (256.2,121) .. controls (264.48,121) and (271.2,127.72) .. (271.2,136) .. controls (271.2,144.28) and (264.48,151) .. (256.2,151) .. controls (247.92,151) and (241.2,144.28) .. (241.2,136) -- cycle ; 
\draw   (242,223.2) .. controls (242,214.92) and (248.72,208.2) .. (257,208.2) .. controls (265.28,208.2) and (272,214.92) .. (272,223.2) .. controls (272,231.48) and (265.28,238.2) .. (257,238.2) .. controls (248.72,238.2) and (242,231.48) .. (242,223.2) -- cycle ; 
\draw   (180,180) .. controls (180,171.72) and (186.72,165) .. (195,165) .. controls (203.28,165) and (210,171.72) .. (210,180) .. controls (210,188.28) and (203.28,195) .. (195,195) .. controls (186.72,195) and (180,188.28) .. (180,180) -- cycle ;
\draw (155.43,160.83) node [anchor=north west][inner sep=0.75pt]  [font=\footnotesize]  {$\clo$};
\draw (246,160.4) node [anchor=north west][inner sep=0.75pt]  [font=\footnotesize]  {$s_1$};
\draw (246,182.4) node [anchor=north west][inner sep=0.75pt]  [font=\footnotesize]  {$s_2$};
\draw (217,165.4) node [anchor=north west][inner sep=0.75pt]  [font=\footnotesize]  {$s_3$};
\draw (273,129.07) node [anchor=north west][inner sep=0.75pt]  [font=\footnotesize]  {$P_{4}$};
\draw (274.11,217.07) node [anchor=north west][inner sep=0.75pt]  [font=\footnotesize]  {$P_{4}$};
\draw (187.29,197.83) node [anchor=north west][inner sep=0.75pt]  [font=\footnotesize]  {$P_{4}$};
\end{tikzpicture}.
\end{align}
As the conformal blocks are now in the same  channel, we can compute the overlap \eqref{eq:genus3overlap}. The delta functions in the inner product \eqref{eq:innerproduct} set the following internal momenta to be equal,
\begin{equation}
    s_1 = s_2 = P_\clo, \quad s_3 = P, \quad P' = P_4 = P_3=P_2=P_1,
\end{equation}
so there is only an integral over $P$ left. Including the correct normalization, we find
\begin{equation}
\begin{split}
\braket{\mathcal{F}^{\,\mathcal{C}_{\mathsf{L}}}_{3,1}}{\mathcal{F}^{\,\mathcal{C}_{\mathsf{R}}}_{3,1}
} = \int_0^\infty \dd P\,\fker{\bbi}{P}{P_\clo}{P_\clo}{P_\clo}{P_\clo}\mathbb{B}^P_{P_\clo P_\clo}\,\fker{P_\clo}{P_4}{P_\clo}{P_4}{P_4}{P}\,&\fker{\bbi}{P_\clo}{P_4}{P_4}{P_4}{P_4} \fker{\bbi}{P_\clo}{P_4}{P_4}{P_4}{P_4} \fker{P_4}{P}{P_4}{P_4}{P_\clo}{P_\clo} \\[0.8em]&\times\frac{\delta(P_1-P_4)\delta(P_2-P_4)\delta(P_3-P_4)}{\rho_{3,1}(\mathbf{P})}.
\end{split}
\end{equation}
The normalization is determined by assigning a factor of $\rho_0$ to each internal line and a factor of $C_0$ to each vertex of the conformal block diagram, giving
\vspace{0.2cm}
\begin{equation}
    \rho_{3,1}(\mathbf{P}) =\rho_0(P)  \rho_0(P_\clo)^2 \rho_0(P_4)^4\, C_0(P_4,P_4,P_\clo)^3 C_0(P_4,P_4,P)C_0(P_\clo,P_\clo,P).\vspace{0.2cm}
\end{equation}
The result can be simplified significantly by swapping the internal indices of the fusion kernel $\mathbb{F}_{P_4 P} \to \mathbb{F}_{P P_4} $ using the formula \eqref{app:swap}. The integral over $P$ can then be performed by combining the hexagon and pentagon equation, see equation \eqref{app:hexapenta}.

 Substituting this result for the overlap into the wormhole amplitude \eqref{eq:WHfourboundary}, we obtain:
 \vspace{0.2cm}
\begin{equation}
     \ket{Z_{\mathrm{Vir}}(M_4)\vphantom{F^2}} = \e^{-2\pi i h_\clo} \int_0^\infty \dd P_4\,C_0(P_4,P_4,P_\clo)^2 \, \fker{P_4}{P_4}{P_{\clo}}{P_4}{P_{\clo}}{P_4} \,\ket{\mathcal{F}_{1,1}(P_4;P_\clo)\vphantom{F^2}}^{\otimes 4}.
\end{equation}

As a final step, we take the inner product with $\bra{\tau_1}\otimes\dots\otimes\bra{\tau_4}$ and multiply by the right-moving sector. We also assume that the spin $h_\clo -\bar h_\clo$ of the external operator is an integer. This then leads to the 3D gravity partition function (before summing over modular images):
\begin{equation}\label{eq:4bdywormhole}
Z_{\mathrm{grav}}\!\!\left(
\!\!\!
\vcenter{\hbox{\begin{tikzpicture}
\node at (0,0) {\includegraphics{img/4bdyWH.pdf}};
\end{tikzpicture}}}
\!\!\!\right) 
=
\left|\int_0^\infty \dd P_4\,C_0(P_4,P_4,P_\clo)^2 \, \fker{P_4}{P_4}{P_{\clo}}{P_4}{P_{\clo}}{P_4} \,\prod_{i=1}^4 \mathcal{F}_{1,1}(P_4;P_\clo;\tau_i)\right|^2. 
\end{equation}
The picture on the left-hand side is a cartoon of the wormhole topology described in this section. Namely, the boundary consists of four once-punctured tori and the Wilson lines cross inside the topologically non-trivial 3D bulk. 

As before, the full gravitational answer includes a sum over boundary mapping class group transformations, modulo bulk diffeomorphisms. Again, it is not immediately clear what is the precise group of large bulk diffeomorphisms to quotient by. This point will be addressed more generally in the discussion section. 

\subsubsection*{CFT$_2$ prediction}
We will now show that the partition function $Z_{\mathrm{grav}}(M_4)$ is indeed the one predicted by the OPE ensemble of the boundary CFT$_2$. Consider the product of four torus one-point functions.
Expanding in conformal blocks, the OPE average becomes
\begin{equation}
    \overline{\langle\clo\rangle_{\tau_1}\langle\clo\rangle_{\tau_2}\langle\clo\rangle_{\tau_3} \langle\clo\rangle_{\tau_4}} = \sum_{1,2,3,4} \overline{C_{1\clo 1}C_{2\clo 2}C_{3\clo 3}C_{4\clo 4}}\,\prod_{i=1}^4\Big|\mathcal{F}_{1,1}(P_i;P_\clo;\tau_i)\Big|^2.
\end{equation}
The fully connected, non-Gaussian, cyclic contraction derived in Section \ref{sec:4-pt} sets all indices equal, $P_1=P_2=P_3=P_4$. Using the asymptotic formula \eqref{eq:nongauss} for the quartic moment, we obtain the generalized ETH prediction\footnote{Here we used the fact that $J_4$ is an integer to discard the phase $(-1)^{4J_4} =1$ present in \eqref{eq:nongauss}.} 
\begin{equation}
    \overline{\langle\clo\rangle_{\tau_1}\langle\clo\rangle_{\tau_2}\langle\clo\rangle_{\tau_3} \langle\clo\rangle_{\tau_4}}\,\big\vert_{\mathrm{cyclic}} = \sum_{4} \,\Big|
    C_0(P_4,P_4,P_{\clo})^2
    \frac{\fker{P_4}{P_4}{P_{\clo}}{P_4}{P_{\clo}}{P_4}}
    {\rho_0(P_4)}
    \prod_{i=1}^4\mathcal{F}_{1,1}(P_4,P_\clo;\tau_i)\Big|^2.
\end{equation}

The contribution from the identity in the sum over $h_4,\bar h_4$ evaluates to zero, since $C_0(\bbi,\bbi,P_\clo) = 0$. If we assume the spectrum only contains the identity plus heavy states, we can take the continuum limit 
\begin{equation}
    \sum_{4\neq \bbi} \to \int_0^\infty \dd P_4 \dd \bar P_4 \,\rho_0(P_4,\bar P_4)
\end{equation}
with $\rho_0$ the universal Cardy density of states. Doing so precisely reproduces the gravity answer obtained in \eqref{eq:4bdywormhole} for the four-boundary wormhole.   

\subsection{A recipe for arbitrary \textit{n}}

The previous two sections give a recipe for how to generalize the construction to torus one-point wormholes with an arbitrary number of boundary components. The product of one-point functions gives rise to the OPE average 
\begin{equation}
    \overline{C_{1\clo 1}C_{2\clo 2}\cdots C_{n\clo n}}.
\end{equation}
The fully cyclic contraction \eqref{eq:nthcyclic} sets all internal indices $1,\dots n$ to be equal. On the gravity side, this means we should get a single integral over the Liouville momentum $P$. To achieve this, we mimic the strategy of the previous sections. Schematically, the recipe is:
\begin{enumerate}
    \item Take a pair of genus $n-1$ handlebodies.
    \item Drill out $n-1$ once-punctured tori from one handlebody, and a single once-punctured torus from the other. The single torus traverses all handles, so that the conformal block inner product sets all internal momenta equal.
    \item Let the external Wilson lines traverse some topologically non-trivial path in the interior of the compression body. When $n$ is odd, allow the Wilson lines to interact in a three-point vertex. 
    \item Glue the genus $n-1$ boundary surfaces, possibly after some twist $\gamma$. 
\end{enumerate}
We invite the reader to work out an example to make this schematic recipe more concrete. However, 
already at $n=5$ and $n=6$ there are many more choices and the calculations quickly get involved. 

\section{Generalized OPE randomness hypothesis}
\label{sec:genOPE}

In the previous subsection, we have restricted our attention to once-punctured torus wormholes, which capture the statistical moments of heavy-heavy-light OPE coefficients. As explained in the introduction, these statistics precisely coincide with the generalized eigenstate thermalization hypothesis (ETH) for the matrix elements of the light external operator $\clo$ between high-energy eigenstates. 

It was conjectured in \cite{Belin:2020hea} that OPE coefficients with only heavy indices should also be distributed randomly. This was dubbed the `OPE randomness hypothesis' and can be seen as a generalization of ETH to chaotic CFT$_2$. In that work, only a Gaussian ansatz for the statistics of $C_{\mathrm{HHH}}$ was explored. However, by now it is clear that non-Gaussianities should be included also for all-heavy OPE statistics, which can lead to contributions to CFT observables whose order competes with the Gaussian contractions \cite{Belin:2021ryy}. 

An example where the non-Gaussian statistics contribute at the same order as the Gaussian contraction is found in the second moment of the genus-two partition function.  Subtracting the disconnected contribution, we will study the variance
\vspace{0.2cm}
\begin{equation}\label{eq:genus2statistics}
    \overline{Z_{g=2}(\Omega,\bar\Omega)Z_{g=2}(\Omega',\bar\Omega')} - \overline{Z_{g=2}(\Omega,\bar\Omega)}\cdot\overline{Z_{g=2}(\Omega',\bar\Omega')} \vspace{0.2cm}.
\end{equation}
If we denote the moduli $\Omega$ by three `inverse temperatures' $\beta_1,\beta_2,\beta_3$ and perform the analytic continuation $\beta_i \to \beta_i\pm iT$, the resulting average is an example of a \emph{generalized spectral form factor}, introduced in \cite{Belin:2021ibv}. As we will review in Section \ref{sec:genus2}, the variance of the genus-two partition function receives a leading non-Gaussian contribution from a contraction that was derived in \cite{Belin:2021ryy} from crossing symmetry at genus three. We refer to this non-Gaussianity  as the `$6j$ contraction' because of its close connection to the Virasoro $6j$ symbol \cite{Belin:2023efa}. 

More generally, one can consider arbitrary products of genus $g$ partition functions: these probe the higher moments of the heavy-heavy-heavy OPE coefficients, and their non-factorization signals the presence of higher non-Gaussianities in their statistics.  Such non-Gaussianities in turn imply the existence of new Euclidean wormholes in AdS$_3$ gravity. These will be constructed explicitly in Section \ref{sec:nG-genus2} for the second moment of the genus-two partition function.

\subsection{The 6\textit{j} contraction}\label{sec:genus2}

As a first place where non-Gaussian statistics for {$C_{\mathrm{HHH}}$} may appear, consider the product of genus-two partition functions, both expanded in the sunset channel. This contains four heavy-heavy-heavy OPE coefficients. The OPE average leads to a non-factorized answer:
\vspace{2mm}
\begin{align}\label{eq:genus2product}
    \overline{Z_{g=2}(\Omega,\bar\Omega)Z_{g=2}(\Omega',\bar\Omega')} &= \sum_{1,2,3}\sum_{1',2',3'}\overline{C_{123}C_{123}^*C_{1'2'3'}C^*_{1'2'3'}}\,|\mathcal{F}_{2,0}^S(\mathbf{P};\Omega)\mathcal{F}_{2,0}^S(\mathbf{P}';\Omega')|^2 \\[1em]
    &=\sum_{1,2,3}\sum_{1',2',3'}(-1)^{\sum_{i=1}^3(J_i+J_i')}\overline{C_{123}^2C^2_{1'2'3'}}\,|\mathcal{F}_{2,0}^S(\mathbf{P};\Omega)\mathcal{F}_{2,0}^S(\mathbf{P}';\Omega')|^2.
\end{align}
In the second line, we used the transformation rules for OPE coefficients under permutation of the indices and their reality conditions. The typicality assumption from Section \ref{sec:typicality} asserts that all possible index contractions should contribute to this average. Among these, there are the contributions from the identity in the intermediate channels, which trivially give a factorized contribution. For the non-identity heavy states, there are also the Gaussian contractions, which set {\small $123 = 1'2'3'$} and lead to non-factorization.\footnote{As shown in \cite{Chandra:2022bqq,Collier:2023fwi}, applying the Gaussian contraction to \eqref{eq:genus2product} and making the continuum approximation for the high-energy spectrum gives the partition function of the Maldacena-Maoz wormhole, whose topology is $\Sigma_{2,0}\times I$.}

However, there are also many non-Gaussian contractions. One important non-Gaussianity was called the `skyline' contraction in \cite{Belin:2021ryy}. It is a connected contraction, proportional to the $6j$ symbol of the Virasoro algebra \cite{Ponsot:1999uf,Collier:2023fwi,Belin:2023efa}, so we will refer to it here as the `$6j$ contraction':
\vspace{2mm}
\begin{equation}\label{eq:6jcontraction}
       \overline{C_{123}C_{456}C_{1'2'3'}C_{4'5'6'}}  \,\big\vert_{6j} =  \delta_{14}\delta_{21'}\delta_{34'}\delta_{56'}\delta_{62'}\delta_{3'5'}  \sqrt{\mathsf{C}_{123}\mathsf{C}_{156}\mathsf{C}_{263'}\mathsf{C}_{33'5}} \,\begin{Bmatrix}
        1 & 2 & 3 \\ 3' & 5 & 6
    \end{Bmatrix}_{6j}.\vspace{2mm}
\end{equation}
Here we used the useful shorthand notation borrowed from \cite{Collier:2024mgv},\vspace{2mm}
\begin{equation}\label{eq:shorthand}
    \mathsf{C}_{ijk} \coloneqq C_0(P_i,P_j,P_k)C_0(\bar P_i,\bar P_j,\bar P_k),\vspace{2mm}
\end{equation} 
and the $6j$ symbol for the Virasoro algebra $\mathrm{Vir}\times \overline{\mathrm{Vir}}$ is defined in Appendix \ref{app:kernels}. The above contraction can be derived in a way similar to the general  procedure of Section \ref{sec:typicality}, using modular invariance and crossing symmetry of the genus-three partition function. The relevant crossing moves are shown in Figure 4 of \cite{Belin:2021ryy}.

We can use the relations between the $6j$ symbol, $\rho_0$, $C_0$ and $\mathbb{F}$ summarized in Appendix \ref{app:kernels} to write the $6j$ contraction in terms of crossing kernels only:
\vspace{3mm}
\begin{equation}\label{eq:6jcon}
\overline{C_{123}C_{156}C_{264}C_{345}}  \,\big\vert_{6j} = \e^{-\sum_{i=1}^3 S_0(P_i,\bar P_i)} \left | \fkerbig{\bbi}{P_1}{P_5}{P_6}{P_6}{P_5} \fkerbig{\bbi}{P_3}{P_5}{P_4}{P_4}{P_5}\fkerbig{P_5}{P_2}{P_4}{P_6}{P_1}{P_3} \right |^2\vspace{3mm} 
\end{equation}
where the microcanonical entropy is  $S_0(P,\bar P) = \log \rho_0(P,\bar P)$. This rewriting demonstrates that the entropic suppression of the $6j$ contraction is stronger than the Gaussian contraction. Nonetheless, the $6j$ contraction may contribute at the same order as the Gaussian contraction \emph{if} the pattern of delta functions sets less internal indices to be equal.

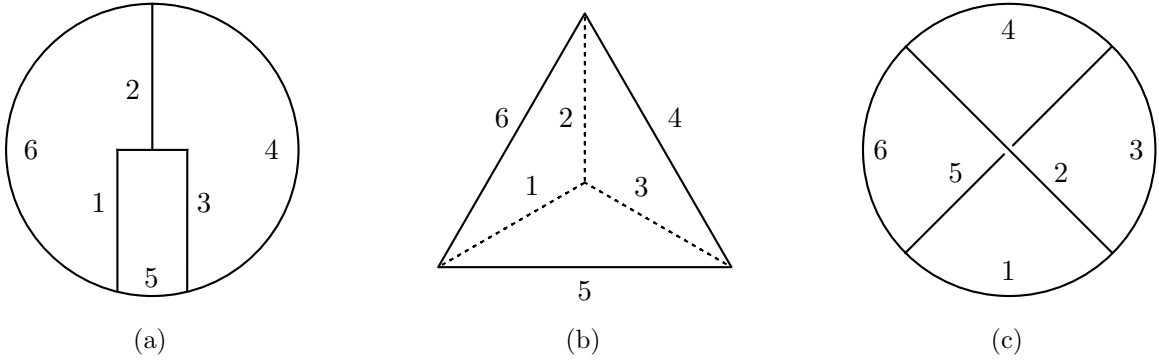
\begin{figure}
    \centering
    \begin{subfigure}
    {0.31\textwidth}
    \centering \begin{tikzpicture}\,\node at (0,0) {\includegraphics{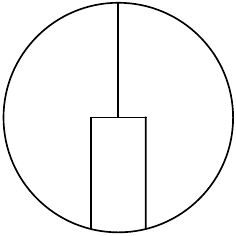}};
    \node at (-1.6,0){$6$};
    \node at (-0.7,-0.7) {$1$};
    \node at (-0.25,0.8) {$2$};
    \node at (0.7,-0.7) {$3$};
    \node at (1.6,0) {$4$};
    \node at (0,-1.7) {$5$};
    \end{tikzpicture} 
    \caption{}
     \label{6jfig:a}
  \end{subfigure}
  \hspace*{\fill}   
  \begin{subfigure}{0.31\textwidth}
    \centering \begin{tikzpicture}\,\node at (0,0) {\includegraphics{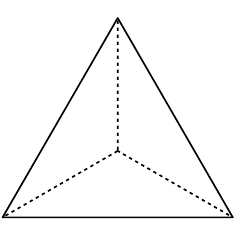}};
    \node at (-1.1,0.3){$6$};
    \node at (-0.7 ,-0.6) {$1$};
    \node at (-0.25,0.3) {$2$};
    \node at (0.75,-0.62) {$3$};
    \node at (1.2,0.3) {$4$};
    \node at (0,-2) {$5$};
    \end{tikzpicture}
    \caption{}
   \label{6jfig:b}
  \end{subfigure}%
  \hspace*{\fill}  
  \begin{subfigure}{0.31\textwidth} \centering
    \begin{tikzpicture}\,\node at (0,0) {\includegraphics{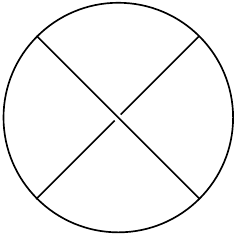}};
    \node at (-1.7,0){$6$};
    \node at (-0.7,-0.3) {$5$};
    \node at (0,1.6) {$4$};
    \node at (0.7,-0.3) {$2$};
    \node at (1.7,0) {$3$};
    \node at (0,-1.6) {$1$};
    \end{tikzpicture}
    \caption{}
    \label{6jfig:c}
  \end{subfigure}
    \caption{{\small The `skyline' channel decomposition of the genus-three partition function (a) can be thought of, diagrammatically, as a three-dimensional tetrahedron (b) with edges labelled by Liouville momenta $P_i$, $i=1,\dots,6$. The smooth function appearing in the quartic non-Gaussianity \eqref{eq:6jcontraction} has the same symmetries as this tetrahedron. The diagram can equivalently be drawn as a crossing partition of an `exchange' diagram (c) between four operators, modulo phases.}}
    \label{fig:6jfig}
\end{figure}

A second feature of the $6j$ contraction is that the combination of crossing kernels in \eqref{eq:6jcon} enjoys a full tetrahedral symmetry under permuting the momenta $P_i$. This is illustrated in Figure \ref{fig:6jfig}. As explained in \cite{Eberhardt:2023mrq}, the tetrahedral property is actually a special case of the more general pentagon equation imposed by the Moore-Seiberg consistency conditions. 

Thirdly, we note that the formula \eqref{eq:6jcontraction} coincides with equation \eqref{eq:nongauss} derived in Section \ref{sec:4-pt}, if we analytically continue $P_2$ and $P_5$ to the conical defect regime. This is surprising, because the two results were derived from different crossing equations: one from the genus-three partition function, and the other from the torus four-point function.\footnote{Diagrammatically, the $6j$ contraction and thermal four-point function may be related by deforming the skyline diagram to an `exchange' diagram with a single crossing, as shown in Figure \ref{fig:6jfig}. To get from diagram \ref{6jfig:b} to \ref{6jfig:c}, one has to perform two permutations on the OPE coefficients of \eqref{eq:6jcontraction}, which explains the phase factor in equation \eqref{eq:nongauss}. Cutting the edges $2$ and $5$ then gives the OPE diagram of the torus with four punctures.} A similar phenomenon arises for the quadratic moment, where the crossing equations on the four-holed sphere, two-holed torus and genus-two surface all lead to analytic continuations of the same universal function $C_0(P_1,P_2,P_3)$ \cite{Collier:2019weq}.

Let us now apply the $6j$ contraction to the averaged product of genus-two partition functions:
\vspace{2mm}
 \begin{align}
     &\overline{Z_{g=2}(\Omega,\bar\Omega)Z_{g=2}(\Omega',\bar\Omega')} \big \vert_{6j}\; \nonumber \\[1em] &\quad = \sum_{1,2,3}\sum_{1',2',3'}(-1)^{\sum_i(J_i+J_i')}\, \overline{C_{\textcolor{blue}{1} 2\textcolor{teal}{3}}C_{\textcolor{blue}{1'}2'{3'}}C_{2{3}{1}}C_{\textcolor{teal}{3'}{1'}2'}}\big \vert_{6j}\, |\mathcal{F}_{2,0}^S(\mathbf{P};\Omega)\mathcal{F}_{2,0}^S(\mathbf{P}';\Omega')|^2
     \\[0.5em]&\quad= \sum_{1,2,3,4} (-1)^{s}\left|\frac{C_0(P_1,P_4,P_3)^2}{\rho_0(P_2)}\fkerbig{P_4}{P_2}{P_1}{P_3}{P_1}{P_3}\,\mathcal{F}_{2,0}^S(P_1,P_2,P_3;\Omega)\mathcal{F}_{2,0}^S(P_1,P_4,P_3;\Omega')\right|^2\label{eq:WHnG}
 \end{align}
 where in the first equality we used the cyclic permutation symmetry $C_{123}=C_{231}=C_{312}$.
 In the second equality we defined {\small $s=2J_1+2J_3+J_2+J_{2'}$} $\in \mathbb{Z}$ and relabelled $2'\to 4$. 
The $6j$ contraction has set $\textcolor{blue}{1} = \textcolor{blue}{1'}$ and $\textcolor{teal}{3} = \textcolor{teal}{3'}$, so there are four sums remaining. In the continuum approximation {\small $\sum_h \to \int \dd P \dd \bar P\,\rho_0(P,\bar P)$} we  see that one of the factors of the density of states is cancelled by the factor of {\small $\rho_0(P_2)$} in the denominator of the $6j$ contraction. So this particular contraction is competing with the Gaussian contraction. A rough estimation of the size of this non-Gaussianity was given in \cite{Belin:2021ryy}, and it was indeed found that it is of the same order as the Maldacena-Maoz contribution. Therefore, a wormhole saddle should exist in 3D gravity that captures this behaviour. We will now show how to construct this wormhole.

\subsection{Non-Gaussian genus-two wormhole}\label{sec:nG-genus2}
There is a large class of Euclidean wormholes in 3D whose boundary consists of a pair of genus-two surfaces. These genus-two wormholes are distinct from the usual Maldacena-Maoz wormhole, and capture the non-Gaussian statistics of $(C_{\mathrm{HHH}})^4$. We first describe their topology and compute their exact partition function, and then we match these to the CFT$_2$ prediction.

\begin{figure}
    \centering
\begin{tikzpicture}
\node at (0,0) {\includegraphics{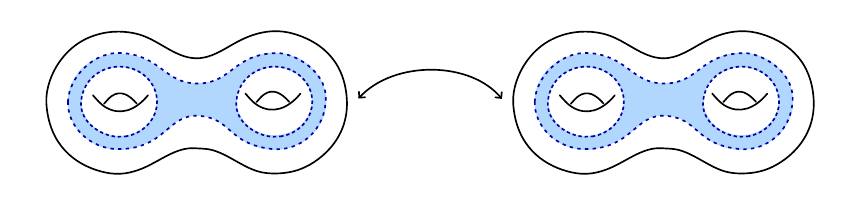}};
\node at (0,1) {$\gamma$};
\end{tikzpicture}
\caption{{\small Heegaard splitting of the non-Gaussian genus-two wormhole.}}
\label{fig:genus2identify2}
\end{figure}
Consider the genus-two Heegaard splitting illustrated in Figure \ref{fig:genus2identify2}. 
The construction is by now familiar: we take a pair of compression bodies and identify the inner boundaries after acting with an element $\gamma$ from the Moore-Seiberg groupoid on the genus-two splitting surface: 
\begin{equation}
    \Sigma_{2,0}^{\mathsf{L},\mathrm{inner}} \sim_\gamma \overline{\Sigma}_{2,0}^{\mathsf{R},\mathrm{inner}}.
\end{equation}
This results in a two-boundary wormhole $M_\gamma$, whose  Virasoro TQFT partition function is a state in the tensor product {\small $\mathcal{H}_{\mathrm{Vir}}(\Sigma_{2,0})\otimes \mathcal{H}_{\mathrm{Vir}}(\Sigma_{2,0})$}. The state is determined by the matrix element of the operator {\small $\mathbb{U}(\gamma)$} that implements the Moore-Seiberg transformation, 
\vspace{0.2cm}
\begin{equation}\label{eq:nGwormhole}
   \ket{Z_{\mathrm{Vir}\vphantom{F^2}}(M_\gamma)} = \int_0^\infty \dd^3 P \dd^3 P\dash \bra{\mathcal{F}^{\,\mathrm{inner}}_{2,0}(\mathbf{P}\dash)}\mathbb{U}(\gamma)\ket{\mathcal{F}^{\,\mathrm{inner}}_{2,0}(\mathbf{P})} \,\ket{\widehat{\mathcal{F}}^{\,\mathrm{outer}}_{2,0}(\mathbf{P})} \otimes \ket{\widehat{\mathcal{F}}^{\,\mathrm{outer}}_{2,0}(\mathbf{P}\dash)}.\vspace{0.2cm}
\end{equation}
Here we denoted $\dd^3P = \dd P_1\dd P_2\dd P_3$ and $\mathbf{P} = (P_1,P_2,P_3)$. The states are genus-two conformal blocks in the sunset channel. The normalization of the hatted state is as in \eqref{eq:normnotation}, with
\vspace{0.2cm}
\begin{equation}
    \rho_{2,0}(\mathbf{P}) = \rho_0(P_1)\rho_0(P_2)\rho_0(P_3)C_0(P_1,P_2,P_3)^2.
\end{equation}

Note that the structure of the partition function is the same as the example of the torus one-point wormhole in Section \ref{sec:onepointWH}. However, a crucial difference is that the homeomorphism $\gamma$ does not have to be an element in the mapping class group of the boundary genus-two surface: it can be any element of the Moore-Seiberg group (which is in general larger than the mapping class group). In particular, $\gamma$ is allowed to change the type of `stick figure' diagram of the conformal block, for example taking the sunset to the dumbbell channel:
\vspace{0.2cm}
\begin{equation}\label{eq:sunset_to_dumbbell}
     \begin{tikzpicture}[x=0.75pt,y=0.75pt,yscale=-1,xscale=1,baseline={([yshift=-2ex]current bounding box.center)}]
\draw   (251.67,230.33) .. controls (251.67,219.29) and (260.62,210.33) .. (271.67,210.33) .. controls (282.71,210.33) and (291.67,219.29) .. (291.67,230.33) .. controls (291.67,241.38) and (282.71,250.33) .. (271.67,250.33) .. controls (260.62,250.33) and (251.67,241.38) .. (251.67,230.33) -- cycle ;
\draw    (291.67,230.33) -- (331.67,230.33) ;
\draw   (331.67,230.33) .. controls (331.67,219.29) and (340.62,210.33) .. (351.67,210.33) .. controls (362.71,210.33) and (371.67,219.29) .. (371.67,230.33) .. controls (371.67,241.38) and (362.71,250.33) .. (351.67,250.33) .. controls (340.62,250.33) and (331.67,241.38) .. (331.67,230.33) -- cycle ;
\draw   (88,230) .. controls (88,218.95) and (110.39,210) .. (138,210) .. controls (165.61,210) and (188,218.95) .. (188,230) .. controls (188,241.05) and (165.61,250) .. (138,250) .. controls (110.39,250) and (88,241.05) .. (88,230) -- cycle ;
\draw    (200,230) -- (237,230) ;
\draw [shift={(240,230)}, rotate = 180] [fill={rgb, 255:red, 0; green, 0; blue, 0 }  ][line width=0.08]  [draw opacity=0] (6.25,-3) -- (0,0) -- (6.25,3) -- cycle    ; 
\draw    (138,210) -- (138,250) ;
\draw (201.67,210.73) node [anchor=north west][inner sep=0.75pt]  [font=\footnotesize]  {$\gamma =\mathbb{F}$};
\end{tikzpicture}.\vspace{0.2cm}
\end{equation}
In other words, the wormhole $M_\gamma$ need not arise as a modular image in the (relative) boundary modular sum, as it was the case for the torus one-point wormhole. 

The above procedure gives an infinite family of wormhole partition functions, one for each choice of $\gamma$, which should all contribute to the quartic OPE average. Our task is therefore to find the $\gamma$ that matches the leading approximation to {\small $C_{\mathrm{HHH}}^4$} predicted by the boundary ensemble. It turns out that the 
Moore-Seiberg element that reproduces the boundary computation in the previous section is given by one fusion move and two braids,
\vspace{1mm}
\begin{equation}
    \gamma = \mathbb{B}\circ \mathbb{F}\circ \mathbb{B}.\vspace{1mm}
\end{equation}
The fusion move acts on the middle leg of the sunset conformal block as in \eqref{eq:sunset_to_dumbbell}. The braiding moves perform a half-twist also around the middle leg of the sunset diagram. 

Using the inner product \eqref{eq:innerproduct}, the matrix element of $\mathbb{U}(\gamma)$ can be evaluated explicitly:
\vspace{0.2cm}
\tikzset{every picture/.style={line width=0.75pt}}        
\begin{align}
\bra{\mathcal{F}^{\,\mathrm{inner}}_{2,0}}\mathbb{U}(\gamma)\ket{\mathcal{F}^{\,\mathrm{inner}}_{2,0}} &=
    \bra{\begin{tikzpicture}[x=0.75pt,y=0.75pt,yscale=-0.8,xscale=0.8,baseline={([yshift=-.5ex]current bounding box.center)}]
\draw   (70,90) .. controls (70,78.95) and (87.91,70) .. (110,70) .. controls (132.09,70) and (150,78.95) .. (150,90) .. controls (150,101.05) and (132.09,110) .. (110,110) .. controls (87.91,110) and (70,101.05) .. (70,90) -- cycle ;
\draw    (110,70) -- (110,110) ;
\draw (72.2,82.0) node [anchor=north west][inner sep=0.75pt]  [font=\scriptsize]  {${1}'$};
\draw (93.2,82.0) node [anchor=north west][inner sep=0.75pt]  [font=\scriptsize]  {${2}'$};
\draw (130.8,82.0) node [anchor=north west][inner sep=0.75pt]  [font=\scriptsize]  {${3}'$};
\end{tikzpicture}}\mathbb{B}\circ \mathbb{F}\circ \mathbb{B}\ket{\begin{tikzpicture}[x=0.75pt,y=0.75pt,yscale=-0.8,xscale=0.8,baseline={([yshift=-.5ex]current bounding box.center)}]
\draw   (70,90) .. controls (70,78.95) and (87.91,70) .. (110,70) .. controls (132.09,70) and (150,78.95) .. (150,90) .. controls (150,101.05) and (132.09,110) .. (110,110) .. controls (87.91,110) and (70,101.05) .. (70,90) -- cycle ;
\draw    (110,70) -- (110,110) ;
\draw (72.2,84.0) node [anchor=north west][inner sep=0.75pt]  [font=\scriptsize]  {${1}$};
\draw (93.2,84.0) node [anchor=north west][inner sep=0.75pt]  [font=\scriptsize]  {${2}$};
\draw (130.8,84.0) node [anchor=north west][inner sep=0.75pt]  [font=\scriptsize]  {${3}$};
\end{tikzpicture}} \\[1.5em]
& \hspace{-2cm}=\int_0^\infty\dd P_4 \,\mathbb{B}_{P_4}^{P_1P_3} \,\fkerbig{P_2}{P_4}{P_1}{P_3}{P_1}{P_3} \mathbb{B}_{P_2}^{P_1P_3} \frac{\delta(P_1'-P_1)\delta(P_2'-P_4)\delta(P_3'-P_3)}{\rho_0(P_1')\rho_0(P_2')\rho_0(P_3')C_0(P_1',P_2',P_3')^2}.\\[-0.8em] \nonumber
\end{align}
 We now substitute our result into \eqref{eq:nGwormhole} to obtain the Virasoro TQFT partition function. From there, the full 3D gravity partition function is calculated by taking the product with the right-moving sector and going to the wavefunction basis $\bra{ \Omega}\otimes \bra{\Omega'}$. Doing so gives
\begin{equation}\label{eq:nonGaussPartFn}
\begin{split}
    Z_{\mathrm{grav}}(M_\gamma) = \Bigg |\int_0^\infty \dd^4 P\,\rho_0(P_1)\rho_0(P_2)\rho_0(P_3) \,&C_0(P_1,P_2,P_3)^2\,\mathbb{B}_{P_2}^{P_1P_3}\,\fker{P_2}{P_4}{P_1}{P_3}{P_1}{P_3} \,\mathbb{B}_{P_4}^{P_1P_3}\\ & \times \mathcal{F}_{2,0}^S(P_1,P_2,P_3;\Omega)\mathcal{F}_{2,0}^S(P_1,P_4,P_3;\Omega')\Bigg|^2.
\end{split}
\end{equation} 

Let us compare this answer to the boundary computation \eqref{eq:WHnG}. Firstly,
 the braiding phases, together with their right-moving counterpart, give the total phase factor 
 \vspace{2mm}
\begin{equation}
  \mathbb{B}_{P_2}^{P_1P_3}\mathbb{B}_{P_4}^{P_1P_3} \,\bar{\mathbb{B}}_{\bar P_2}^{\bar P_1\bar P_3}\bar{\mathbb{B}}_{\bar P_4}^{\bar P_1\bar P_3} =  \e^{\pi i( J_2 + J_4)} \e^{-\pi i (2J_1 + 2 J_3)}.\vspace{2mm}
\end{equation}
This precisely reproduces the phase factor $(-1)^s$ in \eqref{eq:WHnG}, if we assume the spin of the heavy operators to be quantized. As a last step, we swap the internal indices of the $\mathbb{F}$ kernel using the formula \eqref{app:swap}. Then we see that the gravity partition function exactly matches the continuum limit of the averaged product of genus-two partition functions \eqref{eq:WHnG} in the CFT$_2$. 

In conclusion, we have been able to match a particular non-Gaussian contraction, which was predicted by typicality, to a novel Euclidean wormhole. This contraction  highlights the importance of non-Gaussianities, as gravity naturally captures their contribution in the sum over topologies. Even though in the microcanonical ensemble these non-Gaussianities are exponentially suppressed in the $e^{-S}$ expansion, their imprint on the variance of CFT observables can contribute at leading order due to the non-trivial index structure.

\section{OPE statistics and light matter}
\label{sec:lightmatter}

So far, we have discussed the gravitational interpretation of OPE statistics coming from crossing equations where the identity dominates in some channel. However, there are crossing equations where the identity exchange vanishes in the microcanonical window. In these circumstances, one must consider non-identity exchanges of light matter.  In this section, we comment on the effects of massive particle exchanges on the statistics of OPE coefficients, and give a geometrical interpretation using the Virasoro TQFT.

\begin{figure}
\centering
\begin{tikzpicture}
\node at (0,0) [midway]{\includegraphics{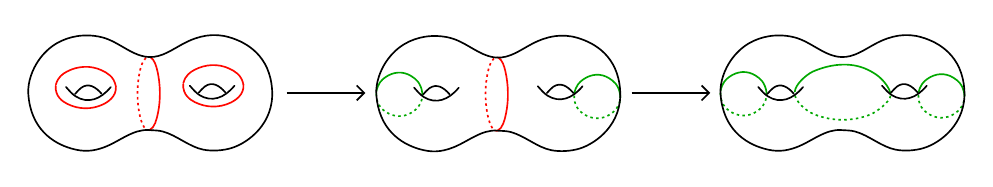}};
\node at (-3,0.3) {$\bbs\,\bbs$};
\node at (3,0.3) {$\bbf$};
\end{tikzpicture}
\caption{{\small Crossing moves relating the genus-two block in the dumbbell (left) and  sunset (right) channels.}}
\label{fig:gen2cros}
\end{figure}

We will narrow our focus to a single subthreshold operator, $\chi$, and assume this operator is sufficiently massive to create a conical defect in the bulk. For simplicity, our discussion will be centered on the genus-two crossing equation resulting from the sequence of moves in Figure \ref{fig:gen2cros}. 
Following the arguments of Sections \ref{sec:3pointfunction} and \ref{sec:4-pt}, the sequence of moves in Figure \ref{fig:gen2cros} leads to the following equation for the density of OPE coefficients 
\begin{equation}
\label{eq:crosEqND}
    \mean{C_{112}C_{233}} = \frac{1}{\rho_0(\bfP,\bar \bfP)} 
    \sum_{1'2'3'} C_{1'2'3'}C^*_{1'2'3'}
    \left| \sker{P_1'}{P_1}{P_2}
    \sker{P_3'}{P_3}{P_2}
    \fker{P_2'}{P_2}{P_3'}{P_3'}{P_1'}{P_1'}\right|^2.
\end{equation}
Note that this is the opposite order of the sequence of moves that was used to derive the $C_0$ formula from the genus-two partition function \cite{Collier:2019weq}.
The exchange of the identity $P_{1,2,3}'=\bbi$ in this sum gives the contraction 
\begin{equation}
\label{eq:identitycont}
\mean{C_{112}C_{233}} \supset \frac{\delta(h_2)}{\rho_0(h_2)}\frac{\delta(\bar{h}_2)}{\rho_0(\bar{h}_2)},
\end{equation}
which follows from $\fker{\bbi}{2}{\bbi}{\bbi}{\bbi}{\bbi} = \delta(P_2-\tfrac{iQ}{2})$. For $h_2$ above the black hole threshold, this contribution always vanishes. So the leading term in the heavy limit, where $P_{1,2,3} = P + \delta_{1,2,3}$ and $P\rightarrow\infty$, is given by the exchange of the lightest scalar above the vacuum, $3',1' = \chi$ and $2'=\bbi$\footnote{See \cite{Belin:2021ibv} for more details on this analysis.}. This term leads to the inclusion of the following contraction in the statistics,
\begin{equation}
\label{eq:nonUsta}
    \mean{C_{112}C_{233}} \supset \left|\frac{
    \sker{P_\chi}{P_3}{P_2}
    \sker{P_\chi}{P_1}{P_2}
    }{\rho_0(P_1)\rho_0(P_3)} C_0(P_2,P_\chi,P_\chi)\right|^2.\vspace{1mm}
\end{equation}
A few things to note about this contraction are the following. In the heavy limit where $P_{1,2,3} \rightarrow \infty$ and the differences $P_i-P_j$ are fixed, this moment scales as (see Eq.(3.14) in \cite{Belin:2021ryy}):
\begin{equation}
   \mean{C_{112}C_{233}} \sim \e^{-\frac{3}{2}S_0(P,\bar P)-4 \pi (\alpha_\chi P + \bar \alpha_\chi \bar P)}.\vspace{2mm}
\end{equation}
This is to be compared to the scaling of the contraction \eqref{eq:non-diagonal} that comes from an identity exchange in a different channel. In this heavy limit, the contraction \eqref{eq:non-diagonal} is of the order of $\e^{-2 S_0(P,\bar P)}$, which is exponentially suppressed with respect to \eqref{eq:nonUsta}. This means that the leading statistics in the microcanonical window in this heavy limit are given the exchange of the $\chi$ operator.

In the Virasoro TQFT, considering massive particles in the bulk allows us to include Wilson lines that do not extend to the asymptotic boundary. Instead, they wrap around some of the non-trivial cycles in the manifold. One of these geometries is dual to the contraction given in \eqref{eq:nonUsta}:
\begin{equation}
\label{eq:wormholepic}
\mean{C_{112}C_{233}}\supset
\left|
\vcenter{\hbox{\begin{tikzpicture}
\node at (0,0) {\includegraphics{img/wilsonlines.pdf}};
\node at (-1.26,0.65) {\scriptsize{$1$}};
\node at (0,0.37) {\scriptsize{$\chi$}};
\node at (0,-0.37) {\scriptsize{$2$}};
\node at (1.26,0.65) {\scriptsize{$3$}};
\end{tikzpicture}}}\right|^2.
\end{equation}
This is a wormhole between two 3-punctured spheres, with a collection of Wilson lines creating conical defects in the bulk. The Wilson lines labelled by Liouville momenta $P_1,P_2,P_3$ are anchored to the boundary, whereas the Wilson loop $\chi$ wraps around the Wilson lines $1$ and $3$. 

To evaluate this amplitude in gravity, we first insert fictitious identity lines and solve for the resulting tangle using crossing transformations: 
\begin{equation}\!\!\!
\vcenter{\hbox{\begin{tikzpicture}
\node at (0,0) {\includegraphics{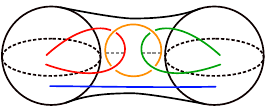}};
\node at (-0.32,0.22) {\scriptsize{$\bbi$}};
\node at (0.33,0.22) {\scriptsize{$\bbi$}};
\end{tikzpicture}}} 
\!=\! 
\int_0^\infty\dd P_s\,\dd P_t\;
\fker{\bbi}{P_s}{P_1}{P_\chi}{P_\chi}{P_1}
\fker{\bbi}{P_t}{P_3}{P_\chi}{P_\chi}{P_3}
\vcenter{\hbox{\begin{tikzpicture}
\node at (0,0) {\includegraphics{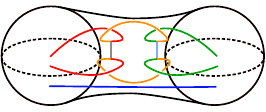}};
\node at (-0.25,0.1) {\scriptsize{$s$}};
\node at (0.25,0.1) {\scriptsize{$t$}};
\end{tikzpicture}}}. 
\end{equation}
Topologically, what this equation is saying is that we cut along two 
4-punctured spheres, one surrounding the tangle of momenta $P_{\chi}$ and $P_1$, and the other surrounding $P_{\chi}$ and $P_3$. This creates a pair of states of four-point identity blocks in the Virasoro TQFT. We then apply a fusion move on each identity block and glue back the resulting balls into the wormhole geometry.

We can undo the braiding of Wilson lines using the following relation 
\begin{equation}
\vcenter{\hbox{\begin{tikzpicture}
\node at (0,0) {\includegraphics{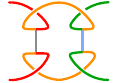}};
\node at (-0.25,0) {$s$};
\node at (0.25,0) {$t$};
\end{tikzpicture}}} 
=
\e^{2\pi i \left(h_s-h_t\right)}
\vcenter{\hbox{\begin{tikzpicture}
\node at (0,0) {\includegraphics{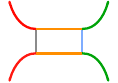}};
\node at (-0.25,0) {$s$};
\node at (0.25,0) {$t$};
\end{tikzpicture}}}.
\end{equation}
After these two moves, we can identify the expression inside the integral as the following wormhole amplitude 
\begin{equation}
\vcenter{\hbox{\begin{tikzpicture}
\node at (0,0) {\includegraphics{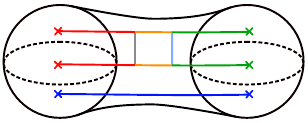}};
\node at (-1.26,0.65) {\scriptsize{$1$}};
\node at (0,0.37) {\scriptsize{$\chi$}};
\node at (0,-0.37) {\scriptsize{$2$}};
\node at (1.26,0.65) {\scriptsize{$3$}};
\node at (-0.5,0.2) {$s$};
\node at (0.5,0.2) {$t$};
\end{tikzpicture}}}
=
\frac{
\fker{P_s}{P_2}{P_\chi}{P_\chi}{P_1}{P_1}
\fker{P_t}{P_2}{P_\chi}{P_\chi}{P_3}{P_3}
}{
\rho_0(P_2)^2C_0(P_\chi,P_2,P_\chi)
}
\ket{\mathcal{F}_{0,3}}
\otimes
\ket{\mathcal{F}_{0,3}}.
\end{equation}
The next step is to solve for the integrals over the momenta $P_t$ and $P_s$. These two integrals can be solved by using the consistency condition at genus one and two punctures \eqref{eq:relgen1} and the complex conjugate of this expression.\footnote{By taking the complex conjugate of \eqref{eq:relgen1} and using the fact that $\sker{P_{\chi}}{P_1}{P_2}^* = \e^{-\pi i h_2}\sker{P_{\chi}}{P_1}{P_2}$, we can solve for the integral weighted by the phase $\e^{-2\pi h_t}$.} The answer is
\begin{equation}
    \int_0^\infty \dd P \,
    \e^{2\pi i h}\,
    \fker{\bbi}{P}{P_{1,3}}{P_\chi}{P_\chi}{P_{1,3}}
    \fker{P}{P_2}{P_\chi}{P_\chi}{P_{1,3}}{P_{1,3}} = \frac{\e^{2\pi i h_{\chi} - \pi i h_2}}{\rho_0(P_{1,3})}
    \fker{\bbi}{P_2}{P_\chi}{P_\chi}{P_\chi}{P_\chi}
    \sker{P_\chi}{P_{1,3}}{P_2}.
\end{equation}
The final result is the following amplitude for the wormhole, 
\begin{equation}
\left|
\vcenter{\hbox{\begin{tikzpicture}
\node at (0,0) {\includegraphics{img/wilsonlines.pdf}};
\end{tikzpicture}}} \right|^2
= \e^{-\pi i(h_2-\bar h_2)} 
\left|\frac{
\sker{P_\chi}{P_3}{P_2}
\sker{P_\chi}{P_1}{P_2}
}{\rho_0(P_1)\rho_0(P_3)}C_0(P_2,P_\chi,P_\chi)\right|^2.
\end{equation}
The reality condition on the OPE coefficient $C_{112} = C_{112} = (-1)^{J_2} C_{121}$ implies that $J_2 =  h_2-\bar h_2$ is an even integer, meaning that $(-1)^{J_2} = 1$, matching exactly to \eqref{eq:nonUsta}.

\begin{figure}
    \centering
    \begin{tikzpicture}
    \node at (0,0) {\includegraphics{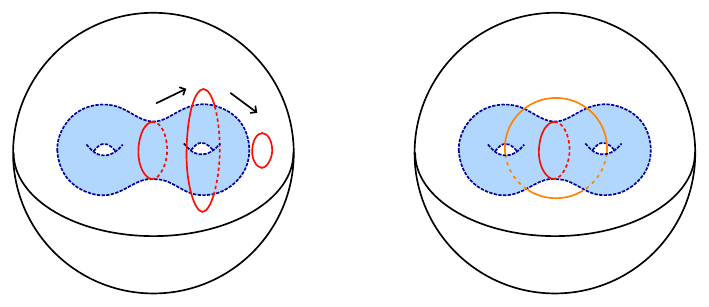}};
    \node at (-1.4,2) {$S^3$};
    \node at (-3.43,-0) {$P_2$};
    \node at (3.4,-0) {$P_2$};
    \node at (3.5,1.3) {$P_\chi$};
    \node at (-5.8,2.5) {a)};
    \node at (1,2.5) {b)};
    \end{tikzpicture}
    \caption{{\small A handlebody with a genus-two boundary whose filling is the outside region of the three-sphere. On the left, the cycle labeled by $P_2$ is contractible in the bulk. On the right, the same cycle is confined by the Wilson line $P_\chi$.}}
    \label{fig:gravInt}
\end{figure}
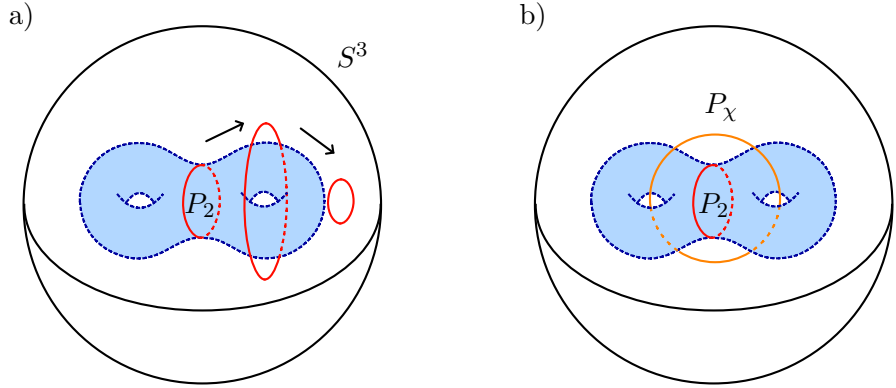

The above derivation is a `microcanonical' bulk interpretation for the statistics of a product of OPE coefficients $C_{122}C_{233}$. However, we can also study the canonical version of the above argument, in which the OPE coefficients are summed over in the genus-two partition function. This can be seen geometrically as the thickening of the Wilson lines in the amplitude \eqref{eq:wormholepic}, which become cylinders, and lead to a handlebody with a genus-two boundary. We can picture a handlebody as a filling of a genus-two surface, where a set of cycles becomes contractible. Equivalently, a handlebody can be seen as the \emph{complement} of the filled-in genus-two surface in the three-sphere $S^3$. This is illustrated in figure \ref{fig:gravInt}a). It is clear from this figure that the $P_2$ cycle is contractible in the bulk. 

Any cycle in the genus two surface that is contractible in the bulk corresponds to an identity exchange in the boundary partition function. Moreover, we cannot have a non-zero momentum state $P\neq \bbi$ propagating through the corresponding channel. However, if we modify the bulk geometry to include a massive point particle, the fundamental group is modified accordingly and the $P_2$ cycle becomes non-contractible. This is shown in Figure \ref{fig:gravInt}b). In this sense, we view the massive particle as confining the $P_2$ cycle and thereby allowing high temperatures in this channel. This gives a gravitational interpretation of the dependence of $P_\chi$ in the statistics of the dumbbell OPE contraction $\overline{C_{122}C_{233}}$. Note that this dependence is only through the conformal dimension of $\chi$, and does not depend on the OPE coefficient $C_{H\chi\chi}$ as it is the case for the thermal one-point function \cite{Kraus_2017}.


\subsection{Geometric insight into the corrections}
\label{sec:geoint}

Now, we would like to offer a geometric understanding for the validity of the crossing equations.  So far, we have only explored crossing in the limit where the identity exchange dominates. The corrections to the identity exchange are encoded in the asymptotics of the crossing kernel via the relation
\begin{equation}
    \frac{
    \mathbb{K}_{\bfP',\bfP}
    }{\mathbb{K}_{\bbi,\bfP}} \rightarrow 0,\quad \text{as}\quad \bfP \rightarrow \infty, \quad \bfP' \neq \bbi 
\end{equation}
in a given large-$\bfp$ limit. The rate at which this ratio goes to zero is usually exponentially fast in $|\bfp|$, rendering the corrections to be `non-perturbative'. In this section, we would like to point out that there is a simple geometric picture that aligns with the asymptotic properties of crossing kernels and gives a physical interpretation to the formulas we have been discussing until now. 

For simplicity, we restrict our discussion to the genus-two partition function and the crossing equation in \eqref{eq:crosEqND}. When integrating the density of OPE coefficients $\mean{C_{112}C_{233}}$ against the dumbbell channel conformal blocks, the two contractions in \eqref{eq:identitycont} and \eqref{eq:nonUsta} yield the vacuum block in the sunset channel and the block with momenta $P_1 = P_3 = P_\chi$, 
\begin{align}
Z_{g=2}(\Omega,\bar \Omega) &= \int\dd^3P \; \rho_0(\bfP,\bfp') \; \mean{C_{112}C_{233}}\;\mathcal{F}_{0,2}^{\mathcal{D}}(\bfP; \Omega)
\bar{\mathcal{F}}_{2,0}^{\mathcal{D}}
(\bar \bfP; \bar \Omega) \\[1em]\notag&=
\mathcal{F}_{0,2}^{\mathcal{S}}(\bbi,\bbi,\bbi; \Omega)
\bar{\mathcal{F}}_{2,0}^{\mathcal{S}}
(\bbi,\bbi,\bbi; \bar \Omega) 
+
\mathcal{F}_{0,2}^{\mathcal{S}}(P_\chi,\bbi,P_{\chi}; \Omega)
\bar{\mathcal{F}}_{2,0}^{\mathcal{S}}
(\bar P_\chi,\bbi,\bar P_{\chi}; \bar \Omega) + \cdots\,.
\end{align}
Intuitively, we understand that the genus-two partition function is well approximated by these two blocks, albeit the second block is exponentially suppressed with respect to the first one, whenever we are in a corner of moduli space  $\Omega$ where the genus-two surface looks like an elongated union of three cylinders:
\begin{equation}
\label{eq:diagsun}
    \vcenter{
    \hbox{
    \begin{tikzpicture}
    \node at (0,0) {\includegraphics{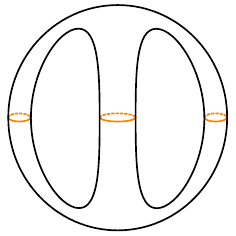}};
    \node at (0,-2.4) {$\Omega = (L_1,L_2,L_3;\tau_1,\tau_2,\tau_3)$};
    \node at (-0.8,0) {$L_2,\tau_2$};
    \node at (-2.4,0) {$L_1,\tau_1$};
    \node at (2.4,0) {$L_3,\tau_3$};
    \end{tikzpicture}
    }}.
\end{equation}
Roughly speaking, one can think of each of these cylinders as carrying a modulus $\beta_i$ for $i=1,2,3$ describing the length of the cylinder which corresponds to Euclidean time evolution. When the length of the cylinder goes to infinity, this effectively zooms into very low temperatures, $\beta\rightarrow\infty$,  and the operator $\e^{-\beta H}$ creates an exponential hierarchy weighted by the energy of the states.
Here, to be more precise, we will parametrize the moduli $\Omega$ using the Fenchel–Nielsen coordinates of Teichmüller space. These coordinates specify the lengths of the cuffs $L_{1,2,3}$ and the gluing twists $\tau_{1,2,3}$ of the pair-of-pants decomposition. These coordinates are not unique as there are multiple ways to decompose a Riemann surface into pairs of pants. For example, we could also have chosen the lengths $\tilde L_{1,2,3}$ and twists $\tilde \tau_{1,2,3}$ associated to the dumbbell channel, 
\begin{equation}
\label{eq:diadum}
    \vcenter{
    \hbox{
    \begin{tikzpicture}
    \node at (0,0) {\includegraphics{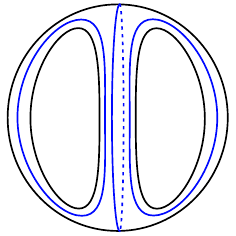}};
    \node at (0,-2.4) {$\tilde \Omega = (\tilde L_1,\tilde L_2,\tilde L_3;\tilde \tau_1,\tilde \tau_2,\tilde\tau_3)$};
    \node at (0,2.4) {$\tilde L_2,\tilde\tau_2$};
    \node at (-2.4,0) {$\tilde L_1,\tilde\tau_1$};
    \node at (2.4,0) {$\tilde L_3, \tilde\tau_3$};
    \end{tikzpicture}
    }}.
\end{equation}

Remarkably, the maps that relate the coordinates $\Omega$ and $\tilde \Omega$ are known in closed form. Similar to how crossing kernels are constructed, maps that connect two coordinate systems in Teichmüller space can be derived from two basic relations found by Okai \cite{Takayuki1993EffectsOA}:\footnote{In the original work of \cite{Takayuki1993EffectsOA} there is a misprint in the formula for $\tau'$ for the one-holed torus. The correct formula for the torus can be found in \cite{2020arXiv201011806E}. Additionally, there is a misprint for $\tau_a$ in the sphere four-point function as presented in  \cite{2020arXiv201011806E}. Formula  (5.20) of that paper is missing two apostrophes in the analogue of $A_{ij}(\ell)$.}

\begin{itemize}
    \item[1.] The first relation connects the lengths $L_a,L_b$ and twists $\tau_a,\tau_b$ of the four-holed sphere, 
    
\begin{equation}
\nonumber
    \vcenter{
    \hbox{
    \begin{tikzpicture}
        \node at (0,0) {\includegraphics{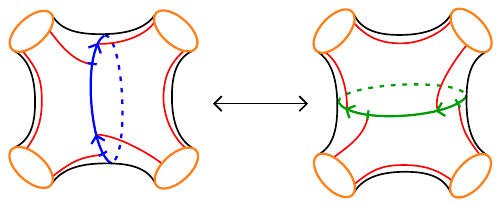}};
        \node at (-4,-1.5) {$L_4$};
        \node at (-4,1.7) {$L_1$};
        \node at (-0.9,1.7) {$L_2$};
        \node at (-0.9,-1.5) {$L_3$};
        \node at (4.2,-1.5) {$L_4$};
        \node at (4.2,1.7) {$L_1$};
        \node at (1,1.7) {$L_2$};
        \node at (1,-1.5) {$L_3$};
        \node at (2.6,0.05) {$L_a$};
        \node at (3.25,-0.5) {$\tau_a$};
        \node at (-3,0) {$L_b$};
        \node at (-3,-0.7) {$\tau_b$};
    \end{tikzpicture}
    }
    }
\end{equation}
\begin{align}
\label{eq:penner1}
\ell_a (\ell_b^2-1) &= \ell_1 \ell_2 + \ell_3\ell_4 + \ell_b 
\left(
\ell_1 \ell_3 + \ell_2\ell_4
\right)
+\cosh(\tau_b)A_{14}(\ell_b)A_{23}(\ell_b),
\\\label{eq:penner2}
\cosh(\tau_{a}) & = \frac{
(\ell_a^2-1)\ell_b - \ell_1\ell_4 - \ell_2\ell_3 
-\ell_a\left(\ell_1\ell_3+\ell_2\ell_4\right)
}{A_{12}(\ell_a) A_{34}(\ell_a)},
\end{align}
where
\begin{equation}
\ell_i \coloneqq \cosh \frac{L_i}{2} \quad \text{and} \quad A_{ij}(\ell) \coloneqq \sqrt{\ell^2 + \ell_i^2+\ell_j^2
+2\ell_i\,\ell_j\,\ell-1}\,.
\end{equation}
With the sign of $\tau_a$ is given by $-\text{sgn}(\tau_b)$.
\item[2.] The second relation connects the lengths and twists of the punctured torus,
\begin{equation}
\nonumber
\vcenter{
\hbox{
\begin{tikzpicture}
    \node at (0,0) {\includegraphics{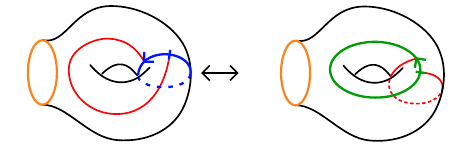}};
    \node at (-3.8,0) {$L_0$};
    \node at (0.5,0) {$L_0$};
    \node at (-1.1,-0.45) {$L$};
    \node at (-1.4,0.6) {$\tau$};
    \node at (2.4,0.8) {$L'$};
    \node at (3.4,0.35) {$\tau'$};
\end{tikzpicture}
}
}
\end{equation} where
\begin{align}
\label{eq:pennertorus}
    \ell' &= \cosh(\frac{\tau}{2}) \sqrt{
    \frac{\ell_0+2\ell^2-1}{2(\ell^2-1)},
    }\\ \label{eq:pennertorus2}
   \cosh(\frac{\tau'}{2}) &= \ell
   \sqrt{
   \frac{\cosh^2(\frac{\tau}{2})(\ell_0+ 2\ell^2-1)-2 (\ell^2-1)}{\cosh^2(\frac{\tau}{2})(\ell_0 + 2\ell^2-1) + (\ell^2-1)(\ell_0-1)}
   }.
\end{align}
Again, $\text{sgn}(\tau') = - \text{sgn}(\tau)$, and the curly notation $\ell$ indicates the hyperbolic cosine of half the length, $\ell = \cosh(L/2)$ as before. 
\end{itemize}

\noindent Using these two relations, we can write down the change of coordinates between any two pair-of-pants decompositions. Here we are going to focus on the case where all twists are zero. (One can check that the formulas above allow for $\tau$ and $\tau'$ to be zero simultaneously.)

One of the properties of the modular $\sker{\alpha_\chi}{P}{P_0}$ in the limit where $P,P_0 \rightarrow \infty$ and $P_0 = x P$, with fixed $x$, is the following asymptotic formula\footnote{We thank Henry Maxfield for discussions on this formula.}
\begin{equation}
    \label{eq:srela}
    \frac{\sker{\alpha_\chi,\,}{P'}{x P'}}{\sker{\bbi}{P'}{\bbi}} \sim \e^{-2\pi(2-x) \alpha_\chi P'}, \quad 0\leq x\leq 2, \,P'\rightarrow\infty.
    \end{equation}
We discuss this formula in Appendix \ref{app:kernels}. 
From \eqref{eq:srela}, one can see that as $x\rightarrow 2$, we lose the exponential hierarchy in the crossing equation that this kernel controls, as states with positive momentum $\alpha$ are no longer exponentially suppressed. This is evident in the asymptotic formula \eqref{eq:crosEqND} for the mean $\mean{C_{112}C_{233}}$, where the exponential suppression is governed by this kernel. The failure of this hierarchy simply means that the sunset identity channel is no longer the dominant contribution to the genus-two partition function in this regime of moduli space, and thus, one cannot ignore the effects of light operators in the crossing equation.

We can explain the breakdown of the crossing equation \eqref{eq:crosEqND} geometrically using Okai's formulas. The asymptotic formula for the dumbbell channel coefficients $\mean{C_{112}C_{233}}$ applies when $\tilde L_{1,2,3} \rightarrow\infty$ and $L_{1,2,3}\rightarrow 0 $. That is, heavy states in the dumbbell channel but light states in the sunset channel dominate the amplitude. More precisely, we are going to consider the region of moduli space where the lengths $\tilde L_1,\tilde L_2$ and $\tilde L_3$ are large, maintaining a constant ratio $\tilde L_2/\tilde L_{1,3} = x$. Using formula \eqref{eq:pennertorus}, we can derive a relationship between $L_1$ and $\tilde L_{1},\tilde L_2$ when $\tau_{1,2,3} = \tilde \tau_{1,2,3}=0$, 
\begin{equation}
    \cosh(\frac{L_1}{2})^2 =
    \frac{\cosh(\frac{\tilde L_2}{2})
    +2\cosh(\frac{\tilde L_1}{2})^2 -1
    }{
    2\cosh(\frac{\tilde L_1}{2})^2-2
    }.
\end{equation}
As $L_1\rightarrow0$ and $\tilde L_{1,2}\rightarrow\infty$, this equation reduces to
\begin{equation}
  1 + \order{L_1^2} = \e^{(\frac{x}{2}-1)\tilde L_1} + 1 + \order{\e^{-\tilde L_1}},
\end{equation}
which can only be satisfied when $x < 2$. This gives a geometrical explanation of the asymptotic behaviour of the modular S-kernel \eqref{eq:srela}. Since the S-kernel determines the exponential hierarchy between the identity and light states, Okai's formulas give a precise regime of the parameters where we can trust the identity approximation. 
It would be interesting to further explore these formulas and use them, for example, to understand the phase diagrams of higher genus Riemann surfaces. These diagrams have been investigated numerically, as in the case of the genus-two partition function \cite{Maxfield:2016mwh}. However, Okai's formulas offer a simple geometric perspective.

\section{Discussion}
\label{sec:discu}

In this paper we explored the consequences of typicality and the ETH in the context of AdS$_3$/CFT$_2$. The typicality assumption discussed in Section \ref{sec:typicality}, together with the consistency conditions of the CFT$_2$ give a complete description of the moments of the ensemble of OPE coefficients. Notably, the ensemble of OPE coefficients necessarily contains non-Gaussian contractions in order to satisfy crossing symmetry at higher genus. We presented several analytic formulas for the moments of this ensemble and matched each of these contractions to an on-shell bulk topology in 3D gravity. The crucial technical tool was the Virasoro TQFT \cite{Collier:2023fwi,Collier:2024mgv}, which allowed us to compute the full gravity partition function of each topology.

In the last part of this paper, we studied the effects of light matter on the statistics of the CFT. We showed that exchanges beyond the identity can be modeled in the Virasoro TQFT by considering Wilson lines in the bulk that do not reach the asymptotic boundary. We also gave a geometric interpretation for the asymptotic expansion of the modular S-kernel by analyzing the shape of the torus with one boundary using the Fenchel–Nielsen coordinates of Teichmüller space. 

We conclude here with some open questions and future directions. 

\paragraph{The sum over saddles:} 
The full set of crossing equations provides an infinite family of constraints on the ensemble of OPE data. Each of these crossing equations can be studied in a limit where the identity dominates in some channel. The expectation is that a single crossing equation, evaluated on the identity contribution, corresponds in the bulk to a single on-shell geometry. In sections \ref{sec:gravity} and \ref{sec:genOPE}, we gave examples of this correspondence, between particular crossing equations of thermal correlators and distinct multi-boundary wormhole saddles. However, an important open problem is to prove or disprove such a one-to-one mapping in generality. 
	
 The difficulty lies in the fact that the sum over bulk 3-manifolds with arbitrary boundary components has not been classified. This should be contrasted to the Maloney-Witten-Keller partition function \cite{Maloney_2010,Keller:2014xba}, for which the full sum over 3-manifolds with a single torus boundary is known in the form of a modular sum. Already in a simple example we can see that this story does not generalize to more complicated boundary conditions. Namely, suppose the boundary consists of 2 once-punctured tori. In Section \ref{sec:onepointWH}, we described a $PSL(2,\mathbb{Z})$ family of Euclidean wormholes filling in the bulk, which arose as relative modular images of the Maldacena-Maoz wormhole. We can easily construct an example of a hyperbolic 3-manifold with the same asymptotic boundary conditions that is not part of this $PSL(2,\mathbb{Z})$ family. An example can be constructed by the following genus-two Heegaard splitting 
\begin{equation}
    \vcenter{
    \hbox{
    \begin{tikzpicture}
    \node at (0,0) {\includegraphics{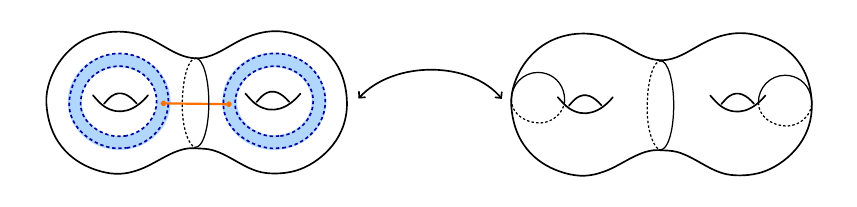}};
    \node at (0,1) {$\gamma$};
    \end{tikzpicture}
    }}
\end{equation}
where $\gamma$ is a (pseudo-Anosov) element of the genus-two mapping class group that is not an element of the $PSL(2,\mathbb{Z})$ subgroup. Moreover, one could go on and construct higher genus Heegaard splittings, with increasingly complicated gluing maps $\gamma \in \MCG(\Sigma_{g,0})$.

This example shows another important technical obstacle for enumerating the sum over saddles. Namely, it is well known that the Heegaard decomposition of a 3-manifold is not unique. Since we want to count each on-shell topology only once, it is important to have a method to distinguish whether two Heegaard splittings give rise to the same topology. This would amount to finding a variant of the Kirby calculus \cite{rolfsen2003knots}, adapted to the language of the Virasoro TQFT. 

\paragraph{The leading saddle point:}
We would like to point out that for multi-boundary wormholes, there is no simple way to determine which geometries are the leading contributions at large $c$, in the sum over saddles. For closed (or cusped) hyperbolic 3-manifolds there is the volume conjecture, see \cite{Kashaev:1996kc,murakami2011introduction,Witten:2010cx,Collier:2024mgv}, which posits that the on-shell action is given by the central charge times the hyperbolic volume of $M$. Therefore, at large $c$, these contributions are ordered according to their volume, and for closed, once-cusped and twice-cusped hyperbolic 3-manifolds the smallest-volume manifold is known (respectively: the Weeks manifold, the figure-8 knot complement and the Whitehead link complement). However, for 3-manifolds with more than two boundaries, there is no proof for a manifold of smallest hyperbolic volume. For infinite volume hyperbolic manifolds with asymptotic boundaries, the on-shell action is proportional to the renormalized volume. For more than two asymptotic boundaries, with a set of boundary moduli, it is also not known which 3-manifolds are the dominant saddles. It is likely that different contributions can exchange dominance as the moduli are varied. As a concrete question about the results in our paper, we could ask whether there exists a three-boundary wormhole that is leading over the wormhole we constructed in Section \ref{sec:onepointWH}. The answer is provided by the dual CFT, where a clear hierarchy exists.
\paragraph{Bulk mapping class groups:}
Virasoro TQFT is a useful tool in the computation of 3D gravity path integrals. However, the full gravity answer includes a sum over boundary mapping class group transformations, modulo the mapping class group of the bulk 3-manifold. Notably, the full MCG-sum is not holomorphically factorized. For finite-volume hyperbolic 3-manifolds $M$, the bulk mapping class group is finite, as a consequence of Mostow-Prasad rigidity \cite{gabai}. It coincides with the outer automorphism group of the fundamental group of $M$. However, determining the bulk MCG and the quotient of the boundary MCG is a non-trivial task in general. In particular, we have not been able to determine the correct “modular sum” for the three- and four-boundary wormholes constructed in section 3. Clearly, such a modular completion is necessary in a consistent duality map, since the average is taken over a product of modular covariant  torus one-point functions.

\paragraph{Inclusion of light matter in the spectrum:} In Section \ref{sec:lightmatter}, we initiated the study of the imprints that light matter fields have on the statistics of OPE coefficients. Usually, in pure gravity, such contributions are ignored, and the spectrum is taken to be the vacuum module plus the black hole spectrum. However, it is likely that a pure theory of gravity needs to be amended by the inclusion of sub-threshold states in order to render the density of states non-negative \cite{Benjamin:2020mfz,DiUbaldo:2023hkc}. Such sub-threshold states either create conical defects, when their conformal dimension scales with $c$, or correspond to Wilson lines in the bulk. We showed that the Virasoro TQFT can describe Wilson loops with arbitrary conformal weight, that can wind non-contractible cycles in the bulk 3-manifold without anchoring to the boundary. We expect that this is the general mechanism by which the bulk theory incorporates light matter corrections to the OPE statistics, similar to the `Wilson spool' proposal of \cite{Castro:2023bvo}. However, from the CFT point of view, it is rather difficult to input a light spectrum that is consistent with crossing symmetry. To preserve crossing, one has to include a family of discrete states with weights $\bar h,h< (c-1)/24$ that correspond to the so-called `double-twist' operators, and a continuum of states above  $\bar h,h > (c-1)/24$.  Moreover, it would be interesting to examine the consequences of including light matter in a more conventional sense, like the inclusion of a light scalar field in a non-trivial bulk topology.

\paragraph{Relation to tensor models:} In this paper, we derived the moments of OPE coefficients using only the modular invariance and crossing symmetry of the CFT. It would be interesting to verify if these moments match those of the tensor model described in \cite{Belin:2023efa}. In their work, the authors developed a tensor model intended to represent a set of approximate CFT data. Given that their model is consistent with crossing in a triple scaling limit, we anticipate a match between the two approaches \cite{Jafferis:2024jkb}. However, this match requires a careful analysis of the Schwinger-Dyson equations of the tensor model.  One advantage of the tensor model is that it provides insight into spectral correlations. Unfortunately, these correlations are not sufficiently constrained by crossing and seem to require additional input, either through the non-perturbative completion of these moments via the tensor model or from gravitational off-shell partition functions. Another interesting question is what the tensor model predicts for mixed correlations between the operator average and the spectral average, and whether these correlations can be detected in the bulk.

\paragraph{Using Virasoro TQFT for non-hyperbolic manifolds?}
One open question about the Virasoro TQFT is whether a finite TQFT amplitude always implies the existence of a complete hyperbolic metric in the bulk. It appears that certain non-hyperbolic 3-manifolds \emph{also} admit a finite amplitude. For example, the trefoil knot complement is known to be non-hyperbolic \cite{rolfsen2003knots} and it is homeomorphic to the mapping torus \enlargethispage{1\baselineskip}$\Sigma_{1,1}\times_\varphi S^1$, where the punctured torus is twisted by an element $\varphi = ST$ of the mapping class group $PSL(2,\mathbb{Z})$. The corresponding partition function is a trace in the TQFT Hilbert space, evaluating to\footnote{We thank Lorenz Eberhardt for discussions on this topic.}
\begin{equation}
    Z(\Sigma_{1,1}\times_\varphi S^1) = \int_0^\infty \dd P\; 
    \sker{P}{P}{P_0}\,\e^{-2\pi i P^2}.
\end{equation}
This integral is rendered finite by an $i\epsilon$ rotation of the $P$ contour. This leaves the possibility that there are other `off-shell' topologies that are also accessible using the Virasoro TQFT. 

\paragraph{Off-shell topologies and spectral statistics:}
In this paper, we only considered the statistics of OPE coefficients, ignoring any statistical variants in the distribution of conformal dimensions. Relatedly, we only described on-shell solutions of the gravitational path integral, which in Euclidean signature are hyperbolic 3-manifolds. There is at least some evidence that the spectral statistics can only be captured by going off-shell. For example, in a series of papers \cite{Cotler:2020ugk,Cotler:2021cqa,Cotler:2022rud}
Cotler and Jensen constructed a family of constrained instantons that describe the universal `ramp' in the spectral form factor. This is a hallmark of quantum chaos and presents the CFT analogue of Berry’s diagonal projection, as shown in \cite{DiUbaldo:2023qli}. Although the status of such off-shell contributions and the rules how to incorporate them are unclear, it is evident that non-trivial spectral statistics have to be present. This is simply due to the fact that the microscopic spectrum is discrete, and not $\rho_0(P)$. The outstanding question is whether this discreteness can be detected by the (semi-classical) gravitational path integral, or whether it requires a more complete microscopic description like string theory.

\section*{Acknowledgements}

We would like to thank 
Alexandre Belin,
Scott Collier,
Gabriele Di Ubaldo,
Anatoly Dymarsky,
Lorenz Eberhardt,
Laura Foini,
Daniel Jafferis,
Henry Maxfield, 
Sridip Pal,
Eric Perlmutter,
Martin Sasieta and
Ioannis Tsiares 
for stimulating discussions. 
JdB, DL and BP are supported by the European
Research Council under the European Unions Seventh Framework Programme (FP7/2007-2013),
ERC Grant agreement ADG 834878.

\appendix

\section{Elementary crossing kernels}
\label{app:kernels}

In this appendix, we review the properties of the elementary crossing kernels of Ponsot and Teschner \cite{Ponsot:1999uf,Ponsot:2000mt}. There are four basic moves that act on conformal blocks:
\begin{itemize}[leftmargin=0.5cm]
    \item \textbf{Braiding.}
     The braiding move acts on conformal blocks as a multiplicative phase factor: 
    \begin{equation}\label{eq:app_braid}
        \vcenter{\hbox{
    \begin{tikzpicture}
    \node at (0,0) {\includegraphics{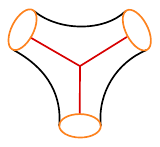}};
    \node at (-1.4,0.9) {\small$P_1$};
    \node at (0,-1.4) {\small$P_3$};
    \node at (1.4,0.9) {\small$P_2$};
    \end{tikzpicture} }} 
    = 
    \mathbb{B}_{P_3}^{P_1P_2} 
    \vcenter{\hbox{
    \begin{tikzpicture}
    \node at (0,0) {\includegraphics{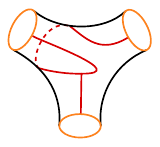}};
    \node at (-1.4,0.9) {\small$P_1$};
    \node at (0,-1.4) {\small$P_3$};
    \node at (1.4,0.9) {\small$P_2$};
    \end{tikzpicture} }} = \mathbb{B}^{P_3}_{P_1P_2}\vcenter{\hbox{
    \begin{tikzpicture}
    \node at (0,0) {\includegraphics{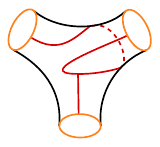}};
    \node at (-1.4,0.9) {\small$P_1$};
    \node at (0,-1.4) {\small$P_3$};
    \node at (1.4,0.9) {\small$P_2$};
    \end{tikzpicture} }}.
    \end{equation}
   In this paper, we define the orientation of this braiding phase as
    \begin{equation}
    \mathbb{B}_{P_1}^{P_2P_3} = e^{\pi i (h_1-h_2-h_3)},\quad \text{and}\quad \mathbb{B}^{P_1}_{P_2P_3} = (\mathbb{B}_{P_1}^{P_2P_3})^*
    \end{equation} 
    where the Liouville momentum is related to the conformal dimension via $h_i = \frac{c-1}{24}+P_i^2$.
    \item\textbf{Dehn twist.} 
    The composition of two braiding moves leads to a Dehn twist: 
    \begin{equation}
        \mathbb{T}_{P_2} \coloneqq \mathbb{B}^{P_1P_2}_{P_3}\mathbb{B}^{P_2P_3}_{P_1} = \e^{-2\pi i h_2}=
    \vcenter{\hbox{
    \begin{tikzpicture}
    \node at (0,0) {\includegraphics{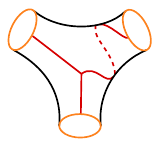}};
    \node at (-1.4,0.9) {\small$P_1$};
    \node at (0,-1.4) {\small$P_3$};
    \node at (1.4,0.9) {\small$P_2$};
    \end{tikzpicture} }}. 
    \end{equation}
    \item \textbf{Fusion.} The Virasoro fusion kernel acts on the four-holed sphere conformal block as
\begin{equation}\label{eq:app_fusion}
    \vcenter{\hbox{
    \begin{tikzpicture}
    \node at (0,0) {\includegraphics{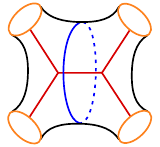}};
    \node at (1.35,-1.15) {\small$P_1$};
    \node at (1.35,1.15) {\small$P_2$};
    \node at (-1.35,1.15) {\small$P_3$};
    \node at (-1.35,-1.15) {\small$P_4$};
    \node at (0,0.3) {\small$P$};
    \end{tikzpicture} }} = \int_0^\infty \dd P' \,\fker{P}{P'}{P_2}{P_3}{P_4}{P_1} \vcenter{\hbox{
    \begin{tikzpicture}
    \node at (0,0) {\includegraphics{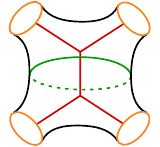}};
    \node at (1.35,-1.15) {\small$P_1$};
    \node at (1.35,1.15) {\small$P_2$};
    \node at (-1.35,1.15) {\small$P_3$};
    \node at (-1.35,-1.15) {\small$P_4$};
    \node at (0.3,0) {\small$P'$};
    \end{tikzpicture} }}
    \end{equation}
    Analytic forms for the fusion kernel can be found in \cite{Eberhardt:2023mrq}. It satisfies the symmetry properties
    \begin{equation}
        \fker{P}{P'}{P_1}{P_2}{P_3}{P_4} =
        \fker{P}{P'}{P_2}{P_1}{P_4}{P_3} =
        \fker{P}{P'}{P_4}{P_3}{P_2}{P_1}.
    \end{equation}
    \item \textbf{Modular kernel.} The modular S-kernel acts on the one-holed torus conformal block as 
    \begin{equation}\label{eq:app_mod}
    \vcenter{\hbox{
    \begin{tikzpicture}
    \node at (0,0) {\includegraphics{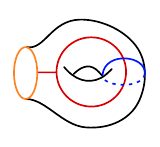}};
    \node at (-1.4,0) {\small$P_0$};
    \node at (1.4,0) {\small$P$};
    \end{tikzpicture} }} = \int_0^\infty \dd P' \,\mathbb{S}_{PP'}[P_0] \vcenter{\hbox{
    \begin{tikzpicture}
    \node at (0,0) {\includegraphics{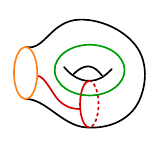}};
    \node at (-1.4,0) {\small$P_0$};
    \node at (0.8,-0.4) {\small $P'$};
    \end{tikzpicture} }}. 
    \end{equation}
    As with the fusion kernel, this kernel is known in closed form as an integral expression \cite{Eberhardt:2023mrq}. 
\end{itemize}

A general element $\gamma$ of the Moore-Seiberg group can be constructed by combining the above basic moves, acting on embedded surfaces $\Sigma_{0,4}$, $\Sigma_{1,1}$ and $\Sigma_{0,3}$ in a given pair-of-pants decomposition of the Riemann surface defining the conformal block.

\subsection{Special values}

The fusion and modular kernels are meromorphic functions of their arguments, with a known pole structure \cite{Collier:2018exn}. At special values, corresponding to degenerate representations of the Virasoro algebra, the kernels simplify. We will only need the expressions for the degeneration of some arguments to the identity $\bbi$, corresponding to $P=\frac{iQ}{2}$. Here we simply state the resulting formulas.
\begin{itemize}
    \item \textbf{Fusion kernel.} Take $\fker{P'}{P}{P_2}{P_3}{P_4}{P_1}$ and set $P'=\bbi$. Also take $P_1=P_2$ and $P_3=P_4$. Then the fusion kernel reduces to the $C_0$ formula:
     \begin{equation}\label{eq:FC_0}
        \fker{\bbi}{P}{P_1}{P_3}{P_3}{P_1} = \rho_0(P)C_0(P,P_1,P_3).
    \end{equation}
    Alternatively, consider the limit where one of the external momenta is taken to the identity, $P_4=\bbi$, and suppose $P_2=P'$. Then the fusion kernel evaluates to a Dirac delta function:
    \begin{equation}
        \fker{P'}{P}{P'}{P_3}{\bbi}{P_1} = \delta(P-P_1).
    \end{equation}
    This should be understood in the sense of distributions inside integral expressions.
    \item \textbf{Modular kernel.} Take $\sker{P'}{P}{P_0}$ and consider the limit $P_0=\bbi$. Then the modular kernel takes a simple form
     \begin{equation}
    \label{eq:specialS}
        \sker{P'}{P}{\bbi} = 2\sqrt{2}\cos(4\pi PP'). 
    \end{equation}
    which is simply the S-kernel for non-degenerate Virasoro characters. 
    
    If we simultaneously want to take $P_0\to \bbi$ and $P'\to\bbi$, we have to be more careful. Namely, the integral representation of the S-kernel only converges when $\frac{1}{2}\Re(\alpha_0)<\Re(\alpha')<\Re(Q - \frac{1}{2}\alpha_0)$. So to study this limit, one instead uses the shift relation satisfied by the S-kernel, see for instance \cite{Nemkov:2015zha} and \cite{Collier:2019weq}.
   Taking $P'= \frac{i}{2}(b^{-1}-\epsilon), \; P_0 = i(Q/2-\epsilon)$ with $\epsilon\rightarrow 0$ of this shift relation gives
    \begin{equation}
        \sker{\bbi}{P}{\bbi} = 4\sqrt{2}\sinh(2\pi b P)\sinh(2\pi b^{-1}P) = \rho_0(P).
    \end{equation}
    \item \textbf{Relation to the 6j symbol.} Combining the degenerate values of the fusion and modular kernel, resp. $C_0$ and $\rho_0$, with the general fusion kernel $\mathbb{F}$, gives a formula for the $6j$ symbol of the Virasoro algebra $\mathrm{Vir}\times\overline{\mathrm{Vir}}$:
    \begin{equation}\label{app:6j}
      \begin{Bmatrix}
        1 & 2 & 3 \\ 4 & 5 & 6
    \end{Bmatrix}_{6j} = \frac{1}{\sqrt{\mathsf{C}_{123}\mathsf{C}_{156}\mathsf{C}_{264}\mathsf{C}_{345}} }\left | C_0(P_1,P_5,P_6)C_0(P_3,P_{5},P_4) \frac{\fker{P_5}{P_2}{P_4}{P_6}{P_1}{P_3}}{\rho_0(P_2)}  \right |^2.
    \end{equation} 
    Here we have included the normalization factors $\mathsf{C}_{123} = |C_0(P_1,P_2,P_3)|^2$, and the absolute value squared signifies the product with the antiholomorphic counterpart. One reason to consider the $6j$ symbol instead of the fusion kernel, is that the $6j$ symbol enjoys tetrahedral symmetry, as indicated by the diagram in Figure \ref{6jfig:b}.
\end{itemize}

\subsection{Integral identities from consistency conditions}\label{app:identities_WH}

We make frequent use of integral identities satisfied by the modular and fusion kernels. These identities can be derived from the Moore-Seiberg consistency conditions, as was extensively reviewed in \cite{Eberhardt:2023mrq}. Here we quote some of these identities for completeness.
\begin{itemize}
    \item \textbf{Pentagon equation:} \begin{equation}
    \label{eq:penta}
    \int_0^\infty \dd P\, \fker{s_1}{P}{P_2}{P_3}{s_2}{P_1}\fker{s_2}{s_3}{P}{P_4}{P_5}{P_1}\fker{P}{s_4}{P_3}{P_4}{s_3}{P_2} =  \fker{s_1}{s_3}{P_2}{s_4}{P_5}{P_1}\fker{s_2}{s_4}{P_3}{P_4}{P_5}{s_1}
\end{equation}
This identity is derived by imposing crossing symmetry of the five-punctured sphere. A consequence of this identity when $s_2 = \bbi$, $s_1=P_3$, $P_4=P_5$ is the tetrahedral symmetry of $\bbf$:\vspace{1mm}
\begin{equation}\label{eq:tetrahedral_sym}
    \fker{\bbi}{s_3}{P_1}{P_4}{P_4}{P_1}
    \fker{P_1}{s_4}{P_3}{P_4}{s_3}{P_2} =
    \fker{P_3}{s_3}{P_2}{s_4}{P_4}{P_1}
    \fker{\bbi}{s_4}{P_3}{P_4}{P_4}{P_3}.
\end{equation}
\item \textbf{Hexagon equation:}
\begin{equation}
    \label{eq:hexa}
    \int_0^\infty \dd P \,\mathbb{B}_{s_1}^{P_3P_4 }\fker{s_1}{P}{P_2}{P_4}{P_3}{P_1} \mathbb{B}_{P}^{P_2P_4} \fker{P}{s_2}{P_1}{P_4}{P_2}{P_3}\mathbb{B}_{s_2}^{P_1P_4} = \fker{s_1}{s_2}{P_2}{P_3}{P_4}{P_1}
\end{equation}
This equation follows from a sequence of braiding and fusion moves of the four-punctured sphere conformal block.
\item \textbf{Swapping internal indices:}
\begin{equation}\label{app:swap}
    \frac{\fker{P}{P'}{P_2}{P_3}{P_4}{P_1}}{\rho_0(P')C_0(P_1,P_4,P')C_0(P_2,P_3,P')} 
 = \frac{\fker{P'}{P}{P_4}{P_3}{P_2}{P_1}}{\rho_0(P)C_0(P_1,P_2,P)C_0(P_3,P_4,P)}.
\end{equation}
This formula follows from a degeneration of the pentagon equation, where one of the external operators is taken to be the identity.
\item \textbf{Relation between $\mathbb{S}$ and $\mathbb{F}$:}
    \begin{equation}
    \label{eq:relgen1}
        \int_0^\infty \dd P \,\fker{\bbi}{P}{P_2}{P_1}{P_1}{P_2} (\mathbb{B}^P_{P_1P_2})^2\fker{P}{P'}{P_2}{P_2}{P_1}{P_1} = \frac{\fker{\bbi}{P'}{P_1}{P_1}{P_1}{P_1}\mathbb{S}_{P_1P_2}[P']}{\rho_0(P_2)}.
    \end{equation}
    This is derived from a crossing equation on the twice-punctured torus. 
\end{itemize}
By combining the hexagon and pentagon equation, we can also evaluate the integral over three fusion kernels and one braid. Namely,
\begin{align}
   &\int_0^\infty \dd P\,\mathbb{B}^P_{P_\clo P_\clo}\,\fker{\bbi}{P}{P_\clo}{P_\clo}{P_\clo}{P_\clo}\,\fker{P_\clo}{P_4}{P_\clo}{P_4}{P_4}{P}\, \fker{P}{P_4}{P_\clo}{P_4}{P_4}{P_\clo} \\[1em] \notag &\stackrel{\mathrm{hexagon}}{=}\int_0^\infty \dd P\dd P'\,\mathbb{B}^P_{P_\clo P_\clo}\,\fker{\bbi}{P}{P_\clo}{P_\clo}{P_\clo}{P_\clo}\,\fker{P_\clo}{P_4}{P_\clo}{P_4}{P_4}{P}\, \mathbb{B}_{P}^{P_\clo P_\clo} \fker{P}{P'}{P_\clo}{P_4}{P_4}{P_\clo} \mathbb{B}_{P'}^{P_4P_\clo}\,\fker{P'}{P_4}{P_4}{P_\clo}{P_4}{P_\clo}\,\mathbb{B}_{P_4}^{P_4 P_\clo} \\[1em]\notag
   &\stackrel{\mathrm{pentagon}}{=} \int_0^\infty \dd P' \,\fker{\bbi}{P_4}{P_\clo}{P'}{P_4}{P_\clo}\fker{P_\clo}{P'}{P_\clo}{P_4}{P_4}{\bbi}\mathbb{B}_{P'}^{P_4P_\clo}\,\fker{P'}{P_4}{P_4}{P_\clo}{P_4}{P_\clo}\,\mathbb{B}_{P_4}^{P_4 P_\clo}.
\end{align}
We can now use the fact that the fusion kernel simplifies to a delta function if one of the external operators is the identity,
\begin{equation}
    \fker{P_\clo}{P'}{P_\clo}{P_4}{P_4}{\bbi} = \delta(P'-P_4),
\end{equation}
as well as the expression for the braiding phase
$
    (\mathbb{B}_{P_4}^{P_4 P_\clo})^2 = e^{-2\pi i\, h_\clo},
$ 
to derive the following integral identity, which is used at several points in the main text:
\begin{itemize}
\item \textbf{Penta-hexagon equation:}
\begin{equation}\label{app:hexapenta}
   \int_0^\infty \dd P\,\mathbb{B}^P_{P_\clo P_\clo}\,\fker{\bbi}{P}{P_\clo}{P_\clo}{P_\clo}{P_\clo}\,\fker{P_\clo}{P_4}{P_\clo}{P_4}{P_4}{P}\, \fker{P}{P_4}{P_\clo}{P_4}{P_4}{P_\clo} = e^{-2\pi i h_\clo} \fker{\bbi}{P_4}{P_\clo}{P_4}{P_4}{P_\clo}\,\fker{P_4}{P_4}{P_4}{P_\clo}{P_4}{P_\clo}.\quad
\end{equation}
\end{itemize}

\subsection{Asymptotic formulas for the modular and fusion kernels}\label{app:asymptotic}

Lastly, we collect several results regarding the asymptotic behaviour of the fusion and modular kernels when a subset of parameters is taken to be heavy. Many of these results can be found in the literature, see for example \cite{Collier:2018exn,Collier:2019weq, Belin:2021ryy,Anous:2021caj}. In the formulas, we will use the convention that $P\rightarrow\infty$, the variables $x_i$ are fixed and positive, and the rest of the parameters are kept fixed in the limit.

\begin{itemize}
    \item \textbf{Fusion kernel.} The formulas that are relevant for this paper are: 
    \begin{align}
    &\log C_0(x_1 P, x_2 P, x_3 P) =  \bigg(
    -4 \sum_{i=1}^3 x_i^2 \log(2 x_i)\, + 
    \\ \notag & \sum_{\epsilon_2,\epsilon_3 = \pm 1} (x_1+\epsilon_2 x_2+\epsilon_3 x_3)^2 \log |x_1+\epsilon_2 x_2 + \epsilon_3 x_3| \bigg) P^2 - \pi Q (x_1\!+\!x_2\!+\!x_3) P + \order{\log P},\\[2em]
    &\log C_0(P_0, x_1 P, x_2 P) = \Big(-4x_1^2\log(2x_1)-4x_2^2\log(2x_2)  \,+ \\[1em]\nonumber  &\qquad 2(x_1+x_2)^2\log(x_1+x_2)+2(x_1-x_2)^2\log(x_1-x_2)\Big)P^2
    -\pi Q(x_1+x_2)P +\order{\log P},
    \\[2em]
     &\log \fker{P_t}{P + \delta_s}{P+\delta_1}{P_2}{P_3}{P + \delta_4} =
     \big[ h_2+h_3-h_t-2(\delta_1-\delta_s)(\delta_4-\delta_s)\big]\log P + \order{1}\\[2em]
     &\log \fker{P\! + \delta_t,}{P\! + \delta_s}{P\! + \!\delta_1}{P_2}{P\! +\!\delta_3}{P_4} = 2(\delta_1+\delta_3-\delta_s-\delta_t)(\delta_s-\delta_t)\log P + \order{1},\\[2em]
     &\log \fker{P\!+\delta_t,}{P\!+\delta_s}{P\!+\!\delta_1}{P\!+\!\delta_2}{P\!+\!\delta_3}{P\!+\!\delta_4} = 2(\delta_s-\delta_t)P \log \frac{27}{16}+\order{1}. 
    \end{align}
    \item \vspace{1mm}\textbf{Modular kernel.} The relevant formulas are \vspace{2mm}
    \begin{align}
    &\log \sker{P'}{P}{P_0} = 2\pi (Q - 2\alpha') P + h_0 \log P + \order{1},\\[2em]
    \label{eq:shierarchy}
     &\log \sker{P'}{P}{x P} = 
    \frac{1}{2}  \Big(-2 x^2 \log (x)+(x-2)^2 \log |x-2|+(x+2)^2 \log (x+2)
    \\[1em] \nonumber
    & \qquad
    -8 \log (2)\Big)P^2 + \pi  (2-x)(Q-2\alpha')P + \left( \frac{7Q^2+1}{6}-4h'\right)\log P +\order{1}, \quad 0<x\leq 2.
\end{align}

\end{itemize}
\pagebreak
\bibliographystyle{ytphys}
\bibliography{ref}
\end{document}